\documentclass[letterpaper,onecolumn,11pt,accepted=2023-03-13]{quantumarticle}
\pdfoutput=1

\usepackage{graphicx,epstopdf}% Include figure files
\usepackage{dcolumn}% Align table columns on decimal point
\usepackage{bm}% bold math
\usepackage{color, xcolor, colortbl}
\usepackage{geometry}
\usepackage{amsmath,amssymb,amsthm}
\usepackage{algorithm}
\usepackage{algorithmic}
\usepackage{bm}
\usepackage[caption=false]{subfig}
\usepackage{appendix}
\usepackage{multirow}
\usepackage{braket}
\usepackage[english]{babel}
\usepackage{hyperref}
\usepackage[capitalize]{cleveref}
\usepackage[sort&compress,square,numbers]{natbib}
\usepackage{adjustbox}
\usepackage{xspace}
\usepackage[roman]{complexity}
\usepackage{qcircuit}
\usepackage{multirow}
\graphicspath{{figure/}}

\usepackage{multirow}
\usepackage{cellspace} 
\usepackage{makecell} 
\setcellgapes{3pt}

\usepackage{pifont}

\renewcommand{\Re}{\operatorname{Re}}

\newcommand{\Tr}{\operatorname{Tr}}

\newcommand{\mc}[1]{\mathcal{#1}}

\newcommand{\wt}[1]{\widetilde{#1}}

\newcommand{\abs}[1]{\left\lvert#1\right\rvert}
\newcommand{\norm}[1]{\left\lVert#1\right\rVert}

\newcommand{\ud}{\,\mathrm{d}}
\newcommand{\Or}{\mathcal{O}}

\newcommand{\RR}{\mathbb{R}}
\newcommand{\CC}{\mathbb{C}}

\newcommand{\rd}{\mathrm{d}}
\newcommand{\dd}{\mathrm{d}}

\newtheorem{thm}{\protect\theoremname}
\theoremstyle{plain}
\newtheorem{lem}[thm]{\protect\lemmaname}
\theoremstyle{plain}

\theoremstyle{plain}
\newtheorem*{lem*}{\protect\lemmaname}
\theoremstyle{plain}

\theoremstyle{plain}
\newtheorem{cor}[thm]{\protect\corollaryname}

\newtheorem{defn}[thm]{\protect\definitionname}

\providecommand{\definitionname}{Definition}
\providecommand{\assumptionname}{Assumption}
\providecommand{\corollaryname}{Corollary}
\providecommand{\lemmaname}{Lemma}
\providecommand{\propositionname}{Proposition}
\providecommand{\remarkname}{Remark}
\providecommand{\theoremname}{Theorem}

\begin{document}

%%%%%%%%%%%%%%%%%%%%%%%%%%%%%%%%%%%%%%%%%%%%%%%%%%%%%%%%%%%%%%%%%%%%%%%%%%%%%%
\title{Time-marching based quantum solvers for time-dependent linear differential equations}% Force line breaks with \\

\author{Di Fang}
\affiliation{Department of Mathematics, University of California, Berkeley, CA 94720, USA}
\affiliation{Simons Institute for the Theory of Computing, University of California, Berkeley, CA 94720, USA}
\affiliation{Challenge Institute for Quantum Computation, University of California, Berkeley, CA 94720, USA}

\author{Lin Lin}
\affiliation{Department of Mathematics, University of California, Berkeley, CA 94720, USA}
\affiliation{Applied Mathematics and Computational Research Division, Lawrence Berkeley National Laboratory, Berkeley, CA 94720, USA}
\affiliation{Challenge Institute for Quantum Computation, University of California, Berkeley, CA 94720, USA}

\author{Yu Tong}
\affiliation{Institute for Quantum Information and Matter, California Institute of Technology, Pasadena, CA 91125, USA}
\affiliation{Department of Mathematics, University of California, Berkeley, CA 94720, USA}

% \date{}

\begin{abstract}
The time-marching strategy, which propagates the solution from one time step to the next, is a natural strategy for solving time-dependent differential equations on classical computers, as well as for solving the Hamiltonian simulation problem on quantum computers. 
For more general homogeneous linear differential equations $\frac{\mathrm{d}}{\mathrm{d} t} |\psi(t)\rangle = A(t) |\psi(t)\rangle, |\psi(0)\rangle = |\psi_0\rangle$, a time-marching based quantum solver can suffer from exponentially vanishing success probability with respect to the number of time steps and is thus considered impractical. We solve this problem by repeatedly invoking a technique called the uniform singular value amplification, and the overall success probability can be lower bounded by a quantity that is independent of the number of time steps. The success probability can be further improved using a compression gadget lemma.
This provides a path of designing quantum differential equation solvers that is alternative to those based on quantum linear systems algorithms (QLSA).
We demonstrate the performance of the time-marching strategy with a high-order integrator based on the truncated Dyson series. The complexity of the algorithm depends linearly on the amplification ratio, which quantifies the deviation from a unitary dynamics. We prove that the linear dependence on the amplification ratio attains the query complexity lower bound and thus cannot be improved in the worst case. This algorithm also surpasses existing QLSA based solvers in three aspects: (1) $A(t)$ does not need to be diagonalizable. (2) $A(t)$ can be non-smooth, and is only of bounded variation. (3) It can use fewer queries to the initial state $|\psi_0\rangle$. 
Finally, we demonstrate the time-marching strategy with a first-order truncated Magnus series, which simplifies the implementation compared to high-order truncated Dyson series approach, while retaining the aforementioned benefits. Our analysis also raises some open questions concerning the differences between time-marching and QLSA based methods for solving differential equations.
\end{abstract}

\maketitle

\tableofcontents

%%%%%%%%%%%%%%%%%%%%%%%%%%%%%%%%%%%%%%%%%%%%%%%%%%%%%%%%%%%%%%%%%%%%%%%%%%%%%%

\section{Introduction} \label{sec:intro}

In modern science and engineering, mathematical descriptions of ``real-world" problems often lead to differential equations, which play a prominent role in many disciplines such as physics, chemistry, biology and economics. 
% Differential equations are among the most important tools to describe and model problems in modern science and engineering 
% Mathematical description of  with wide applications to. 
Efficient simulation of large scale differential equations has therefore served as one of the core tasks in many scientific applications. 
Recent advances of quantum algorithms indicate that quantum computers may significantly accelerate the simulation of such differential equations, particularly for those defined in high dimensional spaces. 
Let $N$ be the number of degrees of freedom (e.g., the number of discretized points in a high dimensional space) which can be very large. For certain tasks, quantum computers can store and manipulate vectors of size $N$ at a cost that scales only as $\polylog(N)$, which leads to potentially significant advantages over classical computers.

In this paper, we focus on the initial value problem of the system of homogeneous linear ordinary differential equations (ODEs) 
\begin{equation} \label{eq:ode_general}
    \frac{\rd}{\rd t} \ket{\psi(t)} = A(t) \ket{\psi(t)}, \quad \ket{\psi(0)} = \ket{\psi_0},
\end{equation}
where $t \in [0, T] \subset \mathbb{R}^+$ is the independent variable with $T$ as the final time, the vector $ \ket{\psi(t)} \in \mathbb{C}^N$ is the dependent variable, the coefficient matrix $A(t) \in \mathbb{C}^{N\times N}$ is a matrix-valued function in $t$, and $\ket{\psi_0} \in \mathbb{C}^N$ is the initial condition. We assume the differential equation has a well-posed solution, which can be implied by, e.g., $A(t)$ being piecewise continuous. % Without loss of generality, we assume $A(t)$ to be piecewise continuous 
% \YT{I think we probably don't need this assumption? And it's not used anywhere?}. 
The coefficients can have jump discontinuity and are not required to be smooth in $t$. More precisely, we only need to assume that $A(t)$ is of bounded variation, i.e., the total variation $V_0^T(A)$ (see \cref{eq:defn_total_variation_first} for definition) is finite.
The task for quantum differential equation solvers is to prepare a quantum state that is proportional to the final solution $\ket{\psi(T)}$ with certain precision.

One prominent example is the Hamiltonian simulation problem, which is a homogeneous linear differential equation governed by $A(t)=-iH(t)$ and $H(t)$ is a Hermitian matrix. 
% This type of dynamics is unitary that matches the nature of quantum mechanics. 
Recent years have witnessed remarkable progresses on designing new algorithms as well as establishing improved theoretical complexity estimate of existing algorithms for both time-independent Hamiltonian simulation~\cite{BerryAhokasCleveSanders2007,BerryChilds2012,BerryCleveGharibian2014,BerryChildsCleveEtAl2014,BerryChildsCleveEtAl2015,BerryChildsKothari2015,LowChuang2017,ChildsMaslovNamEtAl2018,LowWiebe2019,ChildsOstranderSu2019,Campbell2019,Low2019,ChildsSu2019,ChildsSuTranEtAl2020,ChenHuangKuengEtAl2020,SahinogluSomma2020,TongAlbertEtAl2021} and time-dependent ones~\cite{KieferovaSchererBerry2019,LowWiebe2019,BerryChildsSuEtAl2020,WeckerHastingsWiebeEtAl2015,WiebeBerryHoyerSanders2010,AnFangLin2021,AnFangLin2022,ChenKalevHen2021}. These quantum algorithms can be applied, when the underlying dynamics is unitary or can be converted into a unitary dynamics (see e.g., ~\cite{CostaJordanOstrander2019}).
Compared to the Hamiltonian simulation problem whose dynamics is unitary, quantum algorithms for the general linear differential equations are considerably less explored. 
Such algorithms produce a quantum state representing the solution of the differential equation. This is different from having the solution stored in classical memory, but we can still extract information from the quantum state that may not be efficiently obtainable from classical computation (see e.g., ~\cite{ChildsLiu2020}).
%Though one can only extract limited outputs efficiently from the quantum solution, such information may not be efficiently obtained classically (see e.g., ~\cite{ChildsLiu2020,AnFangEtAl2022}). \YT{I don't see why we need this sentence here.}
Unless otherwise specified, we do not consider the special case when $A(t)=A$ is time-independent, in which setting we may directly encode the propagator (i.e., the matrix function $e^{AT}$) using the quantum singular value transformation (QSVT)~\cite{GilyenSuLowEtAl2019} when $A$ is Hermitian or anti-Hermitian, or using a contour integral based strategy for a general $A$ (see \cite{TakahiraOhashiEtAl2021,TongAnWiebeEtAl2021}, as well as \cref{sec:ea_algorithm}).

% Many classical applications, however, involve more general differential equation whose dynamics are not necessarily unitary. 

The \textit{time-marching} strategy is a natural strategy solving time-dependent differential equations, and is adopted by nearly all classical differential equation solvers (see e.g.,~\cite{HairerNorsettWanner1987,HairerLubichWanner2006}). The idea is to divide the entire time interval $[0, T]$ into a number of short line segments separated by the temporal mesh points $0 = t_0 < t_1 < \cdots < t_{L} = T$, and to calculate the solution information at the next time step using the solution information from previous $k$ time steps. 
Methods with $k=1$ are called one-step methods (e.g., Runge-Kutta methods). Methods with $k>1$ are called multi-step methods (e.g., Adams methods).
Multi-step methods require multiple copies of the solution from previous steps, and cannot be directly implemented on quantum computers due to the obstruction of the no-cloning theorem. 
For non-unitary dynamics, even one-step methods lead to severe challenges in the design of quantum algorithms, mainly due to potentially diminishing success probability. In \cref{sec:diff_direct}, we use an illustrative example with the simplest one-step integrator (the forward Euler method) to demonstrate such challenges.
As a result, the time-marching strategy has not been viewed as a practical route for designing quantum differential equation solvers beyond the Hamiltonian simulation problem.

% 
% 
%  At each of the time step, one may use a number of , which is not an issue classically but can hinder the algorithm performance quantumly due to the no-cloning theorem.   For example, suppose at each time step one needs 2 copies of the solution at the previous time step, quantumly one needs to repeat the efforts in preparing the solution at the previous step, leading to 
% % $\Or(2^L)$ re-calculation of the first time-step and 
% a use of $2^L$ copies of the initial state and $\Or(2^L)$ cost in general.
% These copies are typically needed for amplitude amplification procedures at each time step to boost the success probability, when the implementation of the algorithm is not exactly unitary.  

% Let us examine the difficulty due to non-unitarity in a more formal fashion. Essentially, we want to implement a short-time evolution operator $\Xi_l$ at each time step $l$, and $\Xi_l$ can be non-unitary. Suppose we start from a quantum state $\ket{\psi(0)}$, then under the ideal scenario where we have exact eigendecomposition of  $\ket{\psi(T)}$ can be prepared with a  

% \vspace{1em}
% \noindent\textbf{}

\subsection{Related works}\label{sec:related}

The prevailing strategy for designing quantum differential equation solvers to date is to construct a large linear system recording states during the entire history of the evolution, and then to apply the quantum linear systems algorithms (QLSA)~\cite{HarrowHassidimLloyd2009,Ambainis2012,ChildsKothariSomma2017,SubasiSommaOrsucci2019,LinTong2019,AnLin2019,CostaAnSandersEtAl2021} to solve the resulting linear systems of equations.  
One may perform certain amplification procedures to boost the success probability of getting the final solution.
This QLSA-based strategy was first proposed by Berry  \cite{Berry2014}, which successfully avoids the pitfall in \cref{sec:diff_direct}. It has been adopted by various quantum linear differential equation solvers~\cite{BerryChildsOstranderWang2017,LindenMontanaroShao2020,ChildsLiu2020,Krovi2022}, and has been applied to solve nonlinear differential equations using linearization techniques~\cite{LiuKoldenEtAl2021,AnFangEtAl2022,DodinStartsev2020,Joseph2020,LloydDePalmaEtAl2020,EngelSmithParker2021,TronciJoseph2021,JinLiu2022}.

The work \cite{Berry2014} uses multi-step integrators, and the analysis is applicable to time-independent $A$ that is diagonalizable as $A=VDV^{-1}$ with eigenvalues $\lambda_j=D_{jj}$ satisfying
\begin{equation}
    \lvert \arg(-\lambda_j)\rvert \leq \theta_0, \quad 0< \theta_0<\pi/2, \quad \forall j.
\label{eqn:berry_eig_assumption}
\end{equation}
The query complexity of the algorithms scales polynomially in $T$, the spectral norm $\norm{A}$, the inverse precision $\epsilon^{-1}$, and the condition number of the eigenvector matrix $\kappa_{V}:=\norm{V}\norm{V^{-1}}$.
For time-independent $A$, the work \cite{BerryChildsOstranderWang2017} combines a QLSA-based solver with the truncated Taylor series method, and the assumption in \cref{eqn:berry_eig_assumption} was relaxed to $A$ being dissipative (i.e., $\Re(\lambda_j)\le 0, \forall j$). The complexity of this algorithm is also improved to be $\wt{\Or}(T d \norm{A}\kappa_V q \polylog(\epsilon^{-1}))$, where $d$ is the sparsity of the matrix $A$, $\kappa_V$ is the condition number of $V$, and $q= \sup_{t\in [0,T]} \norm{\ket{\psi(t)}}/\norm{\ket{\psi(T)}}$.
For time-dependent linear differential equations, \cite{ChildsLiu2020} combines a QLSA-based solver with a Chebyshev pseudospectral method. It assumes that $A(t)$ is diagonalizable as $A(t)=V(t)D(t)V(t)^{-1}$ and dissipative for all $t\in[0,T]$, and the underlying solution is sufficiently smooth in $t$.
The complexity of the algorithm is $\wt{\Or}(T\alpha \kappa_V q d \polylog(g' g^{-1}\epsilon^{-1} ))$, where $d$ is the sparsity of $A(t)$, $\alpha=\sup_{t\in [0,T]} \norm{A(t)}$, $\kappa_V=\sup_{t\in [0,T]} \norm{\kappa_{V}(t)}$, $g = \norm{\ket{\psi(T)}}$, $g' = \max_{t\in [0, T]} \max_{n \in \mathbb{N}} \norm{\ket{\psi^{(n+1)}(t)}}$, and $q= \sup_{t\in [0,T]} \norm{\ket{\psi(t)}}/\norm{\ket{\psi(T)}}$.
% , and $q= \sup_{t\in [0,T]} \norm{\ket{\psi(t)}}/\norm{\ket{\psi(T)}}$.
The smoothness of the solution implicitly requires that $A(t)$ should be sufficiently smooth (e.g., analytic in $t$).
When the coefficient matrix $A(t)$ is not sufficiently smooth, the polylogarithmic dependence on $g'$ no longer holds, and the complexity of the algorithm depends polynomially on $\epsilon^{-1}$.
% 
%  when . This imposes implicit restriction on  of high regularity (such as analyticity). As other globally approximation techniques, the method is prone to the Gibbs phenomena when discontinuity kicks in. 
% % \LL{ The spectral method cannot be used when $A(t)$ is piecewise smooth (with $O(1)$ number of pieces)?} \YT{The spectral method still works when there are $\Or(1)$ discontinuities, but there could easily be $\Or(N)$ discontinuities because there are this many coefficients in the matrix. Also we could consider singularity of the form $\sin(1/(t-t^*))$ and there isn't a good way for spectral method to deal with it.} 
% The Gibbs phenomena, widely studied in classical numerical analysis, describes the convergence quality issue in recovering point values of a function from its expansion coefficients: near a jump discontinuity, the finite series exhibits an overshoot problem that does not diminish as one increases the number of terms in the series; away from the discontinuity the convergence is slow. It is worth noting that the time-marching strategy is essentially a local approximation that does not suffer from the Gibbs phenomena. 

% involves both the fact that global approximations through a finite series exhibit an overshoot problem that does not go away as one increases the number of term at a jump discontinuity

In all the analysis above, the complexity depends explicitly on the condition number of the eigenvector matrix $\kappa_{V}$, which can be difficult to estimate and may lead to significant overestimation of the cost.
The recent work by Krovi \cite{Krovi2022} replaces the dependence on $\kappa_V$ by the dependence on $\sup_{t\in[0,T]} \norm{e^{At}}$. This is a significant improvement due to the inequality
\[
\sup_{t\in[0,T]} \norm{e^{At}}\le \sup_{t\in[0,T]} \norm{e^{Dt}} \kappa_V.
\]
For instance, consider 
\begin{equation}
A=\begin{pmatrix}
1 & 1\\
0 & 1+\delta
\end{pmatrix}=VDV^{-1}, \quad \mbox{with} \quad V=\begin{pmatrix}
1 & \frac{1}{\delta}\\
0 & 1
\end{pmatrix}, \quad D=\begin{pmatrix}
1 & 0 \\
0 & 1+\delta
\end{pmatrix}.
\end{equation}
Then $\|e^{A}\|=\Or(1)$, but $\kappa_V$ is $\Omega(\delta^{-2})$ which diverges as $\delta\to 0$.
However, the technique in~\cite{Krovi2022} is only applicable when $A$ is time-independent.

% \vspace{1em}
% \noindent\textbf{:}

\subsection{Contribution}

In this work, we propose that the time-marching method \textit{can} become an efficient strategy for solving linear differential equations. 
Our main technical tool (\cref{thm:short_time_to_long_time}) is a method to implement a sequence of non-unitary operations without incurring an exponential overhead. 
\cref{thm:short_time_to_long_time} has two main ingredients. The first is the uniform singular value amplification procedure, which was first developed in \cite{LowChuang2017a} and was refined in~\cite{GilyenSuLowEtAl2019}.
It allows us to amplify the success probability in each time step in a way that is oblivious to the quantum state. 
We find that in our context, the degree of the polynomial used by the uniform singular value amplification procedure in~\cite{GilyenSuLowEtAl2019} exhibits a very large preconstant. 
We use a convex optimization based method to reduce the preconstant by orders of magnitude.
\cref{thm:short_time_to_long_time} also proposes a useful tool called the compression gadget (which improves upon the result developed in \cite{LowWiebe2019}) to coherently combine short-time evolution operators into a long-time one without duplicating ancilla qubits.
Combined with amplitude amplification~\cite{BrassardHoyerMoscaEtAl2002}, this leads to a further quadratic speedup in terms of the query complexity.

Consider a temporal mesh $0 = t_0 < t_1 < \cdots < t_{L}=T$. 
Using the time-marching strategy and the high-order truncated Dyson series algorithms~\cite{BerryChildsCleveEtAl2015,KieferovaSchererBerry2019,LowWiebe2019} for short time evolution, we propose an algorithm to solve \cref{eq:ode_general}, which requires $\wt{\Or}(Q(\alpha T)^2\log(\epsilon^{-1}))$ queries to the coefficient matrix $A(t)$, and $\Or(Q)$ queries to the initial state  $\ket{\psi_0}$ (\cref{thm:truncated_dyson_alg}). 
Here $\alpha=\sup_{t\in [0,T]} \norm{A(t)}$, and
\begin{equation}
    \label{eq:defn_Q_intro}
    Q = \frac{\prod_{l=1}^L\|\mathcal{T}e^{\int_{t_{l-1}}^{t_l}A(t)\dd t}\|}{\|\ket{\psi(T)}\|},
\end{equation}
as a central quantity of this work, is the \textit{amplification ratio}. It quantifies the non-unitarity of the dynamics (e.g., $Q=1$ for unitary dynamics).

Compared to the state-of-the-art results based on the QLSA in \cite{ChildsLiu2020} and \cite{Krovi2022}, our algorithm has several advantages, which we summarize in Table \ref{tab:comparison}. First, our method does not require the diagonalizability of $A(t)$, and the complexity is independent of the condition number $\kappa_V$. This generalizes the result of \cite{Krovi2022} to equations with time-dependent matrix coefficients.

Second, our algorithm also has lower regularity requirement for the coefficient matrix $A(t)$ and the solution $\ket{\psi(t)}$. In \cite{ChildsLiu2020} the query complexity depends on the high-order derivatives of the solution ($g' = \max_{t\in [0, T]} \max_{n \in \mathbb{N}} \norm{\ket{\psi^{(n+1)}(t)}}$ in \cite[Theorem 1]{ChildsLiu2020}), and if the solution is not smooth, the spectral method loses the desirable exponential accuracy. Remarkably, in our method, exponential accuracy is achieved even when the coefficient matrix $A(t)$ is not smooth, thus allowing $\ket{\psi(t)}$ to be non-smooth. 
The results in both \cref{thm:truncated_dyson_alg} and \cref{thm:qhop_alg} are insensitive to the roughness in the coefficients $A(t)$.
Note that $A(t)$ being of bounded variation is a very weak regularity condition, and in particular $V_0^T(A)= \int_0^T \|A'(t)\|\dd t$ when $A$ is differentiable. This also allows for the existence of jump discontinuities in the coefficients.
%, which results in the numerically disastrous Gibbs phenomenon in the spectral method used in \cite{ChildsLiu2020}.

Third, our algorithm may use fewer queries to the initial state, which is advantageous if the initial state preparation is an important factor.  
To simplify the discussion, let us focus on the dependence on $T$ and assume all other quantities such as $\alpha, Q$ to be constants. 
In order to achieve the scaling in \cite[Theorem 1]{ChildsLiu2020} (for other QLSA based differential equation solvers discussed in \cref{sec:related}, the situation is similar), the quantum solver must employ a quantum linear system solver with near optimal query complexities. 
In other words, the complexity of the linear system solver should scale as $\wt{\Or}(\kappa \polylog(\epsilon^{-1}))$, and $\kappa$ is the condition number of the linear system. Moreover, the near optimal quantum algorithms also need to query the initial state for $\wt{\Or}(\kappa)$ times.
Such a dependence is most clearly seen from the perspective of adiabatic based near-optimal quantum linear system solvers~\cite{AnLin2019,LinTong2019,CostaAnSandersEtAl2021}. This is because the construction of the adiabatic Hamiltonian corresponding to the linear system uses the initial state, and thus each query to the adiabatic Hamiltonian also queries the initial state. 
As a result, the quantum differential equation solvers in both \cite{ChildsLiu2020,Krovi2022} need to query the initial state for $\wt{\Or}(\kappa)=\wt{\Or}(T)$ times. The relation between $\kappa$ and $T$ is rooted in the no-fast-forwarding theorem and cannot be generally improved. Our time-marching based algorithm does not rely on such adiabatic constructions, and the query complexity to the initial states does not explicitly depend on $T$. In the case where the time evolution is almost unitary, i.e., $Q=\Or(1)$, and where $T$ is large, this feature can offer significant advantage.

Our method (\cref{thm:truncated_dyson_alg}) also has two drawbacks. The first is that the $T$ dependence in the number of queries to $A(t)$ is sub-optimal. The direct reason is that the uniform singular value amplification procedure becomes increasingly costly as $T$ increases. 
The second is that the cost of our algorithm depends on $Q$ defined in \cref{eq:defn_Q_intro}, while previous QLSA based algorithms depends on $q = \max_{t\in[0,T]} \frac{\norm{\ket{\psi(t)}}}{\norm{\ket{\psi(T)}}}$, which satisfies $q\le Q$. We prove in \cref{thm:lower_bound_Q_dep} that the $\Or(Q)$ dependence attains the query lower bound and cannot be improved in the worst case. 
There exists instances that $Q$ can significantly (even exponentially) overestimate $q$, as will be discussed at the end of \cref{sec:optimality}. However, $q \kappa_V$ and $Q$ may not be directly related in general.

Finally, the implementation of the high-order truncated Dyson series algorithm requires complicated quantum control logic for handling time-ordering operators.
To simplify the implementation, we combine the time-marching strategy with a first-order truncated Magnus series, whose implementation does not require the complex time-clocking quantum control logics.  Though the cost of the resulting algorithm depends on higher powers of  $T$ and $\epsilon^{-1}$ comparing to the high-order integrators, it retains the aforementioned advantages compared to QLSA based solvers. Interestingly, this algorithm also exhibits a commutator scaling for differential equations in the high precision limit (see \cref{thm:qhop_alg}), which can be desirable when the norm of the commutator ${\alpha}_\mathrm{comm} = \sup_{s,\tau \in [0,T]}\norm{[A(s), A(\tau)]}$ is much smaller than $\alpha$.

% number of queries to the coefficient matrix is $\wt{\Or}\left((\overline{\alpha}T)^4  Q^3/\epsilon^2\right)$  where $\overline{\alpha} = \frac{1}{T}\int_0^T \norm{A(s)}\ud s$ is called the $L^1$-norm of the time-dependent matrix $A(t)$, and the number of queries to the initial state is $\Or(Q^2)$ (see \cref{thm:qhop_alg}).

% The number of queries to the coefficient matrix is $\wt{\Or}\left((\overline{\alpha}T)^3  Q^2/\epsilon\right)$, where $\overline{\alpha} = \frac{1}{T}\int_0^T \norm{A(s)}\ud s$ is called the $L^1$-norm of the time-dependent matrix $A(t)$, and the number of queries to the initial state is $\Or(Q)$ (see \cref{thm:qhop_alg}).

\begin{table}[]
    \centering
    \makegapedcells
    \resizebox{1.0\textwidth}{!}{
    \begin{tabular}{c|p{1.7cm}p{2.1cm}p{3.2cm}p{3cm}}
        \hline
        Algorithms & $\kappa_V$ dependence & Requiring smooth $A(t)$ & Queries to $A$ & Queries to $\ket{\psi_0}$  \\
        \hline
        This work (Theorem \ref{thm:truncated_dyson_alg}) & No & No & $\wt{\Or}( T^2  Q \alpha^2)$ & $\Or(Q)$ \\
        % This work (Theorem \ref{thm:qhop_alg}) & No & No & $\wt{\Or}(T^4 )$ & $\Or(1)$ \\
        \cite[Theorem 1]{ChildsLiu2020} & Yes & Yes & $\wt{\Or}\left( Tq\kappa_V \alpha \log\left(g'/g \right)\right)$ & $\wt{\Or}\left( Tq\kappa_V \alpha \log\left(g'/g \right)\right)$ \\
        \cite[Theorem 10]{Krovi2022} & No & - & $\wt{\Or}( Tq \alpha \displaystyle\sup_{t \in [0, T] } \norm{e^{At}})$ & $\wt{\Or}(  Tq \alpha \displaystyle\sup_{t \in [0, T]} \norm{e^{At}})$ \\
        \hline
    \end{tabular}
    }
    \caption{Comparison with state-of-the-art high-order algorithms. We compare the high-order algorithms in terms of whether they require smooth $A(t)$, have $\kappa_V$ dependence, and the $T$ scaling in the number of queries to $A(t)$ and to the initial state in the case when $A(t)$ is smooth.  Here $\alpha=\sup_{t\in [0,T]} \norm{A(t)}$, $Q$ is defined as \eqref{eq:defn_Q_intro}, $\kappa_V= \max_{t\in [0, T]} \kappa_V(t)$ is a uniform upper bound of the condition number of $V(t)$ diagonalizing $A(t)= V(t) \Lambda(t) V^{-1}(t)$ when $A(t)$ is diagonalizable for all time $t\in [0, T]$, $q = \max_{t\in[0,T]} \frac{\norm{\ket{\psi(t)}}}{\norm{\ket{\psi(T)}}}$, $g = \norm{\ket{\psi(T)}}$, $g' = \max_{t\in [0, T]} \max_{n \in \mathbb{N}} \norm{\ket{\psi^{(n+1)}(t)}}$, where $\ket{\psi^{(n+1)}(t)}$ denotes the $(n+1)$-th derivative of $\ket{\psi(t)}$ and $\mathbb{N}$ is the set of natural numbers. The algorithm in \cite[Theorem 10]{Krovi2022} on the last row is designed only for the time-independent case. }
    \label{tab:comparison}
\end{table}
% $V_0^T(A)$ is the total variation defined in \eqref{eq:defn_total_variation_first}, 

\subsection{Challenges in designing time marching based quantum solvers}\label{sec:diff_direct}

% \LL{ Viewed from \cref{eqn:blockencode_onestep}, if the block encoding factor $\alpha_l=1+\Omega(1)$, then the time marching strategy fails outright. However, even if $\alpha_l\approx 1$, we still have a problem as discussed below.} 

% \YT{--------------here I talk about why it is bad to do things like Euler method-----------}

% We consider the time $0< t_1 < \cdots <t_L = T$. Viewed from \cref{eqn:blockencode_onestep}, if the block encoding factor $\alpha_l=1+\Omega(1)$ at each time step $l$, then the time marching strategy fails outright. However, even if $\alpha_l\approx 1$, we still have a problem as discussed below.

The most straightforward way of solving the ODE \eqref{eq:ode_general} is arguably the (forward) Euler method. For simplicity we consider the time-independent case, i.e., $A(t)=A$, and the time step sizes are chosen to be uniform: $t_l-t_{l-1}=T/L$.
We further assume $A$ is a normal matrix and can be unitarily diagonalized. Starting from $\ket{\psi_0}$, at each time step $l$, we go from $\ket{\psi_{l-1}}\approx\ket{\psi(t_{l-1})}$ to $\ket{\psi_{l}}\approx\ket{\psi(t_{l})}$ via
\begin{equation}
\ket{\psi_l} = (I+A(t_l-t_{l-1}))\ket{\psi_{l-1}}.
\end{equation}

We will show that a direct implementation of this method on a quantum computer leads to severe challenges despite its simple appearance. 
Let us first look at how we should implement $\bar{\Xi}_l=I+A(t_l-t_{l-1})$ on a quantum computer.
% \YT{should perhaps introduce block encoding before this point} 
Generally we can assume that $A$ is given through a block encoding (see \cref{sec:block_encoding_and_qsvt} for a short introduction of block encoding and QSVT), with which we can construct a block encoding of $\bar{\Xi}_l$ denoted by $U_l$ through a linear combination of unitaries. This construction, if performed directly using \cite[Lemma 29]{GilyenSuLowEtAl2019}, involves a subnormalization factor of $1+\|A\|(t_l-t_{l-1})=1+\|A\|T/L$. Going from $\ket{\psi_{l-1}}$ to $\ket{\psi_{l}}$, we apply the  block encoding $U_l$, and measure the ancilla qubits. The success of the procedure depends on the measurement outcome, and the success probability is
\begin{equation}
\label{eq:spurious_success_prob}
    \frac{1}{(1+\|A\|T/L)^2} \times \frac{\|\ket{\psi_{l}}\|^2}{\|\ket{\psi_{l-1}}\|^2},
\end{equation}
where the first factor comes from the subnormalization factor discussed above. Since the success at each time step is independent, the total success probability of implementing Euler's method for $L$ steps is
\[
\frac{1}{(1+\|A\|T/L)^{2L}} \times \prod_{l=1}^L\frac{\|\ket{\psi_{l}}\|^2}{\|\ket{\psi_{l-1}}\|^2} \approx e^{-2\|A\|T}\frac{\|\ket{\psi_{L}}\|^2}{\|\ket{\psi_{0}}\|^2}.
\]
For this method to yield a meaningful result, we need $\ket{\psi_L}\approx\ket{\psi(T)}$, and consequently $\|\ket{\psi_L}\|\approx \|\ket{\psi(T)}\|$. The success probability is therefore approximately $e^{-2\|A\|T}\frac{\|\ket{\psi(T)}\|^2}{\|\ket{\psi(0)}\|^2}$, and it takes 
\[
e^{2\|A\|T}\frac{\|\ket{\psi(0)}\|^2}{\|\ket{\psi(T)}\|^2}
\]
trials for the procedure to succeed with $\Omega(1)$ probability.

To see why this is not a reasonable scaling, let us consider the case where $A$ is anti-Hermitian, which yields the Schr\"odinger equation, and we have $\|\ket{\psi(T)}\|=\|\ket{\psi(0)}\|$. The number of trials needed is therefore $e^{2\|A\|T}$, despite the fact that the usual Hamiltonian simulation algorithms generally succeed in one run!

The problem becomes even worse if we want to implement the time step $e^{AT/L}$ with high accuracy, rather than approximating it with $I+AT/L$. For example, we may consider implementing $e^{AT/L}$ through QSVT, if $A$ is either Hermitian or anti-Hermitian. But the subnormalization factor that comes from QSVT has an extra factor of $2$, becoming $2\|e^{AT/L}\|$, due to the summation of the even and odd parts \cite[Theorem 56]{GilyenSuLowEtAl2018arxiv} or the real and imaginary parts \cite[Theorem 58]{GilyenSuLowEtAl2018arxiv}. In the end the number of trials required becomes $4^L\|e^{AT}\|^2\frac{\|\ket{\psi(0)}\|^2}{\|\ket{\psi(T)}\|^2}$, which increases exponentially with respect to the number of segments $L$ even for a finite $T$.
In the anti-Hermitian case, we can use the oblivious amplitude amplification (OAA) \cite[Theorem 15]{GilyenSuLowEtAl2019} to solve this problem, or use the algorithm in \cite{LowChuang2017} to avoid this problem entirely, but both methods work only because of the unitarity of the exact time evolution. For non-unitary dynamics, as OAA is not applicable, we need a different strategy to implement a time-marching based method.

% For Hamiltonian simulation, oblivious amplitude amplification (OAA) can be used without reusing the initial state.
% \DF{poorly written... need to revise...}
% % the best-known implementation of this approach
% However, when the dynamics is not unitary, the implementation of this approach is known to have an exponential small success probability and lead to an exponential cost in the number of time-steps performed~\cite{Berry2014,LeytonOsborne08}. The difficulties encountered by a direct implementation of time-marching will be detailed in \cref{sec:diff_direct}.

% \vspace{1em}
% \noindent\textbf{:}

\subsection{Organization}

The rest of the paper is organized as follows: in Section \ref{sec:overview} we provide an overview of the method, presenting our method (Theorem \ref{thm:short_time_to_long_time}) to link up short-time evolutions into a long-time evolution while keeping the success probability from decaying faster than necessary. In Section \ref{sec:dyson_series_short_time} we will discuss how to implement short-time evolution using the truncated Dyson series algorithm, leading to our first algorithm in Theorem \ref{thm:truncated_dyson_alg}. The amplification ratio $Q$ dependence in this algorithm is shown to be optimal in Section \ref{sec:optimality}. In Section \ref{sec:qHOP_alg} we demonstrate the performance of the time-marching based method with a first-order truncated Magnus series, which simplifies the implementation and also exhibits a commutator scaling in the high precision limit.
%show that the short-time evolution can also be implemented using the low-order short-time evolution techniques, leading to a simpler implementation for the low accuracy regime, with the result summarized in Theorem \ref{thm:qhop_alg}.

% \subsection{Homogeneity} \label{sec:homogeneity}

% \LL{ what is the purpose of discussing time-independent $b$ here?} 
% For the rest of this paper we will focus on solving homogeneous ODEs, i.e., ODEs of the form
% \begin{equation}
% \label{eq:homogeneous_ODE}
%     \frac{\dd}{\dd t}\ket{\psi(t)} = A(t)\ket{\psi(t)}.
% \end{equation}
% Many non-homogeneous ODEs can be turned into homogeneous ones. For example, when the source term $\ket{b(t)}$ is time-independent, i.e., $\ket{b(t)}=\ket{b}$, then the ODE \YT{the non-homogeneous ODE should be defined in previous sections} is equivalent to the following
% \begin{equation}
%     \frac{\dd}{\dd t}
%     \begin{pmatrix} 
%     \ket{\psi(t)} \\
%     (1/c)\ket{b}
%     \end{pmatrix}
%     =
%     \begin{pmatrix}
%     A(t) & cI \\
%     0 & 0
%     \end{pmatrix}
%     \begin{pmatrix} 
%     \ket{\psi(t)} \\
%     (1/c)\ket{b}
%     \end{pmatrix},
% \end{equation}
% for any $c\neq 0$. If we want to solve the original non-homogeneous ODE on a quantum computer, we can first solve the homogeneous one and then perform measurement to recover the solution with some probability.

\section{Overview of the method}
\label{sec:overview}

\subsection{Main idea}\label{sec:main_idea}

Let us first revisit the challenge of implementing the Euler method in \cref{sec:diff_direct}.
The reason that we end up with the exponential overhead $e^{2\|A\|T}$ is that each time step involves a subnormalization factor $1+\|A\|T/L$. Now let us consider, what if the subnormalization factor is $\|I+AT/L\|$ rather than $1+\|A\|T/L$? The subnormalization factor cannot be better than this because we need to encode $I+AT/L$ into a unitary matrix that has spectral norm $1$. Again assume $A$ is a normal matrix. If the subnormalization factor is indeed $\|I+AT/L\|$, then the $e^{2\|A\|T}$ overhead is replaced by
\[
\|I+AT/L\|^{2L}=\|(I+AT/L)^{L}\|^2\approx\|e^{TA}\|^2.
\]
This is a far more reasonable scaling. For instance, if $A$ is anti-Hermitian, then $\|e^{TA}\|=1$, rather than exponentially growing with time.

From the above discussion, we can see that the seemingly subtle difference between the subnormalization factors $1+\|A\|T/L$ and $\|I+AT/L\|$ is in fact crucial. 
To implement $(I+AT/L)\ket{\phi_{l-1}}$ using linear combination of unitaries as discussed in Section \ref{sec:diff_direct} involves a success probability as described in \eqref{eq:spurious_success_prob}, which can be much smaller than the \textit{intrinsic success probability}:
\begin{equation}
    \label{eq:intrinsic_success_prob}
     \frac{1}{\|I+AT/L\|^2} \times \frac{\|\ket{\psi_{l}}\|^2}{\|\ket{\psi_{l-1}}\|^2}.
\end{equation}
The ratio between the intrinsic success probability and \cref{eq:spurious_success_prob} is
\begin{equation}
\label{eqn:ratio_success}
\gamma^2=\left(\frac{1+\|A\|T/L}{\|I+AT/L\|}\right)^2,
\end{equation}
which comes from excessive subnormalization due to the construction of the block encoding. 
This excessive factor can be removed through a technique called the uniform singular value amplification~\cite[Theorem 17]{GilyenSuLowEtAl2019}. 
In a nutshell, the uniform singular value amplification uses an odd polynomial $P(x)$ to approximate a linear function $f(x)=\gamma x$ in an interval $[-{\gamma}^{-1},{\gamma}^{-1}]$ for $\gamma$ defined in \cref{eqn:ratio_success}, and satisfies the norm constraint $\abs{P(x)}\le 1$ for all $x\in[-1,1]$.  
The norm constraint is a crucial requirement for applying QSVT with the polynomial $P$. 
The effect of the uniform singular value amplification is that it approximately multiplies a factor $\gamma$ to the encoded operator, which nearly exactly cancels the excessive subnormalization factor.
One important feature of the uniform singular value amplification is that it is oblivious to the quantum state we want to act on, i.e., $\ket{\psi_{l-1}}$. This means we do not need to repeatedly prepare quantum states from previous time steps in this amplification procedure, and this is the key to avoiding an exponential overhead.
After repeated usage of the uniform singular value amplification, the subnormalization factor scales as $\|e^{TA}\|$ instead of $e^{T\|A\|}$. The same technique also solves the problem with the subnormalization factor being much larger than $1$ as discussed in \cref{sec:diff_direct}.

%coming from QSVT as discussed above. In the QSVT setting, we can improve the subnormalization factor at each time step from $2\|e^{AT/L}\|$ to $\|e^{AT/L}\|$.

%In the above discussion, we implement the short time evolution of a single step using Euler's method. However, in this paper we will mainly discuss implementing the short time evolution using the truncated Dyson series algorithm~\cite{BerryChildsCleveEtAl2015,KieferovaSchererBerry2019,LowWiebe2019} and the quantum highly oscillatory protocol (qHOP)~\cite{AnFangLin2022}, so that we can deal with the time-dependent case which may involve high oscillation. Furthermore, the truncated Dyson series algorithm has the advantage of having only a logarithmic dependence on the precision.

As discussed above, for the time-independent case, assuming $\|\ket{\psi(0)}\|=1$, we need to run the algorithm for $\|e^{AT}\|^2/\|\ket{\psi(T)}\|^2$ times to achieve $\Omega(1)$ overall success probability. In the time-dependent case, this factor takes a slightly more complicated form,
\[
Q^2 = \frac{\prod_{l=1}^L\|\mathcal{T}e^{\int_{t_{l-1}}^{t_l}A(t)\dd t}\|^2}{\|\ket{\psi(T)}\|^2},
\]
where $0=t_0<t_1<\cdots<t_L=T$ are determined by our choice of the temporal mesh. This gives the definition of $Q$ in \cref{eq:defn_Q_intro}

Because this $Q$ dependence comes from the number of trials needed in order to complete the whole procedure successfully with high probability, a natural question is whether the dependence can be improved from $\Or(Q^2)$ to $\Or(Q)$ using amplitude amplification \cite{BrassardHoyerMoscaEtAl2002}. However, a direct application of amplitude amplification comes at a price. As mentioned above, at each time step, we need to perform measurements on the ancilla qubits, and we need to ensure that the measurement results from all the time steps are correct. The direct application of amplitude amplification requires the usage of different ancilla qubits for each time step, and the number of ancilla qubits needed will scale linearly with respect to the number of time steps. 
To reduce the number of ancilla qubits, we employ a simplified version of the compression gadget introduced in \cite{LowWiebe2019} to coherently record the success or failure of each time step. This method ensures that the number of ancilla qubits needed to implement amplitude amplification scales only logarithmically in the number of time steps.

%Our time-marching strategy can be combined with any single step numerical integrators. We will discuss the truncated Dyson series and the quantum highly oscillatory protocol (qHOP)~\cite{AnFangLin2022} as concrete examples.

\subsection{Input Model} \label{sec:input}

To solve the ODE \eqref{eq:ode_general}, we need to store the information of the coefficient matrix $A(t)$ and the initial state $\ket{\psi(0)}$ in the quantum computer. We assume that we have access to a unitary circuit $U_{\mathrm{init}}$ to prepare the initial state, i.e., $U_{\mathrm{init}}\ket{0^n}=\ket{\psi(0)}$, where $n$ is the number of qubits and $2^n = N$. For $A(t)$ this requires some explanation, in particular because we need not just a single matrix but a family of matrices on the time interval $[0,T]$. 

% We write down the definition here for concreteness: %\LL{ State regularity assumptions on $A$ and define TV here?} 
 %\LL{ Maybe write down the definition of block encoding and QSVT} 
 
A similar problem is encountered in the setting of time-dependent Hamiltonian simulation problem, where a \textit{time-dependent matrix encoding} was proposed in \cite{LowWiebe2019} to encode the time-dependent Hamiltonian. We adopt essentially the same idea, and extend it to the non-Hermitian case.

\begin{defn}[Time-dependent matrix encoding]
\label{defn:time_dep_matrix_encoding}
An $(n_q,m,a,b,\alpha,\epsilon)$-$\mathrm{MAT}$ is a unitary that acts on three registers, each containing $n_{q}$, $m$, and $n$ qubits respectively. It satisfies
\begin{equation}
\label{eq:time_dep_matrix_encoding}
(I_{n_{q}}\otimes\bra{0^{m}}\otimes I_n)\mathrm{MAT}(I_{n_{q}}\otimes\ket{0^{m}}\otimes I_n) = \sum_{\gamma=0}^{2^{n_{q}}-1}\ket{\gamma}\bra{\gamma}\otimes \frac{\wt{A}\left( (b-a) \frac{\gamma}{2^{n_{q}}}+a\right)}{\alpha},
\end{equation}
where $\|\wt{A}(t)-A(t)\|\leq \epsilon$ for $t\in[a,b]$.
This  unitary $\mathrm{MAT}$ is called the time-dependent matrix encoding of $A(t)$ on the time interval $[a,b]$.
\end{defn}

Here $n$ is the number of qubits corresponding to the system ($2^n = N$), $m$ is the number of ancilla qubits for block encoding, and $n_q$ qubits are used to store the index of the quadrature points. There are $2^{n_q}$ quadrature points in total.%is the number of qubits corresponding to the quadrature points.  

The total variation of $A(t)$ on the interval $[a,b]$, denoted $V_a^b(A)$, is defined as follows:
\begin{equation}
    \label{eq:defn_total_variation_first}
    V_a^b(A)=\sup_{R\in\mathbb{N}}\sup_{a=t_0<t_1<\cdots <t_R=b} \sum_{j=0}^{R-1}\|A(t_{j+1})-A(t_j)\|.
\end{equation}
The set of all functions of bounded variation is denoted by
\begin{equation}
BV([a,b]):=\set{A|V_a^b(A)<\infty}.
\label{eqn:bounded_variation}
\end{equation}
Our algorithm only requires that the total variation $V_0^T(A)$ of the coefficient matrix $A(t)$ is finite, i.e., $A\in BV([0,T])$. 

%since $V_0^T(A)=\int_0^T\|A'(t)\|\dd t$ for $A$ that is differentiable, and piece-wise differentiable functions also satisfy this condition.

We need to choose $n_q$ to achieve the required accuracy for performing numerical quadrature, and this will be discussed in detail in \cref{sec:short_time_evolution}. We note that for the sparse matrix input model, which is used in other quantum differential equation solvers~\cite{Berry2014,BerryChildsOstranderWang2017,ChildsLiu2020,Krovi2022} and time-dependent Hamiltonian simulation algorithms~\cite{BerryChildsSuEtAl2020}, one can construct an efficient time-dependent matrix encoding. Thus, our algorithms also apply to the case of sparse matrices, which will be discussed in more details in \cref{sec:app_SM}.

%The unitary $\mathrm{MAT}$ can be efficiently implemented using the sparse matrix access model (see \cref{sec:app_SM}), and we can take into account the change of the Hamiltonian norm to achieve a $L^1$ norm scaling\LL{why discuss the $L^1$ norm scaling here?} . 

% Within this framework, $U_{\mathrm{init}}$ is used to prepare $\ket{\psi_0}=\ket{\psi(0)}$ to start the whole process. Each $\wt{U}_l$ is constructed from the time-dependent matrix encoding of $A(t)$, through either the truncated Dyson series algorithm (Section \ref{sec:dyson_series_short_time}) or the qHOP algorithm (Section \ref{sec:qHOP_alg}), and a procedure known as uniform singular value amplification proposed in \cite[Theorem 17]{GilyenSuLowEtAl2019} to address the problem discussed in Section \ref{sec:diff_direct}. In the next section we will first focus on how to use the uniform singular value amplification technique.

\subsection{Uniform singular value amplification} \label{sec:usva}

For a given temporal mesh  $0=t_0<t_1<\cdots<t_L=T$, the short time integrator at step $l$ can be abstractly written as 
\begin{equation}
\ket{\psi_{l}}=\bar{\Xi}_l \ket{\psi_{l-1}}, \quad l=1,\ldots,L,
\label{eqn:scheme}
\end{equation}
where $\bar{\Xi}_l$ is an operator approximating the exact evolution operator $\Xi_l=\mathcal{T}e^{\int_{t_{l-1}}^{t_l}A(t)\dd t}$. 
We require that $\bar{\Xi}_l$ is consistent with $\Xi_l$ to precision $\epsilon_l\|\Xi_l\|$, i.e.,
\begin{equation}
    \label{eq:dist_Xi_l_bar_Xi_l}
    \|\Xi_l-\bar{\Xi}_l\|\leq \epsilon_l\|\Xi_l\|.
\end{equation}
We assume that $\bar{\Xi}_l$ is implemented with its $(\alpha_l,m,0)$-block encoding denoted by $U_l$, i.e.,  
\begin{equation}
\alpha_l(\bra{0^m}\otimes I_n)U_l(\ket{0^m}\otimes I_n)=\bar{\Xi}_l.
\label{eqn:blockencode_onestep}
\end{equation}
Due to \cref{eq:dist_Xi_l_bar_Xi_l}, $U_l$ can also be viewed as an  $(\alpha_l,m_l,\epsilon_l\|\Xi_l\|)$-block encoding of $\Xi_l$.

In this section we discuss how to approximately implement a series of non-unitary operations $\Xi_1$, $\Xi_2$, \ldots, $\Xi_L$ sequentially, and boost the success probability using uniform singular value amplification, which is developed in~\cite[Theorem 17]{GilyenSuLowEtAl2019}. The idea is to linearly amplify the singular values of $\bar{\Xi}_l/\alpha_l$ by a factor that is approximately $\gamma'=\alpha_l/\norm{\bar{\Xi}_l}$.
For simplicity, we first assume $\bar{\Xi}_l=\Xi_l$ (i.e., $U_l$ block encodes the exact short time integrator), and study how this technique can help us boost the success probability of a single operation.

% \LL{ Write down the statement of the polynomial approximation used in [46,Thm17] explicitly? We can then cite this in the discussion section say that this lemma.} 

\begin{lem}[Uniform singular value amplification, {\cite[Theorem 17]{GilyenSuLowEtAl2019}}]
\label{lem:uniform_singular_value_amplification}
Let $U$ be an $(\alpha,m,0)$-block encoding of $\Xi$. We can construct a $(\frac{\|\Xi\|}{1-\delta},m+1,\epsilon\|\Xi\|)$-block encoding $\wt{U}$ of $\Xi$, using $\Or(\frac{\alpha}{\delta\|\Xi\|}\log(\frac{\alpha}{\|\Xi\|\epsilon}))$ applications of (controlled-) $U$ and its inverse. 
\end{lem}

\begin{proof}
Consider the singular value decomposition $\Xi=W\Sigma V^{\dagger}$, where $\Sigma=\mathrm{diag}(\sigma_1,\sigma_2,\ldots,\sigma_N)$. Applying QSVT with an odd polynomial gives us the block encoding $\wt{U}$ of a new matrix, which up to rescaling is $\wt{\Xi}=W\wt{\Sigma} V^{\dagger}$, where $\wt{\Sigma}=\mathrm{diag}(\wt{\sigma}_1,\wt{\sigma}_2,\ldots,\wt{\sigma}_N)$. 
We choose the odd polynomial in the same way as in \cite[Theorem 17]{GilyenSuLowEtAl2019}, and we choose $\gamma=\frac{\alpha(1-\delta)}{\|\Xi\|}$. By these choices all singular values of $\Xi$ are contained in the interval $[0,\gamma'^{-1}]=[0,\frac{1-\delta}{\gamma}]$. If we choose the rescaling factor of $\wt{\Xi}$ so that
\[
\wt{\Xi} = \frac{\|\Xi\|}{1-\delta}(\bra{0^{m+1}}\otimes I_n) \wt{U} (\ket{0^{m+1}}\otimes I_n), 
\]
then by \cite[Theorem 17]{GilyenSuLowEtAl2019},
\[
\Big|\frac{\wt{\sigma}_j}{\sigma_j}-1\Big|\leq \epsilon,
\]
which implies
\[
\|\wt{\Xi}-\Xi\|=\|\wt{\Sigma}-\Sigma\|\leq \epsilon\|\Sigma\|=\epsilon\|\Xi\|.
\]
Therefore $\wt{U}$ is a $(\frac{\|\Xi\|}{1-\delta},m+1,\epsilon\|\Xi\|)$-block encoding $\wt{U}$ of $\Xi$. By \cite[Theorem 17]{GilyenSuLowEtAl2019} it requires $\Or(\frac{\alpha}{\delta\|\Xi\|}\log(\frac{\alpha}{\|\Xi\|\epsilon}))$ applications of (controlled-) $U$ and its inverse. 
\end{proof}

\begin{figure}[ht]
    \centerline{
    \Qcircuit @R=1em @C=2em {
    \text{Ancilla}\quad\quad\quad\quad & \multigate{1}{\wt{U}_1} & \meter{} & \multigate{1}{\wt{U}_2} & \meter{} & \qw & \raisebox{0em}{$\cdots$}\qquad& \multigate{1}{\wt{U}_L} &  \meter{} \\
    \text{State}\quad\quad\quad\quad & \ghost{\wt{U}_1}  & \qw  & \ghost{\wt{U}_2}  & \qw & \qw & \raisebox{0em}{$\cdots$}\qquad&  \ghost{\wt{U}_L} & \qw \\
    }
    }
    \caption{Implementing $\wt{\Xi}_L\cdots\wt{\Xi}_2\wt{\Xi}_1$. After we apply each $\wt{U}_l$, we measure the ancilla qubits, and only proceed when the measurement result is all $0$, and otherwise abort the procedure.}
    \label{fig:usva_sequential}
\end{figure}
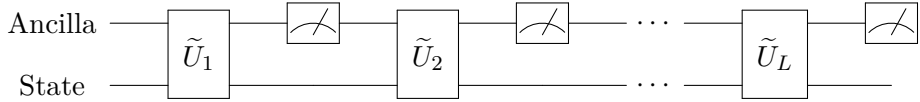

We can now use \cref{lem:uniform_singular_value_amplification} to address the problem discussed in \cref{sec:diff_direct}, and for now neglect all errors in the block encodings. 
The error analysis with the block encoding errors taken into account will be analyzed in \cref{thm:short_time_to_long_time}.
If we directly apply $U_l$ and post-select the measurement results, through the procedure described in Figure \ref{fig:usva_sequential} without the uniform singular value amplification, then the success probability will be
\begin{equation}
\label{eq:succ_prob_wo_usva}
    \frac{\|\Xi_L\cdots\Xi_2\Xi_1\ket{\psi(0)}\|^2}{(\alpha_1\alpha_2\cdots\alpha_L)^2}.
\end{equation}
For each $l$, $\|\Xi_l\|\leq \alpha_l$, and the cumulative difference between $\alpha_1\alpha_2\cdots\alpha_L$ and $\|\Xi_L\|\cdots\|\Xi_2\|\|\Xi_1\|$ can be quite significant, as discussed for the case of Euler's method in Section \ref{sec:diff_direct}. 

With the block encodings $\wt{U}_l$ given by the uniform singular value amplification, we can implement $\wt{\Xi}_L\cdots\wt{\Xi}_2\wt{\Xi}_1$ as depicted in Figure \ref{fig:usva_sequential}. We apply each $\wt{U}_l$ sequentially, measuring the ancilla qubits after each application, only proceeding when the measurement result is all $0$, and otherwise aborting the procedure. The operator $\wt{\Xi}_L\cdots\wt{\Xi}_2\wt{\Xi}_1$ thus implemented will approximate the operator $\Xi_L\cdots\Xi_2\Xi_1$, which is our goal. 
With  with the choice $\delta=1/L$, we have 
\[
\left(1-\delta\right)^L\ge e^{-\frac{\delta L}{1-\delta}}=\Omega(1).
\]
Therefore the success probability is, up to a constant factor,
\begin{equation}
\label{eq:succ_prob_w_usva}
    \frac{\|\Xi_L\cdots\Xi_2\Xi_1\ket{\psi(0)}\|^2}{\|\Xi_L\|^2\cdots\|\Xi_2\|^2\|\Xi_1\|^2}.
\end{equation}
By turning the success probability from \eqref{eq:succ_prob_wo_usva} to \eqref{eq:succ_prob_w_usva} using \cref{lem:uniform_singular_value_amplification}, we address the problem of the vanishing success probability discussed in \cref{sec:diff_direct}.

\begin{figure}
\begin{center}
\subfloat[Polynomial approximation to $(1-\delta)\mathrm{XRect}(x)$]{\includegraphics[width=0.3\textwidth]{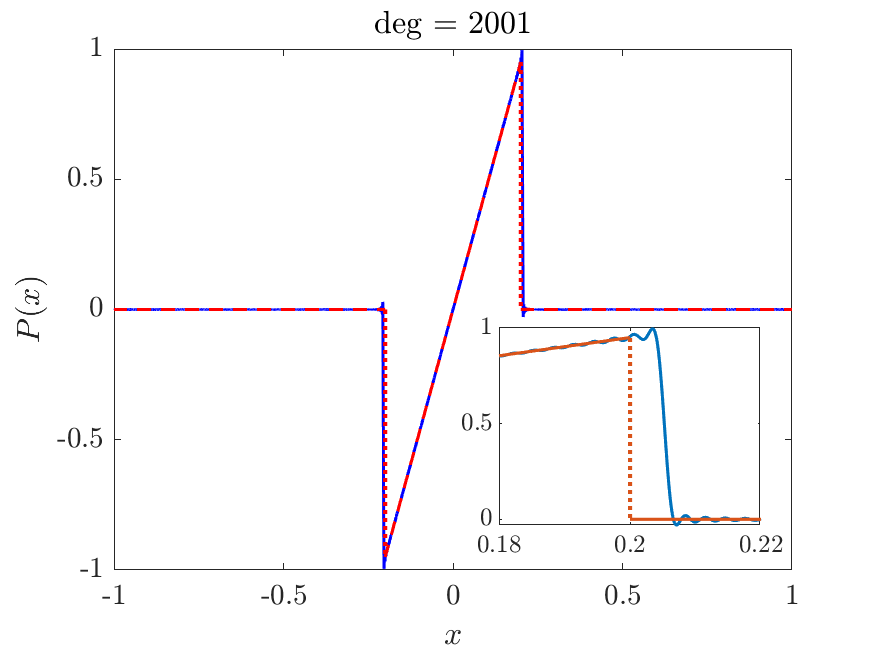}}
\quad
\subfloat[Polynomial from convex optimization]{\includegraphics[width=0.3\textwidth]{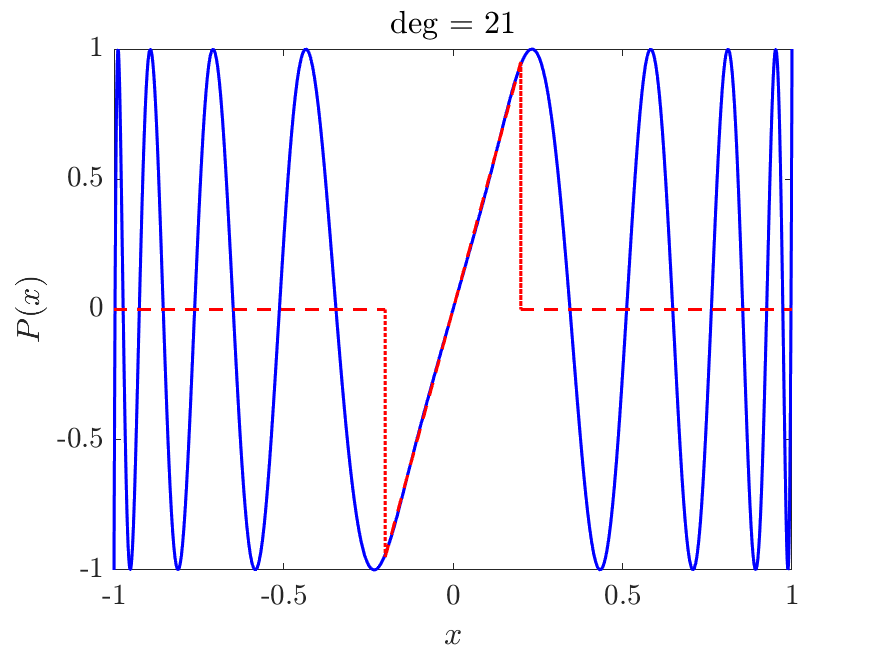}}

\subfloat[Error of polynomial approximation to  $(1-\delta)\mathrm{XRect}(x)$]{\includegraphics[width=0.3\textwidth]{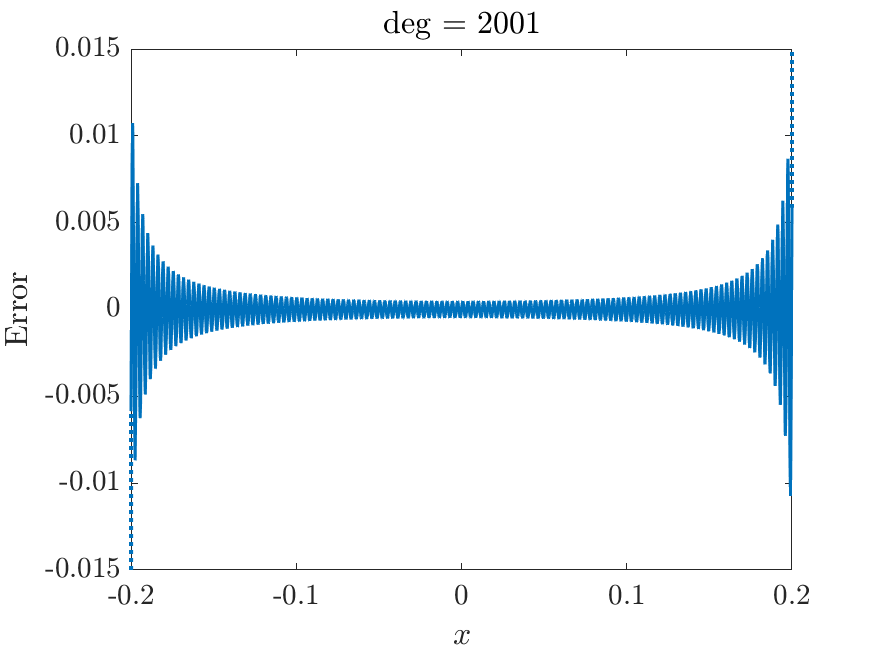}}\quad
\subfloat[Error of polynomial from  convex optimization]{\includegraphics[width=0.3\textwidth]{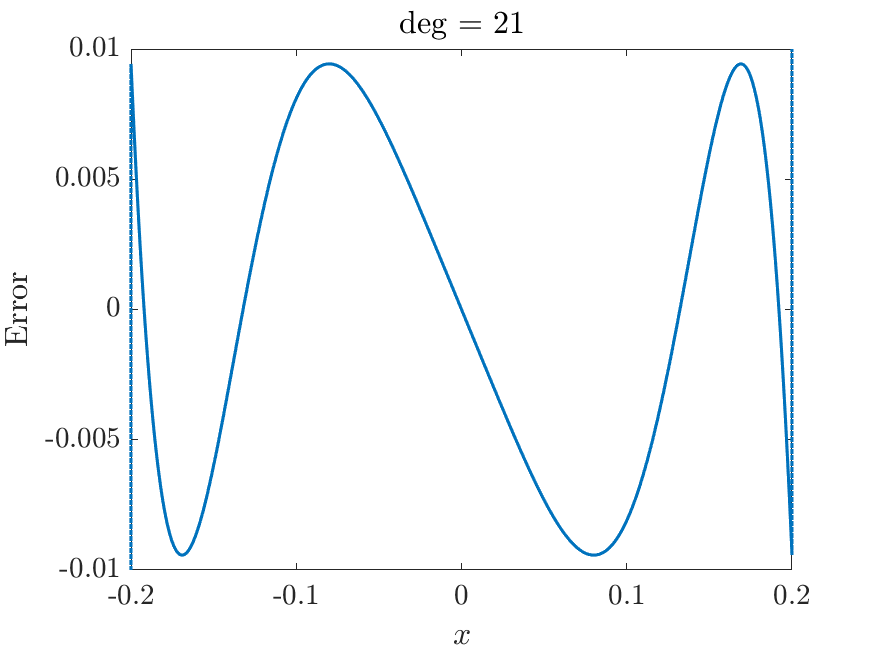}}\quad
\subfloat[Comparison of convergence speed]{\includegraphics[width=0.3\textwidth]{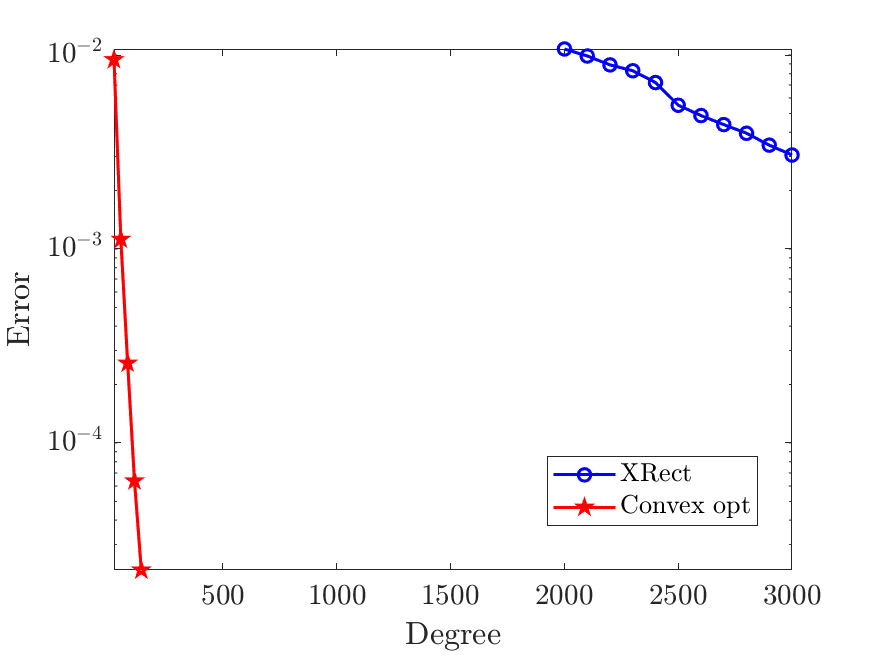}}
\end{center}
\caption{Comparison of the polynomials used for uniform singular value amplification procedure. (a) Polynomial approximating $(1-\delta)\mathrm{XRect}(x)$ used in \cite[Theorem 17]{GilyenSuLowEtAl2019}; (b) Near-optimal polynomial approximation obtained via convex optimization (see \cref{sec:convex}); (c) (d) Errors of the two methods on the interval $\mc{I}=[-{\gamma'}^{-1},{\gamma'}^{-1}]$ with $\gamma'=5, \delta=0.05$; (e) Comparison of the convergence speed of the two methods, measured by the $L^{\infty}$ error $\max_{x\in\mc{I}}|p(x)-(1-\delta)\mathrm{XRect}(x)|$.}
\label{fig:poly_uniform}
\end{figure}

The uniform singular value amplification procedure was first proposed in ~\cite[Theorem 5]{LowChuang2017a}. However, it only constructs an $(2\|\Xi\|,m+1,\epsilon\|\Xi\|)$-block encoding of $\Xi$, and the extra factor $2$ means that the overall success probability of the procedure in \cref{fig:usva_sequential} will still decrease exponentially fast as $\Or(4^{-L})$. This is not acceptable in the context of this paper. Therefore we adopt the version in \cite[Theorem 17]{GilyenSuLowEtAl2019}, which refines the analysis  so that the subnormalization factor is $\|\Xi\|/(1-\delta)$. 

In the abstract form, one needs a polynomial that approximates $\gamma' x$ on the desired interval $\mc{I}=[-{\gamma'}^{-1},{\gamma'}^{-1}]$.
%Let us study the polynomial \REV{construction} used 
Such a polynomial has been constructed by \cite[Theorem 17]{GilyenSuLowEtAl2019}, which we examine in more detail. It first constructs an even polynomial $q(x)$ that approximates an even ``rectangular function'' defined as 
\[
\mathrm{Rect}(x)=\frac12(\mathrm{sgn}({\gamma'}^{-1}-\abs{x})+1), \quad \mathrm{sgn}(x)=\begin{cases}
1, & x>0\\
-1, & x<0\\
0, & x=0
\end{cases}.
\]
The desired odd polynomial is $p(x)=\gamma' x q(x)$, which approximates $\mathrm{XRect}(x):=\gamma 'x \cdot \mathrm{Rect}(x)$, and agrees with the linear function $\gamma' x$ on the interval $[-{\gamma'}^{-1},{\gamma'}^{-1}]$. The polynomial $p(x)$ should also satisfy the \textit{norm constraint}: $\abs{p(x)}\le 1$ for all $x\in[-1,1]$ due to the requirement of QSVT (see \cite[Corollary 5]{GilyenSuLowEtAl2019}). 
However, the function $\mathrm{XRect}(x)$ is discontinuous at $x={\gamma'}^{-1}$, and therefore a polynomial approximation to $\mathrm{XRect}(x)$ exhibits the Gibbs phenomenon, which states that $p(x)$ always overshoots around $x={\gamma'}^{-1}$, even as the polynomial degree increases to infinity (see e.g., \cite{GottliebShu1997} and ~\cite[Chapter 9]{Trefethen2019}).
An example of the Gibbs phenomenon is shown in the inset of \cref{fig:poly_uniform} (a), where the polynomial approximation overshoots $\mathrm{XRect}(x)$. 
As a result, instead of approximating $\mathrm{XRect}(x)$, we can only find a polynomial that approximates the function $(1-\delta)\mathrm{XRect}(x)$ to satisfy the norm constraint. 
It is worth mentioning that although the construction above leads to the desired asymptotic scaling in \cref{lem:uniform_singular_value_amplification}, the preconstant can be quite large, which leads to high polynomial degrees even for moderate values of the parameters $\alpha,\delta,\epsilon$. 
\cref{fig:poly_uniform} (a) shows that for $\gamma'=5, \delta=0.05,\epsilon=0.01$, the required polynomial degree is already as large as $2001$. Reducing to a smaller value $\delta=0.01$ would require a polynomial degree of around $10^4$, which may be too large to be practically useful.

To address this issue, we notice that the uniform singular value amplification only requires us to find a polynomial $p(x)$ that approximates $\gamma' x$ on the desired interval $\mc{I}=[-{\gamma'}^{-1},{\gamma'}^{-1}]$. Outside this interval $\mc{I}$, the value of $p(x)$ can be arbitrary, as long as the norm constraint is satisfied. In particular, $p(x)$ does not need to approximate $\mathrm{XRect}(x)$, which vanishes outside $\mc{I}$.
This allows us to construct a convex optimization based procedure to numerically identify the near-optimal polynomial approximation for the uniform singular value amplification. 
The procedure is detailed in \cref{sec:convex}. For the same parameter setting $\gamma'=5, \delta=0.05,\epsilon=0.01$, the polynomial approximation is given in \cref{fig:poly_uniform} (b) and the polynomial degree is merely $21$. %Furthermore, the near-optimal approximation exhibits two types of \textit{equioscillation} properties. 
\cref{fig:poly_uniform} (e) further shows that as fixing $\gamma', \delta$, both methods converge exponentially with respect to the increase of the polynomial degrees. However, the convergence rate of the convex optimization based method is significantly faster, which reduces the number of queries to $U_l$ by orders of magnitude.

\subsection{Amplitude amplification using  compression gadget} \label{sec:aa_using_compression_gadget}

Since the main concern of running the algorithm in \cref{fig:usva_sequential} is its success probability, it is natural to consider the usage of amplitude amplification \cite{BrassardHoyerMoscaEtAl2002} to reduce the number of repetitions needed to obtain a successful outcome. With the current procedure in \cref{fig:usva_sequential}, however, this results in a large space overhead. Directly applying $\wt{U}_l$ (and hence $\wt{\Xi}_l$) sequentially involves intermediate measurements to determine whether each $\wt{\Xi}_l$ is applied successfully.
We need to record the measurement outcome of each of the $L$ steps, and this means we need to duplicate the ancilla register $L$ times to implement amplitude amplification.  To avoid this overhead, we need to replace the procedure with a fully coherent one, with measurement performed only at the end. This allows us to reduce the $Q$ dependence from $\Or(Q^2)$ to $\Or(Q)$.

Let us first formulate the problem in a more abstract way. We have unitaries $V_1,V_2,\ldots,V_L$, each of which is a $(\alpha'_l,m'_l,0)$-block encoding of a potentially non-unitary operation $\Gamma_l$. The goal is to implement $\Gamma_L\cdots\Gamma_2\Gamma_1$ with amplitude amplification, and without duplicating the ancilla registers.
% Then we can construct an $(\alpha_{\mathrm{comp}},m_{\mathrm{comp}},0)$-block encoding of $\Gamma_L\cdots\Gamma_2\Gamma_1$

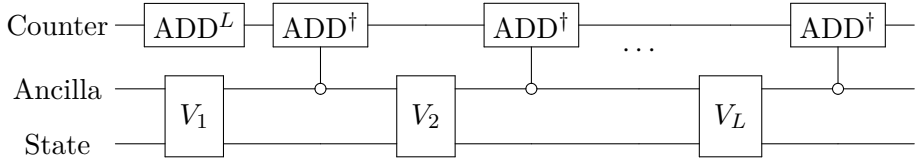
\begin{figure}[ht]
    \centerline{
    \Qcircuit @R=1em @C=1em {
    \text{Counter}\quad\quad\quad\quad & \gate{\text{ADD}^L} & \gate{\text{ADD}^\dagger} & \qw  & \gate{\text{ADD}^\dagger}  & \cds{1}{\cdots} & \qw & \gate{\text{ADD}^\dagger} & \qw \\
    \text{Ancilla}\quad\quad\quad\quad & \multigate{1}{V_1} & \ctrlo{-1} & \multigate{1}{V_2} &  \ctrlo{-1}  & \qw & \multigate{1}{V_L} & \ctrlo{-1} & \qw \\
    \text{State}\quad\quad\quad\quad & \ghost{V_1}  & \qw & \ghost{V_2}   & \qw & \qw & \ghost{V_L} & \qw & \qw \\
    }
    }
    \caption{The simplified compression gadget for coherently applying $\Gamma_L\cdots\Gamma_2\Gamma_1$. The counter register, containing $\lceil\log_2(L)\rceil+1$ qubits, is used for keeping track of whether each $\Gamma_l$ has been applied successfully; the ancilla register, containing $\max_l m'_l$ qubits, is for the ancilla qubits needed in $V_l$'s; the state register stored the quantum state on which we want to apply $\Gamma_L\cdots\Gamma_2\Gamma_1$. $\mathrm{ADD}$ implements addition by $1$ modulo the smallest power of $2$ that is larger than or equal to $2L$. Here each controlled $\mathrm{ADD}^{\dagger}$ is controlled on the state $\ket{0^{m_{\max}}}$.}
    \label{fig:compression_gadget}
\end{figure}

This goal can be achieved using the \textit{compression gadget} in \cref{fig:compression_gadget}, following the idea in Ref.~\cite{LowWiebe2019}. 
In fact we are using a simplified version of the compression gadget in Ref.~\cite{LowWiebe2019}, as the problem we are trying to solve is in some way easier. The main idea is to use a counter register to keep track of how many $\Gamma_l$'s have been applied successfully in a coherent way. This allows us to post-select on the counter register to ensure that all $\Gamma_l$'s have been applied successfully. This result is summarized in the following lemma. Its proof is given in \cref{sec:compression_gadget}.

\begin{lem}[Compression gadget]
\label{lem:compression_gadget}
Suppose we are given unitaries $V_1,V_2,\ldots,V_L$, each of which is a $(\alpha'_l,m'_l,0)$-block encoding of $\Gamma_l$.
Then we can construct a $(\alpha_{\mathrm{comp}},m_{\mathrm{comp}},0)$-block encoding of $\Gamma_L\cdots\Gamma_2\Gamma_1$, where
\[
\alpha_{\mathrm{comp}} = \alpha'_1\alpha'_2\cdots\alpha'_L,\quad m_{\mathrm{comp}} = \max_l m'_l+\lceil\log_2(L)\rceil+1,
\]
using one application of each $V_l$.
\end{lem}

% \YT{This lemma needs to be revised to include error analysis}
% Now we combine Lemma \ref{lem:uniform_singular_value_amplification} and Lemma \ref{lem:compression_gadget} to get the following corollary:
% \begin{cor}
% Suppose we are given $\Xi_l$ through its $(\alpha_l,m_l,0)$-block encoding $U_l$, for $l=1,2,\ldots,L$, then we can construct an $(\alpha_{\mathrm{comp}},m_{\mathrm{comp}},L\epsilon'\prod_{l=1}^L\|\Xi_l\|)$-block encoding of $\Xi_L\cdots\Xi_2\Xi_1$, where
% \[
% \alpha_{\mathrm{comp}} = \frac{\|\Xi_L\|\cdots\|\Xi_2\|\|\Xi_1\|}{(1-\delta)^L},\quad m_{\mathrm{comp}} =\max_l m_l+\lceil\log_2(L)\rceil+2,
% \]
% using $\Or(\frac{\alpha_l}{\delta\|\Xi_l\|}\log(\frac{\alpha_l}{\|\Xi_l\|\epsilon'}))$ applications of each (controlled-) $U_l$ and its inverse.
% \end{cor}

%\subsection{From local error to global error} \label{sec:local_err_to_global_err}

%The following lemma is needed to keep track of the error when applying a sequence of non-unitary operations:

%\begin{lem}\label{lem:hybrid_argument}
%Let $\Gamma_l$ and $\wt{\Gamma}_l$ be operators acting on the same Hilbert space, then 
%\begin{equation*}
%    \|\Gamma_L\cdots \Gamma_2\Gamma_1-\wt{\Gamma}_L\cdots \wt{\Gamma}_2\wt{\Gamma}_1\|\leq \sum_{l=1}^L\prod_{r=l+1}^L\|\Gamma_r\|\|\Gamma_l-\wt{\Gamma}_l\|\prod_{r=1}^{l-1}\|\wt{\Gamma}_r\|.
%\end{equation*}
%\end{lem}

We can now use \cref{lem:compression_gadget} to obtain the following result, taking into account the uniform singular value amplification procedure:

\begin{thm}[Coherent implementation of long-time integrator]
\label{thm:short_time_to_long_time}
Suppose we are given $\Xi_l$ through its $(\alpha_l,m_l,\epsilon_l\|\Xi_l\|)$-block encoding $U_l$, for $l=1,2,\ldots,L$, and $\sum_l\epsilon_l\leq 1/2$, then for any $0<\epsilon'\leq 1/(2L)$, we can construct an $(\alpha_{\mathrm{comp}},m_{\mathrm{comp}},\epsilon_{\mathrm{comp}})$-block encoding of $\Xi_L\cdots\Xi_2\Xi_1$, where
\begin{equation}
\frac{\prod_{l=1}^L\|\Xi_l\|}{2(1-\delta)^L}\leq \alpha_{\mathrm{comp}} \leq \frac{e^{1/2}\prod_{l=1}^L\|\Xi_l\|}{(1-\delta)^L},
\end{equation}
and
\begin{equation}
m_{\mathrm{comp}} =\max_l m_l+\lceil\log_2(L)\rceil+2,\quad
\epsilon_{\mathrm{comp}} =e^{1/2}\left(L\epsilon'+\sum_l\epsilon_l\right)\prod_{l'=1}^L\|\Xi_{l'}\|.
\end{equation}
using $\Or(\frac{\alpha_l}{\delta\|\Xi_l\|}\log(\frac{\alpha_l}{\|\Xi_l\|\epsilon'}))$ applications of each (controlled-) $U_l$ and its inverse.
\end{thm}

\begin{proof}
  We denote by $\bar{\Xi}_l$ the matrix that is exactly encoded in each $U_l$ as in \cref{eqn:blockencode_onestep}. 
  Using Lemma \ref{lem:uniform_singular_value_amplification}, we first construct a $(\frac{\|\bar{\Xi}_l\|}{1-\delta},m_l+1,\epsilon'\|\bar{\Xi}_l\|)$-block encoding $\wt{U}_l$ of each $\bar{\Xi}_l$. Each $\bar{\Xi}_l$ uses $U_l$ $\Or(\frac{\alpha_l}{\delta\|\Xi_l\|}\log(\frac{\alpha_l}{\|\Xi_l\|\epsilon'}))$ times. Here we have used the fact that $\|\bar{\Xi}_l\|=\Theta(\|\Xi_l\|)$ because of \cref{eq:dist_Xi_l_bar_Xi_l}. We denote by $\wt{\Xi}_l$ the matrix that is exactly encoded in $\wt{U}_l$, i.e.,
\[
\frac{\|\bar{\Xi}_l\|}{1-\delta}(\bra{0^{m_l+1}}\otimes I_n)\wt{U}_l(\ket{0^{m_l+1}}\otimes I_n) = \wt{\Xi}_l.
\]
Then
\begin{equation}
    \label{eq:dist_bar_Xi_l_wt_Xi_l}
    \|\bar{\Xi}_l-\wt{\Xi}_l\|\leq \epsilon'\|\bar{\Xi}_l\|.
\end{equation}

Next, we use Lemma \ref{lem:compression_gadget} to combine $\wt{U_l}$'s into a block encoding of $\Xi_L\cdots\Xi_2\Xi_1$. Directly applying Lemma \ref{lem:compression_gadget} yields an $(\alpha_{\mathrm{comp}},m_{\mathrm{comp}},0)$-block encoding of $\wt{\Xi}_L\cdots\wt{\Xi}_2\wt{\Xi}_1$, which we denote by $U_{\mathrm{comp}}$, where
\[
\alpha_{\mathrm{comp}} = \frac{\prod_{l=1}^L\|\bar{\Xi}_l\|}{(1-\delta)^L},\quad m_{\mathrm{comp}} = \max_l m_l+\lceil\log_2(L)\rceil+2.
\]

Noting the fact that
\[
\prod_{l=1}^L(1-\epsilon_l) \geq 1-\sum_l\epsilon_l\geq 1/2,\quad \prod_{l=1}^L(1+\epsilon_l)\leq e^{\sum_l\epsilon_l}\leq e^{1/2},
\]
we can get the upper and lower bounds for $\alpha_{\mathrm{comp}}$ through
\[
\frac{\prod_{l=1}^L\|\Xi_l\|\prod_{l=1}^L(1-\epsilon_l)}{(1-\delta)^L}\leq \alpha_{\mathrm{comp}} \leq \frac{\prod_{l=1}^L\|\Xi_l\|\prod_{l=1}^L(1+\epsilon_l)}{(1-\delta)^L}
\]
To bound the error between $\wt{\Xi}_L\cdots\wt{\Xi}_2\wt{\Xi}_1$ and $\Xi_L\cdots\Xi_2\Xi_1$, we have
\[
\begin{aligned}
&\|\Xi_L\cdots\Xi_2\Xi_1-\wt{\Xi}_L\cdots\wt{\Xi}_2\wt{\Xi}_1\| \\
\leq &\|\Xi_L\cdots\Xi_2\Xi_1-\bar{\Xi}_L\cdots\bar{\Xi}_2\bar{\Xi}_1\| +\|\bar{\Xi}_L\cdots\bar{\Xi}_2\bar{\Xi}_1-\wt{\Xi}_L\cdots\wt{\Xi}_2\wt{\Xi}_1\| \\
\leq& \sum_{l=1}^L\prod_{l'=l+1}^L\|\Xi_{l'}\|\prod_{r=1}^{l-1}\|\bar{\Xi}_r\| \|\Xi_l-\bar{\Xi}_l\|+
\sum_{l=1}^L\prod_{l'=l+1}^L\|\bar{\Xi}_{l'}\|\prod_{r=1}^{l-1}\|\wt{\Xi}_r\| \|\bar{\Xi}_l-\wt{\Xi}_l\|
\\
\leq& \sum_l\epsilon_l\prod_{l'=1}^L(\|\Xi_{l'}\|(1+\epsilon_l)) + L\epsilon'\prod_{l'=1}^L(\|\bar{\Xi}_l\|(1+\epsilon')) \\
\leq& e^{1/2}\left(\sum_l\epsilon_l\prod_{l'}\|\Xi_{l'}\| + L\epsilon'\prod_l\|\Xi_l\|\right),
\end{aligned}
\]
where for the second inequality we have used \cref{eq:dist_Xi_l_bar_Xi_l,eq:dist_bar_Xi_l_wt_Xi_l},  and for the last inequality we have used $\prod_{l=1}^L(1+\epsilon_l)\leq e^{\sum_l\epsilon_l}\leq e^{1/2}$ as well as $(1+\epsilon')^L\leq e^{L\epsilon'}\leq e^{1/2}$.
Therefore we can choose $\epsilon_{\mathrm{comp}}$ as in the statement of the corollary.

\end{proof}

With a coherent implementation of the long-time integrator given as a block encoding in \cref{thm:short_time_to_long_time}, we can readily apply the standard amplitude amplification to boost the success probability and yield a quadratic speedup in terms of the query complexity.

\section{High-order truncated Dyson series approach} \label{sec:dyson_series_short_time}

In this section, we analyze the method described in \cref{sec:overview}, when the short time integrator is implemented using the high-order truncated Dyson series.
In Section \ref{sec:short_time_evolution} we discuss how to implement the short time integrator developed in Ref.~\cite{LowWiebe2019}. Then we use the tools developed in Sections \ref{sec:usva} and \ref{sec:aa_using_compression_gadget} to link different segments of short time evolution into long time evolution in \cref{sec:dyson_long_time}. 
We analyze the success probability in \cref{sec:prob_main_dyson}. 
Our time-marching based strategy can be combined with any input models such as the sparse matrix model \cite{LowWiebe2019,BerryChildsSuEtAl2020,ChildsLiu2020,Krovi2022} and the linear combination of unitaries (LCU) model \cite{LowWiebe2019,BerryChildsSuEtAl2020}.
We discuss in particular the implementation with a sparse matrix input model in \cref{sec:app_SM}.

\subsection{Short time evolution} \label{sec:short_time_evolution}

The truncated Dyson series method implements $\Xi= \mathcal{T}e^{\int_{a}^{b}A(s)\dd s}$ through 
\[
\mathcal{T}e^{\int_{a}^{b}A(s)\dd s} \approx \sum_{k=0}^{K-1}\int_{a}^{b}\dd s_1 \int_{a}^{s_1}\dd s_2\cdots \int_{a}^{s_{k-1}}\dd s_{k} A(s_1)A(s_2)\cdots A(s_k). %+ \Or\left(\frac{(\int_{t_{l-1}}^{t_l}\|A(s)\|\dd s)^K}{K!}\right)
\]
This infinite series can be truncated at order $K$, and the error is upper bounded by $(\int_{a}^{b}\|A(s)\|\dd s)^K/K!$. Therefore if we choose $a$ and $b$ so that $\int_{a}^{b}\|A(s)\|\dd s=\Or(1)$, we can achieve high accuracy with only a few terms.

In order to be robust against error in the coefficient matrix, we need an additional step in the error analysis of the algorithm. If $A(t)$ is accessed through an $(n_q,m,a,b,\alpha,\epsilon)$-$\mathrm{MAT}$ as discussed in \cref{sec:input}, then the above approach will yield a block encoding of a different time evolution operator $\bar{\Xi}\approx\mathcal{T}e^{\int_a^b \wt{A}(t)\dd t}$, where $\wt{A}(t)$ is the time-dependent matrix encoded in $\mathrm{MAT}$ exactly, as defined in \cref{eq:time_dep_matrix_encoding}. By \cref{lem:ODE_error_grow_short_time}, and assume that $(b-a)\alpha=\Or(1)$, then \[
\|\Xi-\bar{\Xi}\|\leq e^{(b-a)\max_{u\in[a,b]}\{\|A(u)\|,\|\wt{A}(u)\|\}}\int_{a}^b \|\wt{A}(u)-A(u)\|\dd u = \Or(\epsilon (b-a)).
\]

With this additional step, using the same algorithm as in \cite[Theorem 3]{LowWiebe2019}, we can encode $\Xi$ with the following costs:
\begin{lem}[Short-time evolution through truncated Dyson series]
\label{lem:short_time_evo}
% Let $\mathrm{MAT}$ be a time-dependent matrix encoding of $A(t)$ on time interval $[a,b]$ as defined in Definition \ref{defn:time_dep_matrix_encoding}. 
Suppose $A\in BV([a,b])$, and is accessed through an $(n_q,m,a,b,\alpha,\epsilon)$-$\mathrm{MAT}$ as defined in Definition \ref{defn:time_dep_matrix_encoding} for some $\epsilon \leq \epsilon'/(2(b-a))$. 
$b-a \leq (2\alpha)^{-1}$,
% $2^{n_q}=\Theta(\frac{(b-a)^2}{\epsilon'}(\braket{\|\dot{A}\|}+\max_{t\in[a,b]}\|A(t)\|^2))$, where $\braket{\|\dot{A}\|}=\frac{1}{b-a}\int_a^b \|\dot{A}(t)\|\dd t$. 
$2^{n_q}=\Theta(\frac{1}{\epsilon'}((b-a)V_a^b(A)+1))$, where $V_a^b(A)$ is the total variation of $A(t)$ on the interval $[a,b]$.
Then we can construct an $(\alpha',m',\epsilon')$-block encoding of $\Xi=\mathcal{T}e^{\int_a^b A(t)\dd t}$ using $\Or(\frac{\log(\epsilon'^{-1})}{\log\log(\epsilon'^{-1})})$ queries to $\mathrm{MAT}$. Here $\alpha'=\Or(1)$ and $m'=\Or(m+n_q)$.  We also use $\Or(m+n_q+\polylog(\alpha\epsilon'^{-1}))$ additional elementary gates.
\end{lem}

%1. The condition $\epsilon \leq \epsilon'/(2(b-a))$ is to ensure that the error caused by inaccurate block encoding itself does not dominate. Also note that this condition is reasonable for sparse matrix input model as this precision can be made arbitrarily small.

In \cite[Theorem 3]{LowWiebe2019}, $n_q$ is chosen to satisfy
\[
2^{n_q} = \Theta\left(\frac{1}{\epsilon'}\left((b-a)\int_a^b\|\dot{A}(t)\|\dd t+(b-a)^2\max_{t\in[a,b]}\|A(t)\|^2\right)\right).
\]
The discussion on the numerical quadrature error in \cref{sec:err_bound_total_variation} shows that we can further relax the regularity condition so that $\int_a^b\|\dot{A}(t)\|\dd t$ can be replaced by $V_a^b(A)$ defined in \cref{eq:defn_total_variation_first}. 
%We made the following changes to this expression in Lemma \ref{lem:short_time_evo}: firstly, . This can be done because in \cite[Theorem 3]{LowWiebe2019}, the reason that $\int_a^b\|\dot{A}(t)\|\dd t$ appears is because it is used in bounding the error that comes from numerical integration. The total variation $V_a^b(A)$ can be used for the same purpose, and it applies to a larger class of functions that may not be continuously differentiable. For more detailed discussion see Appendix \ref{sec:err_bound_total_variation}. In fact, if $A(t)$ is continuously differentiable on $[a,b]$, then $V_a^b(A)=\int_a^b\|\dot{A}(t)\|\dd t$. Secondly, we note that because $\|A(t)\|\leq \alpha$ and $b-a\leq 1/{2\alpha}$, $(b-a)^2\max_{t\in[a,b]}\|A(t)\|^2=\Or(1)$.

\subsection{Block encoding of the long time  evolution operator}
\label{sec:dyson_long_time}
We choose $t_l$'s so that $t_l-t_{l-1}\leq (2 \alpha)^{-1}$. Consequently the total number of segments are $L=\Theta(\alpha T)$. For each segment $[t_{l-1},t_l]$, we construct a time-dependent matrix encoding of $A(t)$ using the sparse matrix oracles in \eqref{eq:sparse_matrix_oracles}, and then implement the short time evolution operator $\Xi_l=\mathcal{T}e^{\int_{t_{l-1}}^{t_l}A(t)\dd t}$.  By \cref{lem:short_time_evo}, this procedure yields an $(\alpha_l,m_l,\epsilon_l\|\Xi_l\|)$-block encoding of $\Xi_l$, which we denote by $U_l$.  Since
\[
e^{-1/2}\leq e^{-\int_{t_{l-1}}^{t_l}\|A(t)\|\dd t} \leq \|\Xi_l\|\leq e^{\int_{t_{l-1}}^{t_l}\|A(t)\|\dd t} \leq e^{1/2}, 
\]
the number of queries $\mathrm{MAT}$ is $\Or\left(\frac{\log(\|\Xi_l\|^{-1}\epsilon_l^{-1})}{\log\log(\|\Xi_l\|^{-1}\epsilon_l^{-1})}\right)=\Or\left(\frac{\log(\epsilon_l^{-1})}{\log\log(\epsilon_l^{-1})}\right)$. Each $\alpha_l=\Or(1)$, and each $m_l$ satisfies
% \[
% m_l = \Or\left(n+\max_l \log\left(\frac{(t_l-t_{l-1})^2}{\epsilon_l}\left(\braket{\|\dot{A}\|}_l+1\right)\right)\right) = \Or\left(n+\max_l\log\left(\frac{\braket{\|\dot{A}\|}_l}{d^2\epsilon_l}\right)\right),
% \]
\[
\begin{aligned}
m_l &= \Or\left(m+\max_l \log\left(\frac{1}{\epsilon_l}\left((t_l-t_{l-1})V_{t_{l-1}}^{t_l}(A)+1\right)\right)\right) \\ 
&\leq \Or\left(m+\max_l \log\left(\frac{1}{\epsilon_l}\left(V_{0}^{T}(A)/\alpha+1\right)\right)\right).
\end{aligned}
\]
We need $\Or\left(m+\max_l \log\left(\frac{1}{\epsilon_l}\left(V_{0}^{T}(A)/\alpha+1\right)\right)\right)$ additional elementary gates and additional $m'=\Or\left(m+\max_l \log\left(\frac{1}{\epsilon_l}\left(V_{0}^{T}(A)/\alpha+1\right)\right)\right)$ ancilla qubits as a working register. The working register starts in state $\ket{0^{m'}}$ and will be returned to $\ket{0^{m'}}$ at each time step, and can therefore be reused.

We introduce the shorthand notation
\begin{equation}
P = \prod_{l=1}^L\|\Xi_l\|,
\end{equation}
and use \cref{thm:short_time_to_long_time} to piece together the short time integrators into a block encoding approximating the long time evolution $\mathcal{T}e^{\int_0^T A(t)\dd t}=\Xi_L\cdots\Xi_2\Xi_1$. This yields an $(\alpha_{\mathrm{comp}},m_{\mathrm{comp}},\epsilon_{\mathrm{comp}})$-block encoding of $\mathcal{T}e^{\int_0^T A(t)\dd t}$, where
\[
\frac{P}{2(1-\delta)^L}\leq \alpha_{\mathrm{comp}} \leq \frac{e^{1/2}P}{(1-\delta)^L},
\]
\[
m_{\mathrm{comp}} =\lceil\log_2(L)\rceil+\Or\left(m+\max_l \log\left(\frac{1}{\epsilon_l}\left(V_{0}^{T}(A)/\alpha+1\right)\right)\right),
\]
and 
\begin{equation} \label{eq:eps_comp_dyson}
\epsilon_{\mathrm{comp}} = e^{1/2}\left(L\epsilon'+\sum_l\epsilon_l\right)P.
\end{equation}
In this block encoding we use each $U_l$ $\Or\left(\frac{1}{\delta}\log\left(\frac{1}{\epsilon'}\right)\right)$ times. 

We now choose $\delta=1/(2L)$, and $\epsilon_l=\Or(\epsilon_{\mathrm{comp}}/(LP))$, $\epsilon'=\Or(\epsilon_{\mathrm{comp}}/(LP))$. Also recall that $L=\Or(\alpha T)$. With this choice of parameters we have
\begin{equation}
\label{eq:alpha_comp_dyson}
    \alpha_{\mathrm{comp}} =\Theta(P)= \Theta\left(\prod_{l=1}^L\|\Xi_l\|\right),
\end{equation}
and
% \begin{equation}
%     \label{eq:m_comp_dyson}
%     m_{\mathrm{comp}} =\lceil\log_2(dT)\rceil+\Or\left(n+\log\left(\frac{\max_l\braket{\|\dot{A}\|}_l T\prod_{l'=1}^L\|\Xi_{l'}\|}{d\epsilon_{\mathrm{comp}}}\right)\right).
% \end{equation}
\begin{equation}
    \label{eq:m_comp_dyson}
    m_{\mathrm{comp}} =\lceil\log_2(\alpha T)\rceil+\Or\left(m+\log\left(\frac{(V_0^T(A)+\alpha)TP}{\epsilon_{\mathrm{comp}}}\right)\right).
\end{equation}
Each $U_l$ is used $\Or\left(\alpha T\log\left(\frac{\alpha TP}{\epsilon_{\mathrm{comp}}}\right)\right)$ times. Given the fact that each $U_l$ uses queries to $\mathrm{MAT}$ $\Or\left(\frac{\log(\epsilon_l^{-1})}{\log\log(\epsilon_l^{-1})}\right)$ times, and that there are $L=\Or(\alpha T)$ of them, the total number of queries to $\mathrm{MAT}$ are
\begin{equation}
\label{eq:num_queries_block_encode_dyson}
    \Or\left(\alpha^2T^2\frac{\log(\alpha T\prod_{l'=1}^L\|\Xi_{l'}\|/\epsilon_{\mathrm{comp}})}{\log\log(\alpha T\prod_{l'=1}^L\|\Xi_{l'}\|/\epsilon_{\mathrm{comp}})}\right).
\end{equation}

% \YT{deal with the $A$ derivative here!}
We summarize the result of the construction of the block encoding for the long-time evolution using truncated Dyson series as follows.
\begin{thm}[Block encoding of long-time Dyson series evolution]
\label{thm:block_encode_dyson}
% $A(t)$ is a $d$-sparse matrix with $\|A(t)\|\leq 1$ for $t\in[0,T]$. 
We assume $A\in BV([0,T])$, and $V_0^T(A)$ is its total variation on $[0,T]$. Suppose $A$ is accessed through an $(n_q, m, a, b, \alpha, \epsilon)$-$\mathrm{MAT}$ as defined in \cref{defn:time_dep_matrix_encoding}. Let $0=t_0<t_1<\cdots<t_L=T$ satisfy $t_l-t_{l-1}\leq 1/(2\alpha)$. 
We denote 
\begin{equation*}
\Xi_l=\mathcal{T}e^{\int_{t_{l-1}}^{t_l} A(t)\dd t}, \quad P = \prod_{l=1}^L\|\Xi_l\|.
\end{equation*}
Then we can construct an $(\alpha_{\mathrm{comp}},m_{\mathrm{comp}},\epsilon_{\mathrm{comp}})$-block encoding of $\mathcal{T}e^{\int_0^T A(t)\dd t}$, where $\alpha_{\mathrm{comp}}$, $m_{\mathrm{comp}}$ and $\epsilon_{\mathrm{comp}}$ are given in \cref{eq:alpha_comp_dyson,eq:m_comp_dyson,eq:eps_comp_dyson}, respectively. In this block encoding the number of times we need to use:
\begin{itemize}
    \item queries to $\mathrm{MAT}$ as given by \cref{eq:num_queries_block_encode_dyson}; 
    \item $m'=\Or\left(m+\polylog(V_0^T(A) d TP \epsilon_{\mathrm{comp}}^{-1})\right)$
additional ancilla qubits that start in, and will be returned to $\ket{0^{m'}}$;
    \item $\Or\left(\alpha^2 T^2 \left(m +\polylog(V_0^T(A) d TP \epsilon_{\mathrm{comp}}^{-1})\right)\right)$ additional elementary gates.
\end{itemize}
 
\end{thm}

% \subsubsection{Success probability and Amplitude Amplification}
\subsection{Success probability and main result for Dyson series approach} \label{sec:prob_main_dyson}
We can now apply the block encoding to an initial state $\ket{\psi(0)}$ to get the final state
$\ket{\psi(T)}=\mathcal{T}e^{\int_{0}^{T}A(t)\dd t}\ket{\psi(0)}$. We assume that $\|\ket{\psi(0)}\|=1$. Directly applying the block encoded time evolution operator will introduce an error, and we want to control the resulting error in the final normalized state. This can be done through the following lemma:
\begin{lem}
\label{lem:normalized_err}
If $\|\ket{\psi}-\ket{\phi}\|\leq \frac{1}{2}\|\ket{\psi}\|$, then 
\[
\left\|\frac{\ket{\psi}}{\|\ket{\psi}\|}-\frac{\ket{\phi}}{\|\ket{\phi}\|}\right\| \leq \frac{4\|\ket{\psi}-\ket{\phi}\|}{\|\ket{\psi}\|}.
\]
\end{lem}
\begin{proof}
Let $\ket{R}=\ket{\psi}-\ket{\phi}$. Then
\begin{equation}
    \label{eq:normalized_error}
    \begin{aligned}
    \left\|\frac{\ket{\psi}}{\|\ket{\psi}\|}-\frac{\ket{\phi}}{\|\ket{\phi}\|}\right\| &\leq \frac{\|\ket{\psi}-\ket{\phi}\|}{\|\ket{\psi}\|} + \|\ket{\phi}\|\Big|\frac{1}{\|\ket{\psi}\|}-\frac{1}{\|\ket{\phi}\|}\Big| \\
    &\leq\frac{\|\ket{R}\|}{\|\ket{\psi}\|} + \frac{(\|\ket{\psi}\|+\|\ket{R}\|)\|\ket{R}\|}{\|\ket{\psi}\|(\|\ket{\psi}\|-\|\ket{R}\|)} \\
    &\leq \frac{4\|\ket{R}\|}{\|\ket{\psi}\|}.
    \end{aligned}
\end{equation}
\end{proof}

We first use Theorem \ref{thm:block_encode_dyson} to construct an $(\alpha_{\mathrm{comp}},m_{\mathrm{comp}},\epsilon_{\mathrm{comp}})$-block encoding of $\Xi=\mathcal{T}e^{\int_{0}^{T}A(t)\dd t}$. We denote this block encoding by $W$.
Let $\wt{\Xi}$ be the time evolution operator that is exactly encoded in $W$. Then if we successfully prepare the state $\ket{\wt{\psi}(T)}=\wt{\Xi}\ket{\psi(0)}$ by applying the block encoding and measuring the ancilla qubits, the error will be
\[
\|\ket{\psi(T)}-\ket{\wt{\psi}(T)}\|\leq \|\Xi-\wt{\Xi}\|\leq \epsilon_{\mathrm{comp}}.
\]
Therefore, by Lemma \ref{lem:normalized_err}, in order to ensure that the error in the normalized state is upper bounded by $\epsilon$, i.e.,
\begin{equation}
    \label{eq:normalized_state_err_criterion}
    \left\|\frac{\ket{\psi(T)}}{\|\ket{\psi(T)}\|}-\frac{\ket{\wt{\psi}(T)}}{\|\ket{\wt{\psi}(T)}\|}\right\| \leq \epsilon,
\end{equation}
it suffices to choose $\epsilon_{\mathrm{comp}}=\Or(\epsilon\|\ket{\psi(T)}\|)$.

Upon measuring the ancilla qubits, if all measurement outcomes are $0$, then we have successfully prepared the state $\ket{\wt{\psi}(T)}$ that approximates the exact solution $\ket{\psi(T)}$. This happens with probability that is at least $\|\ket{\wt{\psi}(T)}\|^2/\alpha_{\mathrm{comp}}^2$. Using the scaling of $\alpha_{\mathrm{comp}}$ in \cref{eq:alpha_comp_dyson}, and the fact that $\|\ket{\wt{\psi}(T)}\|=(1+\Or(\epsilon))\|\ket{\psi(T)}\|$, the success probability is $\Omega(Q^{-2})$, where 
\[
Q = \frac{\prod_{l=1}^L\|\Xi_l\|}{\|\ket{\psi(T)}\|}
\]
is the same as that defined in \cref{eq:defn_Q_intro}.
Suppose that the initial state $\ket{\psi(0)}$ can be prepared using a unitary circuit $U_{\mathrm{init}}$, then we can boost the success probability to $2/3$ with $\Or(Q)$ rounds of amplitude amplification. 
With the above analysis, we can determine the cost of our algorithm, which we state in the following corollary.
\begin{thm}[Time-marching based solver using truncated Dyson series]
 \label{thm:truncated_dyson_alg}
 Let $\ket{\psi(t)}$ be the solution to the problem in \cref{eq:ode_general}.
 Suppose we have a unitary circuit $U_{\mathrm{init}}$ that satisfies $U_{\mathrm{init}}\ket{0^n}=\ket{\psi(0)}$.
 For coefficient matrix $A \in BV([0,T])$, suppose $0=t_0<t_1<\ldots<t_L=T$. For each segment $[t_{l-1},t_l]$ we have a time-dependent matrix encoding of $A(t)$ denoted as $\mathrm{MAT}_l$ that is an $(n_q, m, t_{l-1}, t_{l}, \alpha, \epsilon'')-\mathrm{MAT}$ as defined in \cref{defn:time_dep_matrix_encoding} for some $\epsilon''< {\epsilon}/({2TQ})$, and $t_l-t_{l-1}\leq (2\alpha)^{-1}$ for all $l$.
%  Suppose $0=t_0<t_1<\ldots<t_L=T$, and for each segment $[t_{l-1},t_l]$ we have a time-dependent matrix encoding $\mathrm{MAT}_l$ of $A(t)$. $t_l$'s satisfy $t_l-t_{l-1}\leq (2\alpha)^{-1}$. 
% For $U_{\mathrm{val}}$ the register storing time $t$ contains $n_q$ qubits, with
%  $2^{n_q}=\Theta(\frac{TQ}{d\epsilon}(\braket{\|\dot{A}\|}+\max_{t\in[0,T]}\|A(t)\|^2))$, where $\braket{\|\dot{A}\|}=\frac{1}{T}\int_0^T \|\dot{A}(t)\|\dd t$. 
Then we can prepare, with probability at least $2/3$, a quantum state $\ket{\wt{\psi}(T)}$ that satisfies
 \[
\left\|\frac{\ket{\psi(T)}}{\|\ket{\psi(T)}\|}-\frac{\ket{\wt{\psi}(T)}}{\|\ket{\wt{\psi}(T)}\|}\right\|= \Or(\epsilon),
\]
using 
\[
\Or\left(\alpha^2 T^2 Q\log(\alpha T Q)\frac{\log(\alpha T Q \epsilon^{-1})}{\log\log(\alpha T Q \epsilon^{-1})}\right)
\]
queries to all $\mathrm{MAT}_l$, and $\Or(Q)$ applications of (controlled-) $U_{\mathrm{init}}$ and its inverse. Here $Q$ is defined in \cref{eq:defn_Q_intro}. In total we use $\Or(n+m+\polylog(V_0^T(A) \alpha TQ\epsilon^{-1}))$ qubits, and $\wt{\Or}(\alpha^2T^2 Q(m+\polylog(V_0^T(A)\alpha TQ\epsilon^{-1})))$ additional elementary gates. Success is flagged by the measurement result of a qubit.
\end{thm}

\subsection{Application to sparse matrix input model}\label{sec:app_SM}

In this section, we discuss the complexity of our time-marching strategy using the sparse matrix input model. 
The same input model has also been used in other QLSA-based differential equation solvers~\cite{Berry2014,BerryChildsOstranderWang2017,ChildsLiu2020,Krovi2022} as well as time-dependent Hamiltonian simulation algorithms~\cite{BerryChildsSuEtAl2020}.
We assume that $A(t)$ is a $d$-sparse matrix with $\|A(t)\|\leq 1$, and that the locations of non-zero elements are time-independent. The information of $A(t)$ is given through the following oracles:
\begin{equation}
    \label{eq:sparse_matrix_oracles}
    \begin{aligned}
    & U_{\mathrm{row}}\ket{j,s} = \ket{j,\mathrm{row}(j,s)},\\
    & U_{\mathrm{col}}\ket{j,s} = \ket{j,\mathrm{col}(j,s)},\\ 
    & U_{\mathrm{val}}\ket{t,j,k,z}=\ket{t,j,k,z\oplus A_{jk}(t)}.
    \end{aligned}
\end{equation}
Here $\mathrm{row}(j,s)$ is the row index of the $s$th nonzero element in the $j$th column, $\mathrm{col}(j,s)$ is the column index of the $s$th nonzero element in the $j$th row. 
We can use \cite[Lemma 48]{GilyenSuLowEtAl2018arxiv} to construct $\mathrm{MAT}$ in \cref{defn:time_dep_matrix_encoding}
using the sparse matrix oracles in \cref{eq:sparse_matrix_oracles}. 
To construct a $(n_q,m,a,b,\alpha,\epsilon)$-$\mathrm{MAT}$ we need a single query to both $U_{\mathrm{row}}$ and $U_{\mathrm{col}}$, and two queries to $U_{\mathrm{val}}$.
% This construction uses a single query to both $U_{\mathrm{row}}$ and $U_{\mathrm{col}}$, and two queries to $U_{\mathrm{val}}$. The $\mathrm{MAT}$ constructed this way satisfy
% \[
% (I_{n_{q}}\otimes\bra{0^{n+3}}\otimes I_n)\mathrm{MAT}(I_{n_{q}}\otimes\ket{0^{n+3}}\otimes I_n) = \sum_{\gamma=0}^{2^{n_{q}}-1}\ket{\gamma}\bra{\gamma}\otimes \frac{\wt{A}\left( \frac{\gamma T}{2^{n_{q}}}\right)}{d},
% \]
% where $n+3$ is the number of ancilla qubits needed in \cite[Lemma 48]{GilyenSuLowEtAl2019}, and
% \[
% \|\wt{A}(t)-A(t)\|\leq \epsilon_A.
% \]
% Here this error comes from the inexact implementation of single qubit rotations in \cite[Lemma 48]{GilyenSuLowEtAl2019}. 
The precision parameter $\epsilon$ can be made arbitrarily small, but to keep the error below $\epsilon$ we need $\Or(n+\log^{5/2}(d\epsilon^{-1}))$ additional elementary gates and $\Or(n_b+\log^{5/2}(d\epsilon^{-1}))$ additional ancilla qubits are needed, where $n_b$ is the number of bits to encode the binary $A_{jk}(t)$. These additional ancilla qubits can be reused. 

\begin{cor}[Time-marching based solver using truncated Dyson series with sparse matrix input model]
 \label{cor:truncated_dyson_alg_sparse_access}
 Under the same assumptions in \cref{thm:truncated_dyson_alg}, together with the assumption that $A(t)$ is a $d$-sparse coefficient matrix given by unitaries $U_{\mathrm{row}}$, $U_{\mathrm{col}}$, and $U_{\mathrm{val}}$ in \cref{eq:sparse_matrix_oracles}we can prepare, with probability at least $2/3$, a quantum state $\ket{\wt{\psi}(T)}$ solving the problem in \cref{eq:ode_general} that satisfies
 \[
\left\|\frac{\ket{\psi(T)}}{\|\ket{\psi(T)}\|}-\frac{\ket{\wt{\psi}(T)}}{\|\ket{\wt{\psi}(T)}\|}\right\|= \Or(\epsilon),
\]
using $\Or\left(d^2 T^2 Q\log(d T Q)\frac{\log(d T Q \epsilon^{-1})}{\log\log(d T Q \epsilon^{-1})}\right)$ applications of (controlled-) $U_{\mathrm{row}}$, $U_{\mathrm{col}}$, and $U_{\mathrm{val}}$ and their inverses, and $\Or(Q)$ applications of (controlled-) $U_{\mathrm{init}}$ and its inverse. 
In total we use $\Or(n+\polylog(V_0^T(A)d TQ\epsilon^{-1}))$ qubits, and $\wt{\Or}(d^2T^2 Q(n+\polylog(V_0^T(A)d TQ\epsilon^{-1})))$ additional elementary gates. Success is flagged by the measurement result of a qubit.
\end{cor}

\section{Optimality of the query complexity with respect to $Q$} \label{sec:optimality}

In this section we show that the time-marching based solver using the high-order truncated Dyson series achieves the nearly optimal query complexity with respect to the amplification ratio $Q$.
The optimality is guaranteed by the following lower bound result. 
Although we state this lower bound result in terms of a time-independent ODE, it automatically provides a lower bound for the harder problem of solving a time-dependent ODE in \cref{eq:ode_general}. The block encoding of the coefficient matrix $A$ also automatically provides a time-dependent matrix encoding needed in Theorem \ref{thm:truncated_dyson_alg}.

\begin{thm}
\label{thm:lower_bound_Q_dep}
For any given $N$, there exists a matrix $A\in\CC^{N\times N}$ that can be accessed through its $(1,m,0)$-block encoding $U_A$, a target time $T = \Or(\log(N))$, and an initial state $\ket{\psi(0)}\in\CC^{N\times N}$ with $\|\ket{\psi(0)}\|=1$ such that the following statement holds:
Let $\ket{\psi(t)}$ be the solution of the ODE $\frac{\dd}{\dd t}\ket{\psi(t)}=A\ket{\psi(t)}$ with $0\le t\le T$, i.e., $\ket{\psi(T)}=e^{AT}\ket{\psi(0)}$. Then for any $\theta>0$, there is no quantum algorithm that can prepare a quantum state $\rho$ (allowed to be a mixed state) with $\mathcal{D}(\ket{\psi(T)},\rho)\leq 1/2$, which uses $\Or(Q^{1-\theta}\poly(T))$ queries to $U_A$. 
% there does not exist an algorithm that can prepare a quantum state $\rho$ such that $\mathcal{D}(\ket{\psi(T)},\rho)\leq 1/2$, where $\mathcal{D}(\cdot,\cdot)$ denotes the trace distance between two quantum states, 
Here $\mathcal{D}(\cdot,\cdot)$ denotes the trace distance between two quantum states and
\begin{equation}
Q = \sup_L\sup_{0=t_0<t_1<\cdots<t_L=T}\frac{\prod_{l=0}^{L-1}\|e^{A(t_{l+1}-t_l)}\|}{\|\ket{\psi(T)}\|}.
\label{eqn:Q_lowerbound}
\end{equation}
\end{thm}

\begin{proof}
We assume towards contradiction that such an algorithm exists. Now we apply this hypothetical algorithm to solve the unstructured search problem, in which we are asked to find a marked binary string $\mathrm{targ}$ of length $n$. We denote $N=2^n$. The target $\mathrm{targ}$ is marked by the following oracle $U_{\mathrm{targ}}$:
\[
\begin{aligned}
& U_{\mathrm{targ}}\ket{\mathrm{targ}} = -\ket{\mathrm{targ}} \\
& U_{\mathrm{targ}}\ket{x} = \ket{x},\ x\neq \mathrm{targ}.
\end{aligned}
\]
Here $x$ is any binary string of length $n$ that is different from $\mathrm{targ}$. 

We let $A=-U_{\mathrm{targ}}$, then $-U_{\mathrm{targ}}$ is an $(1,0,0)$-block encoding of $A$. Solving the ODE up to time $T$ yields us a (unnormalized) state
\[
\ket{\psi(T)} = \frac{1}{\sqrt{N}}\left(\sum_{x\neq \mathrm{targ}}e^{-T}\ket{x}+e^T\ket{\mathrm{targ}}\right).
\]
Note that in this quantum state, the amplitude corresponding to the marked element $\mathrm{targ}$ is amplified, while the amplitudes corresponding to all other elements are suppressed. Consequently, the probability getting $\mathrm{targ}$ when measuring the state in the computational basis increases with time. At time $T$, this probability is
\[
\mathrm{Pr}_{\ket{\psi(T)}}(\mathrm{targ})=\frac{|\braket{\psi(T)|\mathrm{targ}}|^2}{\braket{\psi(T)|\psi(T)}} = \frac{1}{(N-1)e^{-4T}+1}.
\]
If we set $T=\frac{1}{4}\log(3(N-1))$, then the above probability will be $3/4$. Suppose we can prepare a state $\rho$ with $\mathcal{D}(\rho,\ket{\psi(T)})\leq 1/2$, then
\[
\mathrm{Pr}_{\rho}(\mathrm{targ}) \geq \mathrm{Pr}_{\ket{\psi(T)}}(\mathrm{targ}) - \mathcal{D}(\rho,\ket{\psi(T)}) \geq 3/4-1/2=1/4.
\]
As a result, once we prepare a copy of $\rho$, we can measure in the computational basis, and with probability at least $1/4$ we will get the marked element $\mathrm{targ}$. We can use one application of $U_{\mathrm{targ}}$ to check whether the measurement output is $\mathrm{targ}$. Therefore in order to find $\mathrm{targ}$ with probability $2/3$, we only need $\Or(1)$ copies of $\rho$.

Let us then look at the cost of preparing $\rho$. With the hypothetical algorithm for solving the ODE, we can prepare $\rho$ using $\Or(Q^{1-\theta}\poly(T))$ queries to $U_A$. Here $T=\frac{1}{4}\log(3(N-1))$ as chosen above, and in this scenario
\[
Q = \sup_L\sup_{0=t_0<t_1<\cdots<t_L=T}\frac{\prod_{l=0}^{L-1}\|e^{A(t_{l+1}-t_l)}\|}{\|\ket{\psi(T)}\|}=\frac{\|e^{TA}\|}{\|\ket{\psi(T)}\|} = \sqrt{\frac{N}{(N-1)e^{-4T}+1}} = \sqrt{\frac{4N}{3}}.
\]
Therefore, we need $\Or(N^{\frac{1-\theta}{2}}\polylog(N))=o(\sqrt{N})$ queries to prepare a single copy of $\rho$. Since only $\Or(1)$ copies are needed, we would then be able to solve the unstructured search problem using only $o(\sqrt{N})$ queries to the oracle $U_{\mathrm{targ}}$. This contradicts the lower bound for the unstructured search problem \cite[Corollary 3.5]{BennettBernstein1997}.
\end{proof}

It is worth noting that the lower bound result in \cref{thm:lower_bound_Q_dep} does not imply that the dependence on $Q$ cannot be improved for specific instances of $A$. For example, consider the following $2\times 2$ non-diagonalizable matrix
\begin{equation}
A=\begin{pmatrix}
i & 1 \\
0 & i
\end{pmatrix}.
\end{equation}
Direct calculation shows that $\norm{e^{AT}}$, and therefore $\norm{\ket{\psi(T)}}$ grows linearly in $T$. However, 
\begin{equation}
\sup_L\sup_{0=t_0<t_1<\cdots<t_L=T}\prod_{l=0}^{L-1}\|e^{A(t_{l+1}-t_l)}\|
\label{eqn:numer_lowerboundQ}
\end{equation}
grows exponentially in $T$, and so is $Q$ in \cref{eqn:Q_lowerbound}.
Applying the result of \cite{Krovi2022}, we know that the cost of the optimal solver should grow as $\poly(T)$ for any $T$. Therefore our time-marching based algorithm is suboptimal in this case. Note that if $A$ is a normal matrix, the quantity in \cref{eqn:numer_lowerboundQ} is equal to $\norm{e^{AT}}$, and this issue does not arise.

\section{Simplified implementation and first-order truncated Magnus series}
\label{sec:qHOP_alg}

The time-marching strategy can be  paired with any reasonable short-time integrators. 
The implementation of the high-order truncated Dyson series algorithms requires complicated quantum control logic for handling time-ordering operators.
In this section, we investigate the performance of time-marching based algorithms with low-order integrators that can be implemented without the explicit treatment of time-ordering operators.
The simplest example is a first-order truncated Dyson series algorithm:
\begin{equation}\label{eq:1st_dyson}
    \Xi = \mathcal{T}e^{\int_{t_j}^{t_{j+1}}A(s)\dd s } \approx
    I + \int_{t_j}^{t_{j+1}}A(s)\dd s.
\end{equation}
We can approximate the integral using numerical quadrature as in the implementation of the high-order truncated Dyson series. 
%This short-time approximation \eqref{eq:1st_dyson} has an error bounded by polynomials of $\alpha$, which gives polynomial dependence in $\alpha$ in the cost of the long-time evolution. 

A closely related first-order integrator takes the form
\begin{equation} \label{eqn:U_short_no_timeordering}
     \Xi =  \mathcal{T}e^{\int_{t_j}^{t_{j+1}}A(s)\dd s } \approx e^{\int_{t_j}^{t_{j+1}}A(s)\dd s },
\end{equation}
i.e., we perform the matrix exponentiation of the time-independent matrix $\int_{t_j}^{t_{j+1}}A(s)\dd s $ directly without further Taylor expansion. This is the first-order truncated Magnus series for approximating the time-ordered integration. In the context of Hamiltonian simulation, this strategy gives rise to the quantum highly oscillatory protocol (qHOP)~\cite{AnFangLin2022}, which exhibits commutator scaling in high-precision limit, can be insensitive to the norm and the variation of $A(t)$, and can lead to second-order convergence (i.e., superconvergence) for certain $A(t)$. 
The rest of the section analyzes the scheme \eqref{eqn:U_short_no_timeordering} for solving the ODE \eqref{eq:ode_general}.  
We verify that this scheme also exhibits the desired commutator scaling. An extreme example where this is useful is when $A(t)$ commutes with itself at any time, then \eqref{eqn:U_short_no_timeordering} becomes exact, which leads to improved accuracy and poly-logarithmic dependence in the query complexity on the precision. For general non-commuting case, we can use higher order truncation of the Magnus series to improve the accuracy dependence while keeping the commutator scaling. However, the  implementation of the quantum circuit can be more complicated.

\subsection{Short-time evolution description}
For each short time evolution, the time-ordered evolution operator is approximated by truncating its first-order Magnus expansion~\eqref{eqn:U_short_no_timeordering},
which is equivalent to directly ignoring the time-ordering operator. The integral can be further approximated using standard Riemann sum with $M$ quadrature points ($M = 2^{n_q}$) as 
\begin{equation} \label{eqn:int_quadrature}
    \int_a^b A(s)\dd s  \approx  \frac{b-a}{M}\sum_{k=0}^{M-1}A\left(a+k(b-a)/M\right). 
\end{equation}
The short-time first-order Magnus evolution operator, denoted as $\bar \Xi$ is thus given as 
\begin{equation}\label{eqn:Mag1}
    \bar \Xi = e^{ \frac{b-a}{M}\sum_{k=0}^{M-1} A\left(a+k(b-a)/M\right)}.
\end{equation}

\begin{figure}
    \centerline{
    \Qcircuit @R=1em @C=1em {
    \text{Ancilla}\quad\quad\quad\quad & \qw & \multigate{2}{\text{MAT}_j} & \qw  & \qw \\
    \text{Control}\quad\quad\quad\quad & \gate{\otimes_q\text{HAD}} & \ghost{\text{MAT}_j} & \gate{\otimes_q\text{HAD}} &  \qw \\
    \text{State}\quad\quad\quad\quad & \qw  & \ghost{\text{MAT}_j} & \qw  & \qw  \\
    }
    }
    \caption{Quantum circuit of implementing a block encoding of the Hamiltonian formulated in \cref{eqn:Mag1_LCU}. 
    % The short-time qHOP evolution operator can then be implemented according to \cref{lem:qsvt1} using the circuit described here as the input block encoding model. 
    Here $\text{HAD}$ represents the single qubit Hadamard gate. }
    \label{fig:qHOP_circuit}
\end{figure}
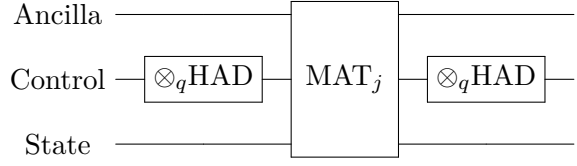

The implementation of the short-time first-order Magnus evolution operator in two steps. The first step is to construct the block encoding of \eqref{eqn:int_quadrature} via applying $\otimes_q \text{HAD}$ on the $n_q$ qubits where $\text{HAD}$ represents the single qubit Hadamard gate, applying MAT and then uncomputing, namely
\begin{align}
    & \quad \left(\bra{0}_m \otimes \bra{0}_q \right) \left(I_m\otimes (\otimes_q\text{HAD})\otimes I_n \right) \text{MAT}_j \left(I_m\otimes (\otimes_q\text{HAD})\otimes I_n \right) \left(\ket{0}_m \otimes \ket{0}_q \right) \nonumber \\
    & = \frac{1}{M \alpha} \sum_{k=0}^{M-1} A\left(a+k(b-a)/M\right) .\label{eqn:Mag1_LCU} 
\end{align}
The quantum circuit of implementing \cref{eqn:Mag1_LCU} is described in \cref{fig:qHOP_circuit}. 

The second step is to implement the time-independent matrix exponential via a contour integral formulation combined with QSVT, as detailed in \cref{sec:app_contour}. 
% Note that for normal $L$, the singular value decomposition (SVD) coincides with the eigenvalue decomposition. Hence when all eigenvalues of $L$ has non-positive real parts, i.e. for dissipative $L$,  one can construct a block encoding of $e^{hL}$ using QSVT \cite[Corollary 64]{GilyenSuLowEtAl2019}. 
The observation is that for general $A$ that are not normal, the singular value transformation no longer agrees with the eigenvalue transformation. Nevertheless, the matrix exponential can be applied by exploring the contour integral formulation proposed in \cite{TongAnWiebeEtAl2021}.  
The idea is to express $e^{A}$ as a contour integral:
\begin{equation}
e^{A}=\frac{1}{2\pi i} \oint_{\Gamma} e^{z} (z-A)^{-1} \ud z,
\end{equation}
where $\Gamma$ is a contour in the complex plane that contains all the eigenvalues of $A$ in its interior. 
We then discretize the integral using numerical quadrature. 
% such as the Gauss-Legendre quadrature formula. 
This procedure turns the matrix exponential into a linear combination of $(z_j- A)^{-1}$, which can be efficiently calculated as matrix inversion problems that does not require the matrix being normal. Here $z_j$'s are some constants determined by the numerical quadrature, which is detailed in \cref{sec:ea_algorithm}. We also remark that there are certain cases of $A$ where $e^{A}$ admits a simpler implementation using QSVT via polynomial approximations of the exponential function~\cite[Corollary 64]{GilyenSuLowEtAl2018arxiv}. Ref.~\cite{TakahiraOhashiEtAl2021} also developed a method to apply a more general class of matrix functions using the contour integral, but they did not consider how to construct a block encoding, instead focusing on preparing the output state.

\subsection{Time discretization error}
We now analyze the time-discretization error for the short-time evolution.
% , the proof of which follows \cite{AnFangLin2022}.
\begin{lem}[Time discretization errors of the first-order Magnus integrator]\label{lem:qHOP_short_error}
    Let $\Xi=\mathcal{T}e^{\int_a^b A(t)\dd t}$ denote the exact evolution operator on the time interval $[a,b]$, and $\bar{\Xi}$ denote the first-order Magnus operator defined in \cref{eqn:Mag1}. Then we have 
    \begin{align}
         \quad \left\|\Xi - \bar{\Xi}\right\| \nonumber 
        &\leq \frac{1}{2}
        \int_{a}^{b}\dd \tau \norm{ \left[\int_{a}^\tau A(s)\dd s , A(\tau) \right] }   e^{\int_a^b \norm{A(s)}\dd s}
        \\ &
        +
        \frac{b-a}{M} V_a^b(A) e^{\int_a^b \norm{A(s) \dd s} + \frac{b-a}{M} V_a^b(A)}.
        % \int_{t_j}^{t_{j+1}} \norm{ \left[\int_{t_j}^\tau L(s)\dd s , L(\tau) \right] }  e^{\int_{\tau}^{t_{j+1}} \mu(L(s)) \dd s} e^{\mu\left(\int_{t_j}^\tau L(s) \dd s\right)} \dd \tau
        % \\ &
        % +
        % \frac{h_j^2}{2M} \max_{s\in[t_j,t_{j+1}]} \norm{L'(s)} e^{\mu\left( \int_{t_j}^{t_{j+1}} L(s) \dd s  \right)+ \frac{h_j^2}{2M} \max_{s\in[t_j,t_{j+1}]} \norm{L'(s)}},
    \end{align}
    % where we recall $\mu$ denotes the logarithmic norm defined as \cref{eq:log_norm}.
\end{lem}
\begin{proof}
   The errors come from two parts: one from dropping the time-ordering operator, while the other from the numerical quadrature. To estimate the error, we introduce the notations of the exact evolution from time $a$ to $t$ as 
   \[
   \mathcal{V}(t, a): = \mathcal{T}e^{\int_a^t A(s)\dd s},
   \]
   so that $\Xi = \mathcal{V}(b,a)$, and the matrix after dropping the time-ordering operator as
    \begin{equation}
        \widetilde{V}(t,a) = e^{\int_{a}^t A(s) \dd s }. 
    \end{equation}
    We first study the approximation error between the two. For any $t \in [a,b]$, by differentiating $\widetilde{V}(t,a)$ with respect to $t$, we have 
    \begin{equation}
     \partial_t \widetilde{V}(t,a) = \int_0^1 e^{ \beta \int_{t_j}^t A(s)\dd s  }  A(t) e^{ (1 - \beta) \int_{t_j}^t A(s)\dd s  } \dd \beta. 
    \end{equation}
    The difference between $\widetilde{V}$ and $\mathcal{V}$ satisfies the differential equation
    \begin{equation}
        \partial_t \left( \widetilde{V} - \mathcal{V}\right) = A(t) \left( \widetilde{V} -\mathcal{V}  \right) +\int_0^1 e^{ \beta \int_{a}^t L(s)\dd s  }  A(t) e^{ (1 - \beta) \int_{a}^t A(s)\dd s  } \dd \beta - A(t) \widetilde{V},
    \end{equation}
    and by the variation of parameter formula~\cite{BrauerNohel2012}, we have 
    \begin{align}
        \widetilde{V}(t,a) - \mathcal{V}(t,a) = \int_{a}^t \mathcal{V}(t, \tau) g(\tau)  \dd \tau,
    \end{align}
    where \begin{align*}
         g(\tau)  := & \int_0^1 e^{ \beta \int_{a}^\tau A(s)\dd s  }  A(\tau) e^{ (1 - \beta) \int_{a}^\tau A(s)\dd s  } \dd \beta  - A(\tau) e^{\int_{a}^\tau A(s) \dd s }
        \\
           = & \int_0^1 g_0(\beta;\tau) - g_0(0; \tau) \dd \beta ,
    \end{align*}
    and \begin{equation*}
        g_0(\beta;\tau)  :=  e^{ \beta \int_{a}^\tau A(s)\dd s  }  A(\tau) e^{ (1 - \beta) \int_{a}^\tau A(s)\dd s  }.
    \end{equation*}
    Applying the fundamental theorem of calculus in terms of $\beta$ yields that
    \begin{align*}
        \norm{g(\tau)} = & \norm{\int_0^1  \int_0^\beta \partial_\gamma g_0(\gamma;\tau) \dd \gamma \dd \beta}
        \\
        = & \norm{\int_0^1 \int_0^\beta e^{\gamma \int_{a}^\tau L(s) \dd s} \left[\int_{a}^\tau A(s)\dd s , A(\tau) \right]  e^{(1-\gamma) \int_{a}^\tau A(s) \dd s} \dd \gamma  \dd \beta}
        \\
        \leq & \int_0^1 \int_0^\beta e^{\gamma \norm{\int_{a}^\tau A(s) \dd s}}\norm{ \left[\int_{a}^\tau A(s)\dd s , A(\tau) \right] }
         e^{(1-\gamma) \norm{\int_{a}^\tau A(s) \dd s}}\dd \gamma  \dd \beta 
         \\
         \leq & \frac{1}{2} \norm{ \left[\int_{a}^\tau A(s)\dd s , A(\tau) \right] } 
         e^{\int_{a}^\tau \norm{A(s)} \dd s}.
        % \\
        % \leq & \int_0^1 \dd \gamma e^{\gamma \mu\left(\int_{a}^\tau A(s) \dd s\right)}\norm{ \left[\int_{a}^\tau A(s)\dd s , A(\tau) \right] } 
        %  e^{(1-\gamma) \mu\left(\int_{a}^\tau A(s) \dd s\right)}
        %  \\
        %  \leq &  \norm{ \left[\int_{a}^\tau A(s)\dd s , L(\tau) \right] } 
        %  e^{\mu\left(\int_{a}^\tau A(s) \dd s\right)},
    \end{align*}
    Note that the exponential factor here is bounded quite loose, and in fact one can get a sharper bound by using the logarithmic norm. But here for our purposes, it is sufficient to estimate at the level of $\norm{A(s)}$. Another useful fact is 
    \begin{equation}\label{eq:norm_time_order_exp}
        \norm{\mathcal{T} e^{\int_{\beta_1}^{\beta_2} A(s) \dd s} } \leq e^{\int_{\beta_1}^{\beta_2}  \norm{A(s)}\dd s},
    \end{equation}
    which can be shown by applying the Gronwall's inequality in terms of $\beta_2$ to
    \[
   \norm{ \Phi(\beta_2, \beta_1) } \leq 1 + \int_{\beta_1}^{\beta_2}  \norm{A(s)} \norm{\Phi(s, \beta_1)} \dd s,
    \]
    where $\Phi(\beta_2, \beta_1) := \mathcal{T} e^{\int_{\beta_1}^{\beta_2}  A(s) \dd s}$ is the fundamental matrix. Thus, the approximation error of the first part can be summarized as
      \begin{align} \label{eq:qhop_err1}
        \norm{\widetilde{V}(b, a)-\Xi} = \norm{\widetilde{V}(b, a) - \mathcal{V}(b, a)} \leq
        \int_{a}^{b}\dd \tau \norm{ \left[\int_{a}^\tau A(s)\dd s , A(\tau) \right] }   e^{ \int_a^b \norm{A(s)} \dd s} .
    \end{align}
    % \begin{align} \label{eq:qhop_err1}
    %     \norm{\Xi - \widetilde{V}(b, a) } = \norm{\widetilde{V}(b, a) - \mathcal{V}(b, a)} \leq
    %     \int_{a}^{b} \norm{ \left[\int_{a}^\tau A(s)\dd s , A(\tau) \right] }  e^{\int_{\tau}^{b} \mu(A(s)) \dd s} e^{\mu\left(\int_{a}^\tau A(s) \dd s\right)} \dd \tau.
    % \end{align}
   
   We now estimate the error induced by the numerical quadrature using the total variation. Following the result in \cref{sec:err_bound_total_variation}, we have
    \begin{equation}
       \norm{E(b, a)} : =  \norm{\int_a^b A(s) \dd s  - \frac{b-a}{M}\sum_{k=0}^{M-1}L(a+k(b-a)/M)} \leq \frac{b-a}{M} V_a^b(A), 
    \end{equation}
    where we recall that $V_a^b(A)$ is the total variation of $A(t)$ on the interval $[a,b]$.
    Furthermore, the variation of parameter formula gives that for square matrices $B$ and $E$,
    % \begin{equation}\label{eq:e_B+E}
    % \norm{e^{B + E}} \leq e^{\mu(B+E)} \leq e^{\mu(B) + \mu(E)} \leq e^{\mu(B) + \norm{E}}.
    % \end{equation}
    % and a priori estimate of $V_1$ is given by
    % \begin{align*}
    %     \norm{V_1(t, t_j)} \leq 
    %     % \sup_{s \in [t, t_j]} 
    %     e^{\mu(\int_{t_j}^{t} L(s) \dd s )}e^{\norm{E(t_{j+1}, t_j)}}.
    % \end{align*}
    % By variation of parameter formula and \eqref{eq:e_B+E}, we have
    \begin{equation*}
    e^{B+E} - e^{B} =\int_0^1 \dd \beta e^{B (1-\beta)} E e^{(B+E)\beta},
    \end{equation*}
    which implies that 
    \begin{equation} \label{eq:err_exp_matrix}
    \norm{e^{B+E} - e^{B}} 
    % &\leq \sup_{\beta \in [0,1]} \norm{E} e^{(1-\beta)\mu(B) } e^{\beta \mu(B) + \beta \norm{E}} 
    \leq \norm{E} e^{\norm{B} +  \norm{E}}.
    \end{equation}
    Therefore, 
    \begin{align} \label{eq:qhop_err2}
        \norm{\widetilde{V}(b, a) -\bar\Xi} \leq \frac{b-a}{M} V_a^b(A) e^{ \int_a^b \norm{A(s)} \dd s + \frac{b-a}{M} V_a^b(A)}.
        % & \frac{h_j^2}{2M} \max_{s\in[t_j,t_{j+1}]} \norm{L'(s)} e^{\mu\left(\int_{t_j}^{t_{j+1}} L(s) \dd s  \right) + \frac{h_j^2}{2M} \max_{s\in[t_j,t_{j+1}]} \norm{L'(s)}}.
    \end{align}
    Combining \eqref{eq:qhop_err1} and \eqref{eq:qhop_err2} yields the desired result.
\end{proof}

\subsection{Algorithm for implementing $e^A$}\label{sec:ea_algorithm}

In order to discuss the complexity of implementing the short-time first-order Magnus evolution operator, we first briefly discuss in this section how to implement a matrix exponential of a general time-independent matrix, namely to implement $f(A) = e^{A}$, when $A$ is given through its $(\alpha,m,0)$-block encoding. The details are laid out in \cref{sec:app_contour}.

The main idea is to use the contour integral representation of a matrix function
\begin{equation}
\label{eq:cauchy_formula_exp}
    f(A) = \frac{1}{2\pi i}\int_{\Gamma} e^z(z-A)^{-1}\dd z,
\end{equation}
where $\Gamma$ is a unit circle with radius $\beta$: $\Gamma=\{z=\beta e^{i\theta}:\theta\in\RR\}$.
This contour integral can be discretized as
\begin{equation}
    \label{eq:trapezoidal_rule_exp}
    f_K(A) = \frac{1}{K}\sum_{k=0}^{K-1}e^{z_k} z_k (z_k-A)^{-1},
\end{equation}
for $z_k=\beta e^{i2\pi k/K}$.
The error that comes from this discretization can be analyzed using Lemma \ref{lem:trapezoidal_rule_error}, which is modified from \cite[Theorem 18.1]{Trefethen2014} and \cite[Proposition 5]{TakahiraOhashiEtAl2021}. We choose $\beta=2\alpha$, $R=4\alpha$, and by Lemma \ref{lem:trapezoidal_rule_error} we have
\begin{equation}
    \label{eq:trapezoidal_rule_err_exp}
    \|f(A)-f_K(A)\| = \Or(e^{4\alpha}2^{-K}).
\end{equation}

Our goal is to construct a block encoding of $e^A$ using \eqref{eq:trapezoidal_rule_exp}. We will proceed as follows: we first construct a block encoding of 
\begin{equation}
\label{eq:defn_Xi}
    \Xi = \sum_{k=0}^{K-1} \ket{k}\bra{k}\otimes (z_k-A),
\end{equation}
where $z_k = \beta e^{i2\pi k/K}$. Inverting $\Xi$ using QSVT gives us
\begin{equation}
\label{eq:Xi_inverse}
    \Xi^{-1} = \sum_{k=0}^{K-1} \ket{k}\bra{k}\otimes (z_k-A)^{-1}.
\end{equation}
Now suppose we can prepare quantum states $\ket{\mathrm{COEF}_{\mathrm{int}}}$, $\ket{\mathrm{COEF}'_{\mathrm{int}}}$ that satisfy
\begin{equation*}
    \ket{\mathrm{COEF}_{\mathrm{int}}} \propto \sum_k \sqrt{e^{z_k} z_k}\ket{k},\quad 
    \ket{\mathrm{COEF}'_{\mathrm{int}}} \propto \sum_k (\sqrt{e^{z_k} z_k})^*\ket{k},
\end{equation*}
then we obtain the desired block encoding
\begin{equation}
\label{eq:LCU_COEF_Xi}
    (\bra{\mathrm{COEF}'_{\mathrm{int}}}\otimes I_n) \Xi^{-1} (\ket{\mathrm{COEF}_{\mathrm{int}}}\otimes I_n) \propto f_K(A)\approx f(A).
\end{equation}
Here for a complex number $z\in\CC$, $\sqrt{z}$ can be any $w\in\CC$ such that $w^2=z$.

To construct this block encoding we  need additional ancilla qubits. The entire circuit acts on five different registers, as shown in Figure \ref{fig:circuit_contour3}. The first register contains one qubit and is used for matrix inversion through QSVT. The second register contains $\log_2(K)$ qubits to encode the $k$ coefficients $e^{z_k}z_k$ in the amplitude. The third register contains one qubit and is used in the block encoding of each $z_k-A$. The fourth register contains $m$ qubits, and this is the ancilla register used in the block encoding $U_A$. The fifth register contains $n$ qubits and is the qubits that $A$ acts on. We will use $I_r$ to denote the identity operator acting on $r$ qubits.

The cost of this implementation can be summarized in the following lemma, whose proof is laid out in \cref{sec:app_contour}.
\begin{lem}[Matrix exponentiation]
\label{lem:block_encoding_expA}
Given a $(\alpha,m,0)$-block encoding $U_A$ of $A$, we can construct a $(4\mathcal{A}/(3\alpha),m+\Or(\log(\alpha)+\log\log(\epsilon^{-1})),\epsilon)$-block encoding of $e^A$, with $\Or(\alpha+\log(\epsilon^{-1}))$ applications of (controlled-) $U_A$ and its inverse, and $\Or(\alpha^2+\log^2(\epsilon^{-1}))$ additional elementary gates. Here
\[
\mathcal{A} = \frac{2\alpha}{K}\sum_{k=0}^K |e^{z_k}|,
\]
where $z_k=2\alpha e^{i2\pi k/K}$.
\end{lem}

\subsection{Short-time complexity of the first-order integrator}

The cost of constructing a block encoding of the short-time first-order Magnus evolution operator is given by the following lemma. 

\begin{lem}[Short-time evolution through the first-order Magnus integrator] \label{lem:qhop_short}
Suppose $A\in BV([a,b])$, and is accessed through an $(n_q,m,a,b,\alpha,\epsilon)$-$\mathrm{MAT}$ as defined in Definition \ref{defn:time_dep_matrix_encoding} for some $\epsilon \leq \epsilon'/(3(b-a))$. Let 
\[
\frac{1}{2}\int_a^b \dd \tau\norm{ \left[\int_{a}^\tau A(s)\dd s, A(\tau) \right] } \leq \beta_0,
\]
and
$b-a \leq \min\left\{  (2\alpha)^{-1}, (2\epsilon)^{-1} \right\} $.
% $2^{n_q}=\Theta(\frac{(b-a)^2}{\epsilon'}(\braket{\|\dot{A}\|}+\max_{t\in[a,b]}\|A(t)\|^2))$, where $\braket{\|\dot{A}\|}=\frac{1}{b-a}\int_a^b \|\dot{A}(t)\|\dd t$. 
% $M = 2^{n_q}=\Theta\left(\frac{4}{\epsilon'}((b-a)V_a^b(A)\right))$, 
Choose 
\[
M = 2^{n_q} = \Theta\left( \frac{(b-a) V_a^b(A) }{\beta_0} \right) ,
% M = 2^{n_q} = \Theta\left( \frac{(b-a) V_a^b(A) }{\max\left\{\int_{a}^{b}\dd \tau \norm{ \left[\int_{a}^\tau A(s)\dd s , A(\tau) \right] }, 1 \right\}} \right) ,
\]
where $V_a^b(A)$ is the total variation of $A(t)$ on the interval $[a,b]$. 
Then we can construct an $(\alpha',m',\delta')$-block encoding of $\Xi=\mathcal{T}e^{\int_a^b A(t)\dd t}$, with
\[
\alpha'=\Or\left(\frac{4\mathcal{A}}{3\alpha(b-a)} \right) = \Or(1), \quad 
\mathcal{A} = \frac{2\alpha (b-a)}{K}\sum_{k=0}^K e^{2\alpha(b-a) \cos({2\pi k/K})} = \Theta( \alpha (b-a))
% \Or(\alpha (b-a)),
\]
\[
m'=m+n_q + \Or\left(\log(\alpha(b-a))+\log\log(1/\epsilon')\right), \quad \delta' = \epsilon' + 4 \beta_0.
\]
%\[
% \delta' = \epsilon(b-a) + \epsilon' + 4 \int_a^b \dd \tau\norm{ \left[\int_{a}^\tau A(s)\dd s ,
% A(\tau) \right] },
% \delta' = e \epsilon(b-a) + \epsilon' + 4 \beta_0,
%\]
This requires
\begin{itemize}
    \item  $\Or\left( \alpha (b-a) + \log(1/\epsilon')\right)$ queries to $\mathrm{MAT}$, 
    \item  $\Or(\alpha (b-a) n_q + n_q\log(1/\epsilon') )$ additional elementary gates.
\end{itemize}

\end{lem}
\begin{proof} 
$\mathrm{MAT}$ is a time-dependent matrix encoding of $A(t)$ with precision $\epsilon$, and is hence a time-dependent matrix encoding of $\tilde{A}(t)$ with precision $0$ by \cref{defn:time_dep_matrix_encoding}. 

We first consider the cost of constructing the block encoding of 
\[
\tilde{S} : = \frac{b-a}{M}\sum_{k=0}^{M-1}\tilde{A}\left(a+k(b-a)/M\right).
\]
Let $n_q = \log_2 M$. As depicted in \eqref{eqn:Mag1_LCU}, one can construct a $\left(\alpha (b-a), m+n_q, 0 \right)$ block encoding of $\tilde{S}$, using $1$ query to $\mathrm{MAT}$ and $n_q$ additional one-qubit gates. 

We then use QSVT and contour integral formulation~\cref{lem:block_encoding_expA} to implement $e^{\tilde{S}}$. Then we obtain the following
\[
\left(\frac{4\mathcal{A}}{3\alpha(b-a)}, m + n_q + \Or\left(\log(\alpha(b-a)) + \log\log(1/\epsilon')\right),  \frac{3-e}{3}\epsilon' \right)
\]
block encoding of $e^{\tilde{S}}$, 
where 
\[
\mathcal{A} = \frac{2\alpha (b-a)}{K}\sum_{k=0}^K e^{2\alpha(b-a) \cos({2\pi k/K})} = \Theta( \alpha (b-a)).
\]
This uses
$\Or\left(\alpha(b-a)) + \log(1/\epsilon') \right)$ queries to the block encoding of $\tilde{A}$ and hence $\Or\left(\alpha(b-a)) + \log(1/\epsilon') \right)$ queries to $\mathrm{MAT}$. The additional elementary gates needed for this procedure is 
\[
\Or\left( \alpha^2(b-a)^2 + n_q \log(1/\epsilon') + n_q \alpha (b-a)
\right) = \Or\left(  n_q \log(1/\epsilon') + n_q \alpha (b-a)
\right).
\]

We now control the error between $e^{\tilde{S}}$ and $\Xi =\mathcal{T}e^{\int_a^b A(t)\dd t}$. Denote
\[
{S} : = \frac{b-a}{M}\sum_{k=0}^{M-1}{A}\left(a+k(b-a)/M\right).
\]
We first consider the error between $e^{\tilde{S}}$ and $e^S$. Notice that
\[
\norm{\tilde{S} - S} \leq (b-a) \epsilon \leq 1/2, 
\quad
\norm{\tilde{S}} \leq (b-a) \alpha \leq 1/2.
\]
By \eqref{eq:err_exp_matrix}, we have
% \eqref{lem:ODE_error_grow_short_time}
\[
\norm{e^{\tilde{S}} - e^S} \leq \norm{\tilde{S} - S} e^{\norm{\tilde{S}} + \norm{\tilde{S} - S}}
\leq e (b-a) \epsilon  < \frac{e}{3}\epsilon'
\]

Choose \[
% M = \Theta\left( \frac{(b-a) V_a^b(A) }{\max \left\{ \int_{a}^{b}\dd \tau \norm{ \left[\int_{a}^\tau A(s)\dd s , A(\tau) \right] } , 1\right\}} \right)
M = \Theta\left( \frac{(b-a) V_a^b(A) }{\beta_0} \right)
\]
so that 
\[
% \frac{b-a}{M} V_a^b(A) = \Theta\left(\int_{a}^{b}\dd \tau \norm{ \left[\int_{a}^\tau A(s)\dd s , A(\tau) \right] }\right).
\frac{b-a}{M} V_a^b(A) = \Theta\left(\beta_0\right).
\]
Thanks to \cref{lem:qHOP_short_error}, the error
\[
% \norm{e^{\tilde{A}} - \Xi} \leq \left(\int_{a}^{b}\dd \tau \norm{ \left[\int_{a}^\tau A(s)\dd s , A(\tau) \right] }\right) e^{(b-a)\sup_{s\in[a,b]} \norm{A(s)}} \left( 1 + e^{\left( \int_{a}^{b}\dd \tau \norm{ \left[\int_{a}^\tau A(s)\dd s , A(\tau) \right] } \right)}\right).
\norm{e^{\tilde{A}} - \Xi} \leq \beta_0 e^{\int_a^b \norm{A(s)} \dd s} \left( 1 + e^{ \frac{1}{2}\int_{a}^{b}\dd \tau \norm{ \left[\int_{a}^\tau A(s)\dd s , A(\tau) \right] } }\right).
\]
% Choose $a$ and $b$ such that 
% \[
% \left(\int_{a}^{b}\dd \tau \norm{ \left[\int_{a}^\tau A(s)\dd s , A(\tau) \right] }\right) = \Or(1)
% \]
By our choice of $a$ and $b$ we have that the right-hand-side is bounded by $4\beta_0$,
as
\[
\frac{1}{2}\int_{a}^{b}\dd \tau \norm{ \left[\int_{a}^\tau A(s)\dd s , A(\tau) \right] } \leq \frac{1}{2}(b-a)^2\alpha^2 \leq \frac{1}{8}.
\]
Therefore, we get a 
\[
\left(\frac{4\mathcal{A}}{3\alpha(b-a)}, m + n_q + \Or(\log(\alpha(b-a)) + \log\log(1/\epsilon'), \epsilon' + 4\beta_0 \right)
\]
block encoding of $\Xi$.

\end{proof}

%\subsection{qHOP approach for long time evolution} \label{sec:qHOP_long_time}

% Following the same procedure as \cref{sec:dyson_long_time}, we provide the complexity analysis for the long-time qHOP based approach. 

\subsection{Block encoding of the long-time evolution operator}
For simplicity, we choose a uniform temporal mesh. Let $t_l = lT/ L$ so that $0 = t_0 < t_1 < \cdots < t_l < \cdots < t_L = T$ where $L$ is the number of time steps.
We perform the short-time first-order Magnus evolution on each interval $[t_l, t_{l+1}]$ and multiply all the evolution together.

By \cref{lem:qhop_short}, each short-time first-order Magnus evolution on $[t_l, t_{l+1}]$ yields a $(\alpha_l, m_l, \epsilon_l \norm{\Xi_l})$-block encoding (denoted $U_l$) of $\Xi_l = \mathcal{T} e^{\int_{t_l}^{t_{l+1}} A(s)\dd s}$, where
\[
\alpha_l = \Or(1), \quad 
%m_l = m + \Or\left(  \log (V_{t_l}^{t_{l+1}}(A)/\beta_l ) + \log(\alpha(t_{l+1}-t_l))+\log\log(1/\delta)\right) 
m_l = m +n_{q,l} + \Or\left( \log(\alpha(t_{l+1}-t_l))+\log\log(1/\delta)\right), \quad \epsilon_l \norm{\Xi_l} = \delta + 4 \beta_l,
\]
and $\beta_l$ is an upper bound of
\[
\frac{1}{2}\int_{t_l}^{t_{l+1}} \dd \tau\norm{ \left[\int_{t_l}^\tau A(s)\dd s, A(\tau) \right] }.
\]
The construction uses $\Or\left( \alpha (t_{l+1} - t_l) + \log(1/\delta)\right)$ queries to $\mathrm{MAT}_l$ and $\Or(\alpha_l (t_{l+1} - t_l) n_{q,l} + n_{q,l}\log(1/\delta) )$ additional elementary gates with $n_{q,l} = \Or \left( \log ( \frac{(t_{l+1} - t_l) V_{t_l}^{t_{l+1}}(A)}{\beta_l} ) \right)$.
    
Applying \cref{thm:short_time_to_long_time}, we get a $(\alpha_{\mathrm{comp}},m_{\mathrm{comp}},\epsilon_{\mathrm{comp}})$-block encoding of $\mathcal{T}e^{\int_0^T A(t)\dd t}$ with
\[
\frac{P}{2(1-\delta)^L}\leq \alpha_{\mathrm{comp}} \leq \frac{e^{1/2}P}{(1-\delta)^L}.
\]
Here
\[
m_{\mathrm{comp}} =\lceil\log_2(L)\rceil+ m+ \max_l n_{q,l} + \max_l \Or\left(\log(\alpha_l(t_{l+1}-t_l))+\log\log(1/\epsilon_l)\right),
\]
and 
\begin{equation} \label{eq:eps_comp_qhop}
\epsilon_{\mathrm{comp}} = e^{1/2}\left(L\epsilon'+ L \delta + 4 \beta_{\mathrm{comp}} \right)P,
\end{equation}
where $\beta_{\mathrm{comp}} = \sum_l \beta_l $ is an upper bound of 
\[
\frac{1}{2}\sum_{l=0}^{L-1}\int_{t_l}^{t_{l+1}} \dd \tau\norm{ \left[\int_{t_l}^\tau A(s)\dd s, A(\tau) \right] }   
\leq 
\frac{1}{4}\sup_{s,\tau \in [0,T]}\norm{[A(s), A(\tau)]} \frac{T^2}{L}.
\]
In this block encoding we use $\Or\left(\frac{1}{\delta}\log\left(\frac{1}{\epsilon'}\right)\right)$ applications of each $U_l$.

Denote the upper bound of the time average $L^1$-scaling $\int_0^T \norm{A(s)}\dd s /T$ as $\overline \alpha$. Thanks to our choice of $t_l$'s, each $\beta_l$ can be upper bounded by $ \overline \alpha ^2 T^2/L^2$.
Choose 
\[
\epsilon'=\Or\left(\frac{\epsilon_{\mathrm{comp}}}{LP} \right),\quad  \delta=\Or\left(\frac{\epsilon_{\mathrm{comp}}}{LP} \right), \quad
L = \Or\left( \max\left\{\frac{{\alpha}_\mathrm{comm} T^2 P}{\epsilon_{\mathrm{comp}}}, \alpha T  \right\}\right), \quad
\delta=\frac{1}{2L},
\] 
where 
\begin{equation} \label{eqn:alpha_comm}
    {\alpha}_\mathrm{comm}: = \sup_{s,\tau \in [0,T]}\norm{[A(s), A(\tau)]}.
\end{equation}
With this choice of parameters we have
\begin{equation}
\label{eq:alpha_comp_qhop}
    \alpha_{\mathrm{comp}} =\Theta(P)= \Theta\left(\prod_{l=1}^L\|\Xi_l\|\right),
\end{equation}
and
% \begin{equation}
%     \label{eq:m_comp_dyson}
%     m_{\mathrm{comp}} =\lceil\log_2(dT)\rceil+\Or\left(n+\log\left(\frac{\max_l\braket{\|\dot{A}\|}_l T\prod_{l'=1}^L\|\Xi_{l'}\|}{d\epsilon_{\mathrm{comp}}}\right)\right).
% \end{equation}
% \begin{align}
%      & m_{\mathrm{comp}} \notag 
%     \\
%     = & m + \Or\left( \log_2(\overline \alpha T P/\epsilon_{\mathrm{comp}}) \right)
%     + \max_l  \Or \left( \log ( \frac{(t_{l+1} - t_l) V_{t_l}^{t_{l+1}}(A)}{\beta_l} ) \right)
%     + \max_l \Or\left(\log(\alpha_l(t_{l+1}-t_l)) \right) \notag
%     \\ \label{eq:m_comp_dyson}
%      = & m + \Or\left( \log_2( V_0^T(A)\alpha T P \epsilon_{\mathrm{comp}}^{-1}  ) \right)
% \end{align}
\begin{equation}%\label{eq:m_comp_dyson}
    m_{\mathrm{comp}} = m + \Or\left( \log_2( V_0^T(A)\alpha T P \epsilon_{\mathrm{comp}}^{-1}  ) \right).
\end{equation}
Each $U_l$ is used 
\[
\Or\left(L\log\left(\frac{LP}{\epsilon_{\mathrm{comp}}}\right)\right)
=  \Or\left( 
% \frac{\overline{\alpha}^2 T^2 P}{\epsilon_{\mathrm{comp}}}  
\max\left\{\frac{{\alpha}_\mathrm{comm} T^2 P}{\epsilon_{\mathrm{comp}}}, \alpha T  \right\}
\log\left(\max\left\{{{\alpha}_\mathrm{comm}  }, \alpha\right\} \frac{ T P}{\epsilon_{\mathrm{comp}}}\right) \right)
\]
 times. Given the fact that each $U_l$ uses queries to $\mathrm{MAT}_l$ $\Or\left( \alpha (t_{l+1} - t_l) + \log\left(\max\left\{{{\alpha}_\mathrm{comm}  }, \alpha\right\} T P/ \epsilon_{\mathrm{comp}} \right)\right)$ times, and that there are $L$ of them, the total number of queries to all $\mathrm{MAT}$ are
% \begin{equation}
% \label{eq:num_queries_block_encode_qHOP}
% % \widetilde{\Or}\left(  
% % \frac{\overline{\alpha}^2 \overline{\alpha}_d T^3 P}{\epsilon_{\mathrm{comp}}}
% % \right),
% \Or\left(\sum_{l = 0}^{L-1} L  \log\left(\frac{LP}{\epsilon_{\mathrm{comp}}}\right) \left(  \alpha (t_{l+1} - t_l) + \log(\frac{\overline{\alpha} T P}{\epsilon_{\mathrm{comp}}})\right) \right)
% = \Or\left( \left( \frac{\overline{\alpha}^2 \alpha T^3 P}{\epsilon_{\mathrm{comp}}} +   \frac{\overline{\alpha}^4  T^4 P^2}{\epsilon_{\mathrm{comp}}^2} \log(\frac{\overline{\alpha} T P}{\epsilon_{\mathrm{comp}}}) \right) \log\left(\frac{\overline{\alpha} T P}{\epsilon_{\mathrm{comp}}}\right)   \right)
% ,
% \end{equation}
\begin{align}
\label{eq:num_queries_block_encode_qHOP}
% \widetilde{\Or}\left(  
% \frac{\overline{\alpha}^2 \overline{\alpha}_d T^3 P}{\epsilon_{\mathrm{comp}}}
% \right),
&\Or\left(\sum_{l = 0}^{L-1} L  \log\left(\frac{LP}{\epsilon_{\mathrm{comp}}}\right) \left(  \alpha (t_{l+1} - t_l) + \log\left( \max\left\{{{\alpha}_\mathrm{comm}  }, \alpha\right\}\frac{ T P}{\epsilon_{\mathrm{comp}}}\right)\right) \right)
\notag \\
=& \Or\Bigg( \bigg( \alpha T \max\left\{\frac{{\alpha}_\mathrm{comm} T^2 P}{\epsilon_{\mathrm{comp}}}, \alpha T  \right\} +   \max\left\{\frac{{\alpha}_\mathrm{comm} T^2 P}{\epsilon_{\mathrm{comp}}}, \alpha T  \right\}^2 \log \left(\max\left\{{{\alpha}_\mathrm{comm}  }, \alpha\right\}\frac{ T P}{\epsilon_{\mathrm{comp}}} \right) \bigg) \times
\\
& \times \log\left( \max\left\{{{\alpha}_\mathrm{comm}  }, \alpha\right\}\frac{ T P}{\epsilon_{\mathrm{comp}}}\right)   \Bigg)
.%\label{eq:num_queries_block_encode_qHOP}
\end{align}
Note that when $\alpha_\mathrm{comm} = 0$, this first-order truncated series implementation in fact has the same complexity scaling as the high-order truncated Dyson series. We remark that it is also possible to get the $L^1$ scaling $\overline{\alpha} = \frac{1}{T}\int_0^T \norm{A(s)}\dd s$, by varying time step sizes in the propagation according to the average performance of the Hamiltonian as in \cite{BerryChildsSuEtAl2020,AnFangLin2022}: Let $0 = t_0 < t_1 < \cdots < t_l < \cdots < t_L = T$ where $L$ is the number of time steps and $t_1,\cdots,t_{L-1}$ are chosen such that 
\begin{equation}\label{eqn:L1_time_steps}
    \int_0^{t_1} \norm{A(s)} \dd s  = \cdots  = \int_{t_l}^{t_{l+1}} \norm{A(s)} \dd s  = \frac{1}{L} \int_{0}^{T} \norm{A(s)} \dd s , \quad 0 \leq l \leq L-1,
\end{equation}
which we shall not discuss more details here.

% Note that it is possible to replace the $\alpha$ dependence by some the discrete $L^1$-norm scaling 
% \begin{equation} \label{eq:alpha_d_def}
%     \overline{\alpha}_d = \frac{1}{T}\left(\sum_{l=0}^{L-1} \alpha_l (t_{l+1} - t_l) \right)
% \end{equation}
% that is a piecewise approximation of the continuous $L^1$ scaling $\overline{\alpha} = \frac{1}{T}\int_0^T \norm{A(s)}\dd s$, if one allows a different subnormalization factor $\alpha_l$ of $\mathrm{MAT}$ at each time interval $[t_{l-1}, t_l]$, in which case the calculation becomes more complicated and will not be discussed here.

\subsection{Success probability and main result paired with  first-order integrator }
The success probability of the first-order Magnus method follows a similar argument as the Dyson series approach as detailed in \cref{sec:prob_main_dyson}. It suffices to choose $\epsilon_{\mathrm{comp}}=\Or(\epsilon\|\ket{\psi(T)}\|)$ and the success probability is at least $\Omega(Q^{-2})$. Suppose that the initial state $\ket{\psi(0)}$ can be prepared using a unitary circuit $U_{\mathrm{init}}$, then we can boost the success probability to $2/3$ with $\Or(Q)$ rounds of amplitude amplification.  We can conclude the result when pairing the time-marching strategy with the first-order Magnus-type integrator for long time evolution as follows.

\begin{thm}[The time-marching differential equation solver paired with the first-order Magnus integrator]
 \label{thm:qhop_alg}
 Let $\ket{\psi(t)}$ be the solution to the ODE $\frac{\dd}{\dd t}\ket{\psi(t)}=A(t)\ket{\psi(t)}$.
 Suppose we have a unitary circuit $U_{\mathrm{init}}$ that satisfies $U_{\mathrm{init}}\ket{0^n}=\ket{\psi(0)}$.
 For a coefficient matrix $A \in BV([0,T])$, suppose $0=t_0<t_1<\ldots<t_L=T$, and for each segment $[t_{l-1},t_l]$ we have a time-dependent matrix encoding of $A(t)$ denoted as $\mathrm{MAT}_l$ that is an $(n_q, m, t_{l-1}, t_{l}, \alpha, \epsilon'')-\mathrm{MAT}$ as defined in \cref{defn:time_dep_matrix_encoding} for some $\epsilon''< {\epsilon}/({2TQ})$.
Let $V_0^T(A)$ be the total variation of $A$ on $[0,T]$ as defined in \eqref{eq:defn_total_variation_first}.
%  Suppose $0=t_0<t_1<\ldots<t_L=T$, and for each segment $[t_{l-1},t_l]$ we have a time-dependent matrix encoding $\mathrm{MAT}_l$ of $A(t)$. $t_l$'s satisfy $t_l-t_{l-1}\leq (2\alpha)^{-1}$. 
% For $U_{\mathrm{val}}$ the register storing time $t$ contains $n_q$ qubits, with
%  $2^{n_q}=\Theta(\frac{TQ}{d\epsilon}(\braket{\|\dot{A}\|}+\max_{t\in[0,T]}\|A(t)\|^2))$, where $\braket{\|\dot{A}\|}=\frac{1}{T}\int_0^T \|\dot{A}(t)\|\dd t$. 
We can prepare, with probability at least $2/3$, a quantum state $\ket{\wt{\psi}(T)}$ that satisfies
 \[
\left\|\frac{\ket{\psi(T)}}{\|\ket{\psi(T)}\|}-\frac{\ket{\wt{\psi}(T)}}{\|\ket{\wt{\psi}(T)}\|}\right\|= \Or(\epsilon),
\]
using
% \[
% \widetilde{\Or}\left( \left( \frac{\overline{\alpha}^2 \alpha T^3 Q}{\epsilon} +   \frac{\overline{\alpha}^4  T^4 Q^2}{\epsilon^2} \log \left(\frac{\overline{\alpha} T Q}{\epsilon} \right) \right) \log\left(\frac{\overline{\alpha} T Q}{\epsilon}\right)   \right)
% \]
% \[
% \Or\left( \frac{{\alpha}_\mathrm{comm}^2  T^4 Q^3}{\epsilon^2} \log^2 \left(\frac{\max\{ \alpha_\mathrm{comm}, \alpha\} T Q}{\epsilon} \right) \right)
% \]
% total number of queries to all $\mathrm{MAT}_l$, and $\Or(Q)$ applications of (controlled-) $U_{\mathrm{init}}$ and its inverse in the high-precision limit. The total number of queries to all $\mathrm{MAT}_l$ for general precision is given by 
% \begin{gather*}
% \Or\Bigg( \bigg( \alpha T \max\left\{\frac{{\alpha}_\mathrm{comm} T^2 Q}{\epsilon}, \alpha T  \right\} +   \max\left\{\frac{{\alpha}_\mathrm{comm} T^2 Q}{\epsilon}, \alpha T  \right\}^2 \log \left(\max\left\{{{\alpha}_\mathrm{comm}  }, \alpha\right\}\frac{ T Q}{\epsilon} \right) \bigg) \times
% \\
%  \times \log\left( \max\left\{{{\alpha}_\mathrm{comm}  }, \alpha\right\}\frac{ T Q}{\epsilon }\right)   \Bigg)
% .
% \end{gather*}
\begin{gather*}
\Or\Bigg( \bigg( \alpha T \max\left\{\frac{{\alpha}_\mathrm{comm} T^2 Q}{\epsilon}, \alpha T  \right\} +   \max\left\{\frac{{\alpha}_\mathrm{comm} T^2 Q}{\epsilon}, \alpha T  \right\}^2 \log \left(\max\left\{{{\alpha}_\mathrm{comm}  }, \alpha\right\}\frac{ T Q}{\epsilon} \right) \bigg) \times
\\
 \times \log\left( \max\left\{{{\alpha}_\mathrm{comm}  }, \alpha\right\}\frac{ T Q}{\epsilon }\right)   \Bigg)
.
\end{gather*}
total number of queries to all $\mathrm{MAT}_l$, and $\Or(Q)$ applications of (controlled-) $U_{\mathrm{init}}$ and its inverse. In particular, in the high-precision limit given a non-zero $\alpha_\mathrm{comm}$, the total number of queries to all $\mathrm{MAT}_l$ can be simplified as
\[
\Or\left( \frac{{\alpha}_\mathrm{comm}^2  T^4 Q^3}{\epsilon^2} \log^2 \left(\frac{\max\{ \alpha_\mathrm{comm}, \alpha\} T Q}{\epsilon} \right) \right).
\]
In total we use $\Or(n+m+\polylog(V_0^T(A) \alpha TQ\epsilon^{-1}))$ qubits.
Here $Q$ is defined as \eqref{eq:defn_Q_intro}, and $\alpha_\mathrm{comm}$ is defined in Eq.~\eqref{eqn:alpha_comm}.
% Here $\overline{\alpha}_d$ represents the discrete $L^1$ scaling as defined in \eqref{eq:alpha_d_def}, and
% \[
% % P = \prod_{l=0}^{L-1}\|\mathcal{T}e^{\int_{t_{l}}^{t_{l+1}}A(t)\dd t}\|,
% % \quad
% Q=\frac{\prod_{l=0}^{L-1}\|\mathcal{T}e^{\int_{t_{l}}^{t_{l+1}}A(t)\dd t}\|}{\|\mathcal{T}e^{\int_{0}^{T}A(t)\dd t}\ket{\psi(0)}\|}.
% \]
% for $0=t_0<t_1<\ldots<t_L=T$ for $t_l$'s satisfying $t_l-t_{l-1}\leq (2\alpha)^{-1}$.
% In total we use $\Or(n+\polylog(V_0^T(A) \alpha TQ\epsilon^{-1}))$ qubits \DF{incorrect now; still need to revise}, and $\wt{\Or}(\alpha^2T^2 Q(n+\polylog(V_0^T(A)\alpha TQ\epsilon^{-1})))$ additional elementary gates. 
Success is flagged by the measurement result of a qubit.
\end{thm}

\section{Discussion} \label{sec:conclusion}

We revisit the time-marching strategy for solving the ODE \eqref{eq:ode_general}, and demonstrate that it can be a useful alternative to QLSA based quantum differential equation solvers.
On a technical level,  our time-marching based algorithm achieves the following tasks for the first time: (1) It has provable performance guarantees for non-diagonalizable, and time-dependent dynamics. (2) It retains high-order accuracy for non-smooth $A(t)$. (3) It can use few queries to the initial state. For inhomogeneous linear differential equations $\frac{\rd}{\rd t} \ket{\psi(t)} = A(t) \ket{\psi(t)}+\ket{b(t)}$, we may use the variation of constants to express the solution as
\begin{equation}
\ket{\psi(t)}=\mathcal{T}e^{\int_0^t A(s)\dd s}\ket{\psi_0}+\int_0^t \mathcal{T}e^{\int_{t'}^t A(s)\dd s}\ket{b(t')}\ud t'.
\end{equation}
After discretization of the integration with respect to $t'$, this is reduced to a series of homogeneous linear differential equations that can be implemented by the time-marching method. The analysis for the inhomogeneous case can thus be similar using the tools developed in this paper.

At a high level, time-marching based methods and QLSA based methods simply amount to two different ways of rearranging the same identities of the form $\ket{\psi_{l}}=\bar{\Xi}_l \ket{\psi_{l-1}},l=1,\ldots,L$. However, there are many differences between these two classes of algorithms, and we do not know whether the gaps can be closed with further improvements, or are intrinsic to each type of the method. 
For instance, it is possible to refine the complexity analysis of QLSA based algorithms so that the cost does not directly depend on the condition number $\kappa_V$, or to improve the algorithm so that it achieves high-order accuracy for non-smooth $A(t)$~\cite{PetroBerryPrivate}. However, it is unclear whether the number of queries to the initial state of any QLSA based algorithm can be independent of $T$.
For time-marching based algorithms, though the dependence on the precision is near-optimal and scales as $\polylog(1/\epsilon)$, the dependence of the query complexity to the input matrix is $\Or(T^2)$. We do not know whether the $\Or(T^2)$ scaling in the query complexity to the matrix can be improved, or whether the dependence on $Q$ can be weakened to $\bar{Q}=\frac{\|\mathcal{T}e^{\int_{0}^{T}A(t)\dd t}\|}{\|\ket{\psi(T)}\|}$. 
It is also possible that another class of algorithms is needed to provide a unifying perspective to the questions above, and to achieve near-optimal complexity with respect to all parameters.

\section*{Acknowledgments} 

This work was partially supported by the NSF Quantum Leap Challenge Institute (QLCI) program under Grant No. OMA-2016245 and NSF DMS-2208416 (D.F.), by the U.S. Department of Energy under the Quantum Systems Accelerator program  under Grant No. DE-AC02-05CH11231 (Y.T.), and by a Google Quantum Research Award (L.L.). L.L. is a Simons Investigator. 
We thank Dong An, Dominic Berry, Andrew Childs and Jin-Peng Liu for discussions.

\bibliographystyle{abbrvurl}
\bibliography{ref_all_doi}

\appendix

\section{Notations}

Throughout the paper, $\norm{A}$ denotes the spectral norm of a matrix $A$, $\norm{v}$ denotes 2-norm of a vector $v$, and $\mathcal{D}(\cdot,\cdot)$ denotes the trace distance between two quantum states defined as
\[
\mathcal{D}(\rho,\sigma) = \frac{1}{2}\Tr\left( \sqrt{(\rho - \sigma)^2}\right).
\]
 We use the following asymptotic notations besides the usual $\Or$ notation: we write $f=\Omega(g)$ if $g=\Or(f)$; $f=\Theta(g)$ if $f=\Or(g)$ and $g=\Or(f)$; $f=\wt{\Or}(g)$ if $f=\Or(g\polylog(g))$. 
For two quantum states $\ket{x}$ and $\ket{y}$, we sometimes write $\ket{x,y}$ to denote $\ket{x}\ket{y}$. The notations regarding each short time evolution can be summarized as follows.
Note that the same $U_l$ can be viewed as a block encoding of $\Xi_\ell$ and $\bar{\Xi}_\ell$, and the difference is only in the error of the block encoding.

\setlength{\tabcolsep}{12pt}
\renewcommand*{\arraystretch}{1.2}

\begin{center}
\resizebox{1.0\textwidth}{!}{
\begin{tabular}{cll}
\hline
Notation & Meaning & Corresponding block encoding\\
\hline
$\Xi_\ell$ & the exact time evolution operator $\mathcal{T} e^{\int_{t_{l-1}}^{t_l} A(t)\dd t}$ & $U_l$ is an $(\alpha_l, m_l, \epsilon_l \norm{\Xi_l})$-block encoding of ${\Xi}_\ell$  
\\
$\bar{\Xi}_\ell$ & an approximation of the exact evolution $\Xi_\ell$ & $U_l$ is an $(\alpha_l, m_l, 0)$-block encoding of $\bar{\Xi}_\ell$ 
\\
$\widetilde{\Xi}_\ell$ & $\bar{\Xi}_\ell$ after uniform singular value amplification & $\widetilde{U}_l$ is a $(\frac{\norm{\bar{\Xi}_\ell}}{1-\delta}, m_l +1,  0)$-block encoding of $\widetilde{\Xi}_\ell$ 
\\
\hline
\end{tabular}
}
\end{center}

\section{Block encoding and quantum singular value transformation}
\label{sec:block_encoding_and_qsvt}

We call a matrix $A\in \CC^{2^n\times 2^n}$ an $n$-qubit matrix or an $n$-qubit operator. In this work we extensively use the technique of block encoding \cite{LowChuang2019hamiltonian,ChakrabortyGilyenJeffery2018,GilyenSuLowEtAl2019}, which is a way of embedding an arbitrary matrix as a sub-matrix of a larger unitary matrix. Using a unitary matrix $U_A$ to encode $A$ as a sub-matrix means that there exist a subnormalization factor $\alpha>0$ such that
\begin{equation}
    U_A = \begin{pmatrix} A/\alpha & \cdot\\ \cdot & \cdot \end{pmatrix},
\end{equation}
where $\cdot$ denotes arbitrary matrix blocks of proper sizes. In general, the matrix that we block encode may only approximate, but is not exactly equal to, $A/\alpha$.  We use the following notation to describe such encodings.

\begin{defn}[Block encoding, \cite{GilyenSuLowEtAl2019}]\label{defn:block_encoding}
An $(m+n)$-qubit unitary operator $U_A$ is called an $(\alpha, m, \epsilon)$-block encoding of an $n$-qubit operator $A$, if 
$
\norm{A-\alpha(\bra{0^m}\otimes I_n) U_A (\ket{0^m}\otimes I_n)}\leq \epsilon.
$
\end{defn}
Here $m$ is the number of ancilla qubits for block encoding, and $\alpha$ is called the subnormalization factor. %The block encoding is a powerful and versatile model, which can be used to efficiently encode density operators, Gram matrices, positive-operator valued measure (POVM), sparse-access matrices, as well as addition and multiplication of block encoded matrices (we refer to \cite{GilyenSuLowEtAl2019} for a detailed illustration of such constructions). 
With the $(\alpha,m,0)$-block encoding $U_A$, we can perform a quantum singular value transformation (QSVT) of $A$ on a quantum computer using the technique developed in \cite{GilyenSuLowEtAl2019}. More precisely, suppose $A$ has a singular value decomposition $A=U\sigma V^{\dagger}$, then for any odd polynomial $f(x)$ with degree $d$ such that $|f(x)|\leq 1$ for $x\in[-1,1]$, we can construct a block encoding of $Uf(\Sigma/\alpha)V^{\dagger}$ , and this block encoding uses $U_A$ $d$ times. This is the main tool for implementing the uniform singular value amplification in Lemma \ref{lem:uniform_singular_value_amplification}. QSVT requires finding a sequence of phase factors corresponding to the polynomial we want to implement. There are classical algorithms capable of doing this \cite{Haah2019} in polynomial time, and later works developed more efficient methods for high-degree polynomials with high precision requirements \cite{ChaoDingGilyenEtAl2020,DongMengWhaleyEtAl2021,WangDongLin2021energy,Ying2022stable}.

\section{Convex optimization based method for uniform singular value amplification}
\label{sec:convex}

In order to approximate an odd target function using an odd polynomial of degree $d$, we can express the target polynomial as the linear combination of Chebyshev polynomials with some unknown coefficients $\{c_k\}$:
\begin{equation}
F(x)=\sum_{k=0}^{(d-1)/2} T_{2k+1}(x) c_k.
\end{equation}
To formulate this as a discrete optimization problem, we first discretize $[-1,1]$ using $M$ grid points (e.g., roots of Chebyshev polynomials $\{x_j=-\cos\frac{j\pi}{M-1}\}_{j=0}^{M-1}$).  We define the coefficient matrix, $A_{jk}=T_{2k+1}(x_j), \quad k=0,\ldots,(d-1)/2$. Then the coefficients for approximating the polynomial used for uniform singular value amplification can be found by solving the following optimization problem
\begin{equation}
\begin{split}
\min_{\{c_k\}} \quad& \max\left\{\max_{x_j\in[0,{\gamma'}^{-1}]} \abs{F(x_j)-(1-\delta)\gamma' x_j}\right\}\\
\text{s.t.} \quad & F(x_j)=\sum_{k} A_{jk} c_k, \quad \abs{F(x_j)}\le c, \quad \forall j=0,\ldots,M-1.
\end{split}
\label{eqn:opt_ground}
\end{equation}
This is a convex optimization problem and can be solved using software packages such as CVX~\cite{cvx}. 
The norm constraint $\abs{F(x)}\le 1$ is relaxed to $\abs{F(x_j)}\le c$ to take into account that the constraint can only be imposed on the sampled points, and the values of $|F(x)|$ may slightly overshoot on $[-1,1]\backslash \{x_j\}_{j=0}^{M-1}$. The effect of this relaxation can be negligible when we choose $c$ to be sufficiently close to $1$ (for instance, $c$ can be $\max\{0.9999,1-0.1\delta\}$). Since \cref{eqn:opt_ground} approximately solves a min-max problem, it achieves the near-optimal solution (in the sense of the $L^{\infty}$ norm) by definition both in the asymptotic and pre-asymptotic regimes.
Once the polynomial $F(x)$ is given, the Chebyshev coefficients can be used as the input to find the phase factors using QSPPACK~\footnote{\url{https://github.com/qsppack/QSPPACK}} with an optimization based method~\cite{DongMengWhaleyEtAl2021}.

\section{The compression gadget}
\label{sec:compression_gadget}

In this section we discuss how to construct the compression gadget that is used in Section \ref{sec:aa_using_compression_gadget} and prove \cref{lem:compression_gadget}. We suppose we are given unitaries $V_1,V_2,\ldots,V_L$, each of which is a $(\alpha'_l,m'_l,0)$-block encoding of a potentially non-unitary operation $\Gamma_l$. Our goal is to construct a block encoding of $\Gamma_L\cdots\Gamma_2\Gamma_1$ without duplicating the ancilla registers in each $V_l$.

To do this we introduce a counter register to count how many times $\Gamma_l$ is applied successfully. This counter register contains $\lceil\log_2(L)\rceil+1$ qubits, and its state encodes a binary that is used to track successful applications of $\Gamma_l$. We introduce a unitary operator $\mathrm{ADD}$ on this register defined through
\[
\mathrm{ADD}\ket{c} = \ket{c+1\mod 2^{\lceil\log_2(L)\rceil+1}}.
\]
This operator performs addition modulo $2^{\lceil\log_2(L)\rceil+1}$. Its inverse perform subtraction. We initialize the counter register in the state $\ket{L}$, and subtract from the number it encodes by applying $\mathrm{ADD}^{\dagger}$ each time $\Gamma_l$ is applied successfully. We keep track of success/failure by applying controlled $\mathrm{ADD}^{\dagger}$ so that the whole process is coherent. In the end if all steps are successful the counter register will be in the state $\ket{0}$. 

The circuit construction for the above coherent procedure is described in Figure \ref{fig:compression_gadget}. We denote $m_{\max}=\max_l m'_l$, and regard each $V_l$ as acting on two registers, one is the ancilla register containing $m_{\max}$ qubits and the other the state register. If $m'_l<m_{\max}$, then we can simply let $m_{\max}-m'_l$ qubits in the ancilla register remain idle. The circuit in \cref{fig:compression_gadget} is in fact a block encoding of $\Gamma_L\cdots\Gamma_2\Gamma_1$. In this way we prove Lemma \ref{lem:compression_gadget}, which we restate here.
\begin{lem*}[Compression gadget]
% \label{lem:compression_gadget}
Suppose we are given unitaries $V_1,V_2,\ldots,V_L$, each of which is a $(\alpha'_l,m'_l,0)$-block encoding of a potentially non-unitary operation $\Gamma_l$.
Then we can construct an $(\alpha_{\mathrm{comp}},m_{\mathrm{comp}},0)$-block encoding of $\Gamma_L\cdots\Gamma_2\Gamma_1$, where
\[
\alpha_{\mathrm{comp}} = \alpha'_1\alpha'_2\cdots\alpha'_L,\quad m_{\mathrm{comp}} = \max_l m'_l+\lceil\log_2(L)\rceil+1.
\]
\end{lem*}

\section{Bounding the error due to the coefficient matrix}
\label{sec:coef_matrix_perturbation}

In this section, we show that small errors in the coefficient matrix can be controlled. 
\begin{lem}
\label{lem:ODE_error_grow_short_time}
Let $\mathcal{V}_A(t,s)=\mathcal{T}e^{\int_s^t A(u)\dd u}$, and $\mathcal{V}_B(t,s)=\mathcal{T}e^{\int_s^t B(u)\dd u}$. Then
\[
\|\mathcal{V}_A(t,s)-\mathcal{V}_B(t,s)\|\leq  e^{(t-s)\max_{u\in[s,t]}\{\|A(u)\|,\|B(u)\|\}}\int_{s}^t \|B(u)-A(u)\|\dd u.
\]
\end{lem}

\begin{proof}
We observe that
\[
\frac{\dd}{\dd t}(\mathcal{V}_A(t,s)-\mathcal{V}_B(t,s)) = A(t)(\mathcal{V}_A(t,s)-\mathcal{V}_B(t,s)) + (A(t)-B(t))\mathcal{V}_B(t).
\]
Using Duhamel's principle we have
\begin{equation}
    \label{eq:Duhamel}
\mathcal{V}_A(t,s)-\mathcal{V}_B(t,s) = \int_s^t \mathcal{V}_A(t,u)(A(t)-B(u))\mathcal{V}_B(u,s)\dd u.
\end{equation}
Because
\[
\begin{aligned}
\|\mathcal{V}_A(t,u)\| &\leq e^{\int_u^t\|A(v)\|\dd v}\leq e^{(t-u)\max_{v\in[u,t]}\|A(v)\|}, \\
\|\mathcal{V}_B(t,u)\| &\leq e^{\int_s^u\|B(v)\|\dd v}\leq e^{(u-s)\max_{v\in[s,u]}\|B(v)\|},
\end{aligned}
\]
Eq.~\eqref{eq:Duhamel} implies
\[
\|\mathcal{V}_A(t,s)-\mathcal{V}_B(t,s)\|\leq \int_s^t e^{(t-u)\max_{v\in[u,t]}\|A(v)\|+(u-s)\max_{v\in[s,u]}\|B(v)\|}\|B(u)-A(u)\|\dd u.
\]
The lemma can then be proved by observing that
\[
(t-u)\max_{v\in[u,t]}\|A(v)\|+(u-s)\max_{v\in[s,u]}\|B(v)\| \leq (t-s)\max_{u\in[s,t]}\{\|A(u)\|,\|B(u)\|\}.
\]
\end{proof}

\section{Bounding numerical integration error using the total variation}
\label{sec:err_bound_total_variation}

In this section, we bound the numerical integration error. The standard numerical quadrature results typically bound the error by the derivative of the matrix $\|\dot{A}\|$, and hence the matrix $A$ needs to be differentiable (see, e.g., \cite[Eq.~(A11)]{LowWiebe2019} or the textbook \cite{BurdenNA}), 
% In Ref.~\cite{LowWiebe2019}, the numerical quadrature error is bounded by the derivative of the Hamiltonian $\|\dot{H}\|$. 
% Note that this results holds for general matrix~\cite{BurdenNA}
% the derivative of the Hamiltonian $\|\dot{H}\|$ is used to bound 
% The numerical integration error can be estimated as
% \[
% \left\|\int_0^{\Delta} A(x_k+s)\dd s - \Delta A(x_k) \right\|\leq \int_0^{\Delta}\int_0^s\|\dot{A}(x_k+y)\|\dd y \dd s \leq \Delta \int_0^{\Delta}\|\dot{A}(x_k+y)\|\dd y.
% \]
% This works for small $\Delta$. 
Here we bound the numerical integration error by the total variation instead to account for non-differentiable cases.

For large interval $[0,T]$, we can partition the interval into segments $0=x_0<x_1<\cdots x_K=T$, $x_k-x_{k-1}=\Delta$, and the difference between the integral and the Riemann sum can be bounded through
\[
\begin{aligned}
\left\|\int_0^{T} A(s)\dd s - \Delta \sum_{k=0}^{K-1} A(x_k) \right\| &\leq \sum_{k=0}^{K-1}\left\|\int_0^{\Delta} A(x_k+y)\dd y - \Delta A(x_k)\right\| \\
&\leq \Delta \sum_{k=0}^{K-1}\int_{x_k}^{x_{k+1}}\|\dot{A}(y)\|\dd y \\
&=\Delta \int_{0}^{T}\|\dot{A}(y)\|\dd y
\end{aligned}
\]

Now we obtain similar bounds using the total variation. First for short time
\[
\left\|\int_0^{\Delta} A(x_k+s)\dd s - \Delta A(x_k) \right\|\leq \int_0^{\Delta} \|A(x_k+s)-A(x_k)\| \dd s \leq \Delta V_{x_k}^{x_k+\Delta}(A),
\]
where the second inequality is because $\|A(x_k+s)-A(x_k)\|\leq V_{x_k}^{x_k+\Delta}(A)$.
For long time
\[
\begin{aligned}
\left\|\int_0^{T} A(s)\dd s - \Delta \sum_{k=0}^{K-1} A(x_k) \right\| &\leq \sum_{k=0}^{K-1}\left\|\int_0^{\Delta} A(x_k+y)\dd y - \Delta A(x_k)\right\| \\
&\leq \Delta \sum_{k=0}^{K-1}V_{x_k}^{x_{k+1}}(A) \\
&=\Delta V_{0}^{T}(A),
\end{aligned}
\]
where we have used the fact that
\[
V_{0}^{T}(A) = \sum_{k=0}^{K-1}V_{x_k}^{x_{k+1}}(A).
\]
% Here we use $A$ to be consistent with Ref.~\cite{LowWiebe2019}, but we do not assume that it is Hermitian.

% \section{Implementing $e^A$ using QSVT and contour integral formulation} 

\section{Circuit construction of the contour integral formulation }\label{sec:app_contour}

In this section, we construct the circuit to implement $e^A$ using QSVT and the contour integral formulation. We devote \cref{sec:inverse_matrix} and \cref{sec:trapezoidal_rule} to the preliminary discussions of the block encoding of the inverse of a matrix and the quadrature discretization errors of the contour integral formulation, respectively. \cref{sec:block_encoding_Xi}-\cref{sec:use_LCU_block_encode_expA} discuss the circuit construction in details and prove \cref{lem:block_encoding_expA}.

\subsection{Block encoding the inverse of a matrix}
\label{sec:inverse_matrix}

This section follows \cite[Appendix B]{TongAnWiebeEtAl2021}. We will discuss how to build a block encoding of the inverse of a matrix $A$, given a block encoding of $A$. 

For an odd function $f$ we define the singular value transformation $f^{\diamond}$ in the following way: if $M=W_M \Sigma_M V_M^{\dagger}$, then $f^{\diamond}(M)=W_M f(\Sigma_M) V_M^{\dagger}$. This transformation can be implemented on a quantum computer using the quantum singular value transformation (QSVT) method developed in Ref.~\cite{GilyenSuLowEtAl2019}.
The matrix inversion can be implemented as a singular value transformation in the following way: when $A=W\Sigma V^{\dagger}$ is invertible, $A^{-1}=V \Sigma^{-1} W^\dagger$. Therefore 
\begin{equation}
(A/\alpha)^{-1}=(f^{\diamond}(A/\alpha))^\dagger
\label{eqn:ainv_qsvt}
\end{equation}
where $f(x)=x^{-1}$. However $f(\cdot)$ here is not bounded by 1 and is in fact singular at $x=0$. Therefore instead of approximating $f(x)=x^{-1}$ on the whole interval $[-1,1]$ we consider an odd polynomial $p(x)$ such that
\[
\left|p(x)-\frac{3\delta}{4x}\right|\leq \epsilon',\quad \forall x\in [-1,-\delta]\cup[\delta,1].
\]
and $|p(x)|\leq 1$ for all $x\in[-1,1]$. The existence of such an odd polynomial of degree $\Or(\frac{1}{\delta}\log(\frac{1}{\epsilon'}))$ is guaranteed by \cite[Corollary 69]{GilyenSuLowEtAl2019}.

Then \cite[Theorem 2]{GilyenSuLowEtAl2019} enables us to implement $(p^{\diamond}(A/\alpha))^\dagger=V p(\Sigma/\alpha)W^\dagger$. We have
\begin{equation}
\label{eq:block_encoding_err_A_inv}
    \|(p^{\diamond}(A/\alpha))^\dagger-(3\delta/4)(A/\alpha)^{-1}\|=\|p(\Sigma/\alpha)-(3\delta/4)(\Sigma/\alpha)^{-1}\|\leq \epsilon',
\end{equation}
if all diagonal elements of $\Sigma/\alpha$, i.e. the singular values of $A/\alpha$, are in the interval $[\delta,1]$. Therefore we want all singular values of $A/\alpha$ to be at least $\delta$ distance away from the origin. We then use QSVT to block encode $(p^{\diamond}(A/\alpha))^\dagger$ given a block encoding of $A$.
%Using QSVT, a $(1,m+1,0)$-block encoding of $(p^{\diamond}(A/\alpha))^\dagger$ can be implemented, and this is at the same time a block encoding of $A^{-1}$. 

We assume $A$ can be accessed by its $(\alpha,m,0)$-block encoding $U_A$. 
The singular values of $A/\alpha$ are contained in $[1/(\alpha\|A^{-1}\|),\|A\|/\alpha]$.
% Let $\kappa$ be the condition number of $A$, then the singular values of $A/\alpha$ are contained in $[\norm{A}/(\alpha\kappa),\norm{A}/\alpha]$. 
Therefore we choose $\delta = 1/(\alpha\|A^{-1}\|)$. Using QSVT, a $(1,m+1,0)$-block encoding of $p^{\diamond}(A/\alpha)$ can be implemented \cite[Theorem 2]{GilyenSuLowEtAl2019}. We denote this block encoding by $\mathcal{U}$. Then by Eq.~\eqref{eq:block_encoding_err_A_inv}
\[
\left\|\frac{4}{3\delta\alpha}(\bra{0^{m+1}\otimes I})\mathcal{U}^\dagger(\ket{0^{m+1}\otimes I})-A^{-1}\right\|
=\left\|\frac{4}{3\delta\alpha}(p^{\diamond}(A/\alpha)^\dagger)-A^{-1}\right\|
\leq \frac{4\epsilon'}{3\delta\alpha}. 
\]
% Therefore $\mathcal{U}^\dagger$ is a  $(4/(3\delta\alpha),m+1,4\epsilon'/(3\delta\alpha))$-block encoding of $A^{-1}$. 
Consequently, by our choice of $\delta$, $\mathcal{U}^\dagger$ is a  $(4\|A^{-1}\|/3,m+1,\epsilon)$-block encoding of $A^{-1}$ where $\epsilon=4\|A^{-1}\|\epsilon'/3$. 
Because the cost of QSVT scales linearly with respect to the degree of the polynomial $p(x)$, the total number of queries to to $U_A$ and its inverse is
\[
\Or\left(\frac{1}{\delta}\log\left(\frac{1}{\epsilon'}\right)\right)
=\Or\left(\alpha\|A^{-1}\|\log\left(\frac{\|A^{-1}\|}{\epsilon}\right)\right).
\]
We summarize the result in the following Lemma:
\begin{lem}
\label{lem:matrix_inversion}
We assume $A$ can be accessed by its $(\alpha,m,0)$-block encoding $U_A$. Then a $(4\|A^{-1}\|/3,m+1,\epsilon)$-block encoding of $A^{-1}$ can be constructed using $\Or\left(\alpha\|A^{-1}\|\log\left(\frac{\|A^{-1}\|}{\epsilon}\right)\right)$ applications of (controlled-) $U_A$ and its inverse.
\end{lem}
Note that in the case where $\|A^{-1}\|$ is unknown, it can be replaced with an upper bound of $\|A^{-1}\|$. 

\subsection{Evaluating contour integrals using the trapezoidal rule}
\label{sec:trapezoidal_rule}

In this section we analyze the error that comes from evaluating contour integrals using the trapezoidal rule. 
We want to evaluate a matrix function $f(A)$. Because of the fact that $A$ may not be a Hermitian matrix, we cannot directly apply QSVT. As a workaround, we use contour integral and linear combination of unitaries (LCU) \cite{ChildsWiebe2012} to implement this matrix function. By Cauchy's formula we have
\begin{equation}
\label{eq:cauchy_formula}
    f(A) = \frac{1}{2\pi i}\int_{\Gamma} f(z)(z-A)^{-1}\dd z,
\end{equation}
where $\Gamma$ is a circle with radius $\beta$: $\Gamma=\{z=\beta e^{i\theta}:\theta\in\RR\}$, and $\|A\|<\beta$.
We need to discretize this integral in actual implementation. To do this we use the trapezoidal rule. If we use $K$ grid points on the unit circle we have
\begin{equation}
    \label{eq:trapezoidal_rule}
    f_K(A) = \frac{1}{K}\sum_{k=0}^{K-1}f(z_k) z_k (z_k-A)^{-1},
\end{equation}
for $z_k=\beta e^{i2\pi k/K}$.

We directly use the result in \cite[Theorem 18.1]{Trefethen2014}, which has been modified in Ref.~\cite{TakahiraOhashiEtAl2021} to a form more suitable for our purposes:
\begin{lem}{\cite[Theorem 18.1]{Trefethen2014}, \cite[Proposition 5]{TakahiraOhashiEtAl2021}}
\label{lem:trapezoidal_rule_error}
For a matrix function $f(A)$ and its discretized version $f_K(A)$ in \eqref{eq:trapezoidal_rule}, if $f(z)$ is analytic in the disc $\{z:|z|< R\}$, then 
\begin{equation}
    \label{eq:trapezoidal_rule_error}
    \|f(A)-f_K(A)\|\leq \frac{\sup_{|z|<R}|f(z)|}{1-\frac{\|A\|}{R}}\left(\frac{1}{1-(\frac{\|A\|}{\beta})^K}\left(\frac{\|A\|}{\beta}\right)^K+\frac{1}{1-(\frac{\beta}{R})^K}\left(\frac{\beta}{R}\right)^K\right).
\end{equation}
\end{lem}

\subsection{Block encoding of $\Xi$}
\label{sec:block_encoding_Xi}

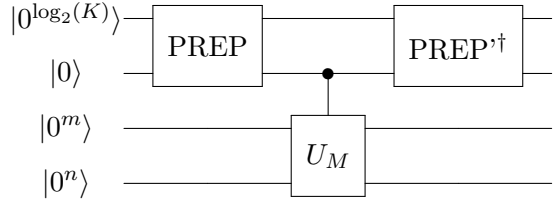
\begin{figure}
    \centerline{
    \Qcircuit @R=1em @C=1em {
     \ket{0^{\log_2(K)}}\quad\quad\quad\quad & \multigate{1}{\text{PREP}} & \qw & \multigate{1}{\text{PREP'}^\dagger}  & \qw \\
     \ket{0} \quad\quad\quad\quad & \ghost{\text{PREP}} & \ctrl{1} & \ghost{\text{PREP'}^\dagger} & \qw\\
     \ket{0^m} \quad\quad\quad\quad & \qw & \multigate{1}{U_M} & \qw & \qw\\
     \ket{0^n} \quad\quad\quad\quad & \qw & \ghost{U_M} & \qw & \qw \\
    }
    }
    \caption{ Quantum circuit of implementing the block encoding of the linear combination for each matrix inversion problem $(z_k - U)$. Here $\text{PREP}$ and $\text{PREP'}$ are defined as \eqref{eq:prep_coef_Xi}.
    % and the select oracle $\text{SEL}_\text{inv}$ is as defined in \cref{fig:circuit_contour2}.
    % Here $\text{HAD}$ represents the single qubit Hadamard gate.  
    }
    \label{fig:circuit_contour1}
\end{figure}

In this section we will only use the second to the fifth registers.
We first define 
\[
\ket{\mathrm{COEF}_k} = \frac{1}{\sqrt{\beta+\alpha}}(\sqrt{z_k}\ket{0}+\sqrt{\alpha}\ket{1}),\quad \ket{\mathrm{COEF}'_k} = \frac{1}{\sqrt{\beta+\alpha}}((\sqrt{z_k})^*\ket{0}-\sqrt{\alpha}\ket{1}).
\]
Because $|z_k|=\beta$ these are normalized quantum states. We can then implement using $\Or(K)$ gates unitaries $\mathrm{PREP}$ and $\mathrm{PREP}'$ such that
\begin{equation}
\label{eq:prep_coef_Xi}
    \mathrm{PREP}\ket{k}\ket{0} = \ket{k}\ket{\mathrm{COEF}_k},\quad 
    \mathrm{PREP}'\ket{k}\ket{0} = \ket{k}\ket{\mathrm{COEF}'_k}.
\end{equation}
These two unitaries will give us the coefficients $z_k$ we need in $\Xi$. Now we can construct a unitary $\mathrm{SEL}$:
\begin{equation}
    \mathrm{SEL} = (\mathrm{PREP}'\otimes I_{m+n})^{\dagger} (I_{\log_2(K)}\otimes cU_A) (\mathrm{PREP}\otimes I_{m+n}),
\end{equation}
where $cU_A=\ket{0}\bra{0}\otimes I+\ket{1}\bra{1}\otimes U_A$ is controlled-$U_A$. The circuit is given in \cref{fig:circuit_contour1}.
One can verify that 
\[
(I_{\log_2(K)}\otimes \bra{0}\otimes \bra{0^m}\otimes I_n)\mathrm{SEL} (I_{\log_2(K)}\otimes \ket{0}\otimes \ket{0^m}\otimes I_n) = \frac{1}{\beta+\alpha}\Xi.
\]
Therefore $\mathrm{SEL}$ is a $(\beta+\alpha,m+1,0)$-block encoding of $\Xi$. Note that $\mathrm{SEL}$ uses $U_A$ only once and $\Or(K)$ additional elementary gates.

\subsection{Block encoding of $\Xi^{-1}$}
\label{sec:block_encoding_Xi_inv}

Now that we have a block encoding of $\Xi$ we will directly apply Lemma \ref{lem:matrix_inversion} to get a block encoding of $\Xi^{-1}$. We need to upper bound $\|\Xi^{-1}\|$. First we have
\[
\|\Xi^{-1}\| = \max_{k}\|(z_k-A)^{-1}\|\leq \max_k |z_k|^{-1}\sum_{r=0}^{\infty}\left(\frac{\|A\|}{|z_k|}\right)^r=\max_k\frac{1}{|z_k|-\|A\|}\leq \alpha^{-1},
\]
where we have used the fact that $\|A\|\leq \alpha$ and $|z_k|=\beta=2\alpha$.
Using this result, and direct calculation of the other parameters in Lemma \ref{lem:matrix_inversion}, it can be seen that we can get a $(4/(3\alpha),m+2,\epsilon')$-block encoding of $\Xi^{-1}$ using $\Or(\log(\alpha^{-1}\epsilon'^{-1}))$ applications of (controlled-) $\mathrm{SEL}$ and its inverse, which translates to the same number of applications of (controlled-) $U_A$ and its inverse. We denote this block encoding of $\Xi^{-1}$ by $\mathrm{SEL}_{\mathrm{inv}}$. It acts on all $5$ registers of the circuit, and the first register contains the ancilla qubit needed for QSVT.

\subsection{Using LCU to block encode $e^A$}
\label{sec:use_LCU_block_encode_expA}

\begin{figure}
    \centerline{
    \Qcircuit @R=1em @C=1em {
    \ket{0} \quad\quad\quad\quad  & \qw & \multigate{4}{\text{SEL}_\text{inv}} & \qw & \qw \\
     \ket{0^{\log_2(K)}} \quad\quad\quad\quad & \gate{\text{PREP}_\text{int}} & \ghost{\text{SEL}_\text{inv}} & \gate{\text{PREP'}^\dagger_\text{int}}  & \qw \\
     \ket{0} \quad\quad\quad\quad & \qw & \ghost{\text{SEL}_\text{inv}} & \qw & \qw\\
     \ket{0^m} \quad\quad\quad\quad & \qw & \ghost{\text{SEL}_\text{inv}} & \qw & \qw\\
     \quad\quad\quad\quad & \qw & \ghost{\text{SEL}_\text{inv}} & \qw & \qw \\
    }
    }
    \caption{ Quantum circuit of implementing the block encoding of the linear combination of the preconditioned matrix inversions. Here $\text{PREP}_\text{int}$ and $\text{PREP'}_\text{int}$ are defined as \eqref{eq:prep_coef_expA} and the select oracle $\text{SEL}_\text{inv}$ is the standard QSVT circuit for the matrix function as discussed in \cref{{sec:block_encoding_Xi_inv}}.
    % \eqref{eq:sel_inv_def}.
    }
    \label{fig:circuit_contour3}
\end{figure}
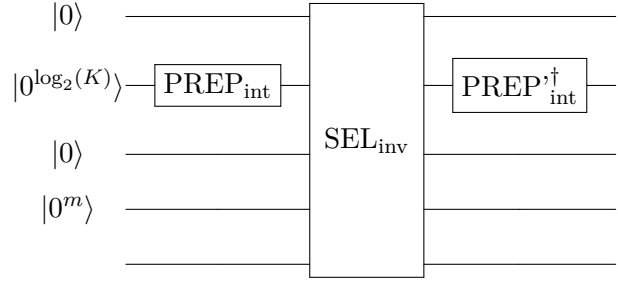

We have now come to the final step in which we construct a block encoding of $e^A$ using LCU. As discussed at the beginning of Section \ref{sec:ea_algorithm}, we need quantum states $\ket{\mathrm{COEF}_{\mathrm{int}}}$, $\ket{\mathrm{COEF}'_{\mathrm{int}}}$ to encode the coefficients. We specify the normalization factor below:
\begin{equation}
    \ket{\mathrm{COEF}_{\mathrm{int}}} = \frac{1}{\mathcal{A}K}\sum_k \sqrt{e^{z_k} z_k}\ket{k},\quad 
    \ket{\mathrm{COEF}'_{\mathrm{int}}} = \frac{1}{\mathcal{A}K} \sum_k (\sqrt{e^{z_k} z_k})^*\ket{k},
\end{equation}
where
\[
\mathcal{A} = \frac{1}{K}\sum_{k=0}^{K-1} |e^{z_k}z_k| = \frac{\beta}{K}\sum_{k=0}^{K-1} |e^{z_k}|.
\]
One can see that $0<\mathcal{A}\leq 2\alpha e^{2\alpha}$. Now we can construct unitaries $\mathrm{PREP}_{\mathrm{int}}$ and $\mathrm{PREP}'_{\mathrm{int}}$ that satisfy
\begin{equation} \label{eq:prep_coef_expA}
    \mathrm{PREP}_{\mathrm{int}}\ket{0^{\log_2(K)}} = \ket{\mathrm{COEF}_{\mathrm{int}}},\quad \mathrm{PREP}'_{\mathrm{int}}\ket{0^{\log_2(K)}} = \ket{\mathrm{COEF}'_{\mathrm{int}}}.
\end{equation}
These two unitaries each uses $\Or(K)$ elementary gates. Then we let
\begin{equation}\label{eq:sel_inv_def}
    \mathcal{U}_{\mathrm{LCU}} = (I_1\otimes \mathrm{PREP}'_{\mathrm{int}} \otimes I_{m+n+1})^{\dagger} \mathrm{SEL}_{\mathrm{inv}} (I_1\otimes \mathrm{PREP}_{\mathrm{int}} \otimes I_{m+n+1}).
\end{equation}
The circuit is given by \cref{fig:circuit_contour3}.
% \YT{need a circuit diagram here}
We can then verify that
\[
\begin{aligned}
& (\bra{0}\otimes \bra{0^{\log_2(K)}}\otimes \bra{0}\otimes \bra{0^m}\otimes I_n) \mathcal{U}_{\mathrm{LCU}} (\ket{0}\otimes \ket{0^{\log_2(K)}}\otimes \ket{0}\otimes \ket{0^m}\otimes I_n) \\
&= \frac{3\alpha}{4\mathcal{A}} f_{K}(A) + \Or(\alpha\epsilon') = \frac{3\alpha}{4\mathcal{A}} (f_{K}(A) + \Or(\mathcal{A}\epsilon')).
\end{aligned}
\]
Therefore $\mathcal{U}_{\mathrm{LCU}}$ is a $(4\mathcal{A}/(3\alpha),m+\log_2(K)+2,\Or(\mathcal{A} \epsilon'))$-block encoding of $f_K(A)$. By \eqref{eq:trapezoidal_rule_err_exp}, it is also a $(4\mathcal{A}/(3\alpha),m+\log_2(K)+2,\Or(\mathcal{A} \epsilon'+e^{4\alpha}2^{-K}))$-block encoding of $f(A)$. In order to get the error to be below $\epsilon$ we can choose $\epsilon'=\Theta(\epsilon/\mathcal{A})$, and $K=\Or(\alpha+\log(\epsilon^{-1}))$. In the whole process (controlled-) $U_A$ and its inverse are used $\Or(\log(\alpha^{-1}\epsilon'^{-1}))=\Or(\log(\mathcal{A}\alpha^{-1}\epsilon^{-1}))=\Or(\alpha+\log(\epsilon^{-1}))$ times. In sum, we get \cref{lem:block_encoding_expA}. 
% We summarize the result as follows

\end{document}